\documentclass[11pt,preprint]{aastex}
\usepackage{graphicx}
\usepackage{amsmath}

\newcommand{\w}{\it WISE}

\begin{document}

\shorttitle{}
\shortauthors{Azimlu M.; Martinez-Galarza J.R. ; Muench A. A.}

\title{A WISE Census of Young Stellar Objects in Perseus OB2 Association}

\author{Mohaddesseh Azimlu \altaffilmark{1};  J. Rafael Martinez-Galarza \altaffilmark{1};  August  A., Muench\altaffilmark{1} }
\altaffiltext{1}{Harvard-Smithsonian Center for Astrophysics, Cambridge, MA 02138}

\begin{abstract}

We have performed a {\it WISE} (Wide-Field Infrared Survey Explorer)  based study to identify and characterize young stellar objects (YSOs) in $12^{\circ} \times 12^{\circ}$ Perseus OB~2 association.   Spectral energy distribution (SED) slope in range of  $3.4-12 \mu$m and a 5$\sigma$ selection criteria were used to select our initial sample. Further manual inspection reduced our final catalog to 156 known and 119 YSO candidate.  The spatial distribution of newly found YSOs all over the field  shows an older generation of star formation which most of its massive members have evolved into main sequence stars.  In contrast, the majority of younger members lie within the Perseus molecular cloud and currently active star forming clusters such as NGC~1333 and IC~348. We also  identified additional 66 point sources  which passed YSO selection criteria but are likely AGB stars. However their spatial distribution suggests that they may contain a fraction of the YSOs.  Comparing our results with the commonly used color-color selections,  we found that while color selection method fails in picking up bright but evolved weak disks, our SED fitting method can identify such sources, including transitional disks. In addition we have less contamination with background sources such as galaxies, but in a price of loosing fainter ($J_{mag}>12$) YSOs. Finally we employed a Bayesian Monte Carlo SED  fitting method  to determine the characteristics of each YSO candidate. Distribution of SED slopes and model driven  age and mass confirms separated YSO populations with suggested  three age groups of younger than 1~Myr old, $1-5$~Myr old, and older than 5~Myrs  which agrees with the age of Per OB~2 association and currently star forming sites within the cloud.

\end{abstract}

\section{INTRODUCTION} 
\label{introduction}

Perseus is one of the most interesting and well studied star forming complexes, containing massive stars (Per OB2 association), young stellar clusters (e.g. IC348 and NGC1333) and several dense star forming cores within Perseus molecular cloud (e.g. \citet{kirk2006,ridge2006,rosolowsky2008}).
Perseus OB2 is the second closest OB association to the Sun (after  Scorpius-Centaurus association) at a distance of $\sim 300~$pc. \citet{dezeeuw99} estimated that Per OB2 is approximately 6~Myr old while other studies also suggest different generation of star formation with an age less than 15 Myr (e.g. \citet{bally2008} and references therein).

Per OB2 contains both massive and intermediate mass stars while low mass to intermediate mass stars are currently forming within the Perseus molecular cloud
 (e.g. \citet{enoch2009,arce2010,sadavoy2014}).
  Therefore identifying the members of the association and determining their characteristics such as  their age and mass  will provide critical information on the  process of different generations of star formation in this region. 

We constructed a panchromatic census of young stellar objects (YSOs) in Per OB2 using Wide-field Infrared Survey Explorer ($\w$) point source catalogue. We also employed all available optical and infrared all sky surveys including $\it Spitzer$, 2MASS and SDSS to construct spectral energy distribution (SED) of each source in order to determine its physical parameters. We have performed a $\w$ based survey to identify young stellar objects throughout a $12^{\circ} \times 12^{\circ}$ area covering the Per OB2 association.

The Perseus molecular cloud covers a $2^{\circ}\times6^{\circ}$  ($\sim10\times30$ pc)  area and was observed with the  {\it Spitzer} space observatory in both IRAC ($3.6-8.0~\mu$m) and MIPS ($24-160~\mu$m) as part of ``From Molecular Cores to Planet-Forming Disks" (c2d) Legacy Survey \citep{jorgensen2006,rebull2007}. We selected the c2d covered region as footprint, but did not limit our survey  to the c2d area. c2d only covers very young star forming regions such as IC348 and NGC1333 clusters. To study the  complete population of Perseus OB2 association  we need to track all the members of  the original ensemble, in particular the entire  OB association which has  blown up the initial molecular cloud and other young stellar objects, now spread into a larger area.  
Assuming an average speed of $1-2$~km~s$^{-1}$ for stars, members  can travel more than 20~pc in 10~Myrs in each direction which converts to an extended area of $\sim 10^\circ$  from the cloud center at $250-300$~pc distance.

To be complete, we selected a  $12^\circ \times 12^\circ$ region ($50^{\circ}<$RA$<64^{\circ}$ and $26^{\circ}<$Dec$<38^{\circ}$) around the Perseus c2d footprint which completely covers the Perseus OB2 association. This region slightly  overlaps with Taurus star forming region in South-West and may contain older members of Pleiades in South as well. Figure \ref{regions} shows the c2d footprint, and the area selected around it in our survey. The Blue dots present $\w$ point sources. 
Unfortunately we do not have the spectral or dynamical data for each source, hence cannot confirm if each individual source in our final catalog is associated with the Per OB2 or the molecular cloud, but the luminosity and other parameters matches with the properties we expect for YSOs at the estimated distances. In addition, we are looking out of  the Galactic plane and no other known background star forming region is along the line of sight. Foreground young stars, if any, would be much brighter than our criteria.

Section \ref{archive_data} explains the data collection and data reduction procedures. In section \ref{selection} we describe our YSO candidate selection method and compare the results with other YSO identification schemes. 
Section \ref{candidate_analysis} contains the analysis of  our sample of YSO candidates. In Section \ref{modeling} we describe our SED fitting modeling and the results. We summarize this study in Section \ref{summary}.

\section{Archival Data and Data Reduction}
\label{archive_data}
To identify young stars we established a multi-wavelength database of point sources in Per  OB2 star forming region, using all available large sky surveys. New YSO candidates then were selected through evidence for infrared excess in their spectral energy distributions (SED); specifically, we used power-law fits to the longer infrared wavelengths covered by the $\w$ all sky survey. In section \ref{selection} we detail the candidate selection procedure.

To summarize, $\w$ is an all sky survey in four  infrared wavebands  3.4, 4.6, 12, and 22 $\mu$m (hereafter W1, W2, W3 and W4) with an angular resolution of 6-6.5$''$ for W1 to W3 and  12.0$''$ for W4. The $\w$ depth is not homogenous, but  achieved 5$\sigma$ point source sensitivities of 0.068, 0.098, 0.86 and 5.4 mJy (16.6, 15.6, 11.3, 8.0 Vega mag) in unconfused regions in the ecliptic plane for  the four bands respectively. Our point source database was initially constructed using the $\w$ preliminary data release \citep{wright2010}, but we upgraded to the final source release in March 2012. The $12^\circ \times 12^\circ$ selected region in  our census of Per OB2 includes more than 1.6 million $\w$ point sources (Figure \ref{regions}).

After an initial inspection we realized that W4 sensitivity is very low for faint sources and in particular any W4 profile magnitude $>$8 should be considered with caution. We therefore limited our search for infrared excess to the range of $3-12$~$\mu$m, and restricted our source list to have S/N$>$7 in the three corresponding W1, W2, and W3 bands.  This criteria decreased the number of candidates to 65,662. We did not use the $\w$ catalog flags to further filter our candidate list because we found the flags to be unreliable; in particular many sources were not flagged as extended or questionable when in fact we found that no point source actually existed.  Instead all of the final candidates were visually inspected to remove contaminated or false detections.

Our primary goal was the building of composite spectral energy distributions (SED) using archival data products.  We initially considered $\w$ as the base catalog to astrometrically match with other data sets, but the $\w$ beam is significantly larger than other relevant all sky surveys. We also decided to use a near-IR band as a tracer of the stellar spectral type.  Therefore we adopted the 2MASS \citep{2mass} catalog and coordinates as the base catalog. Requiring our sources to have a 2MASS companion reduced the size of our sample to 48,692.

\subsection{Additional ancillary data}
\begin{figure}
\begin{center}
\includegraphics[width=0.7\columnwidth]{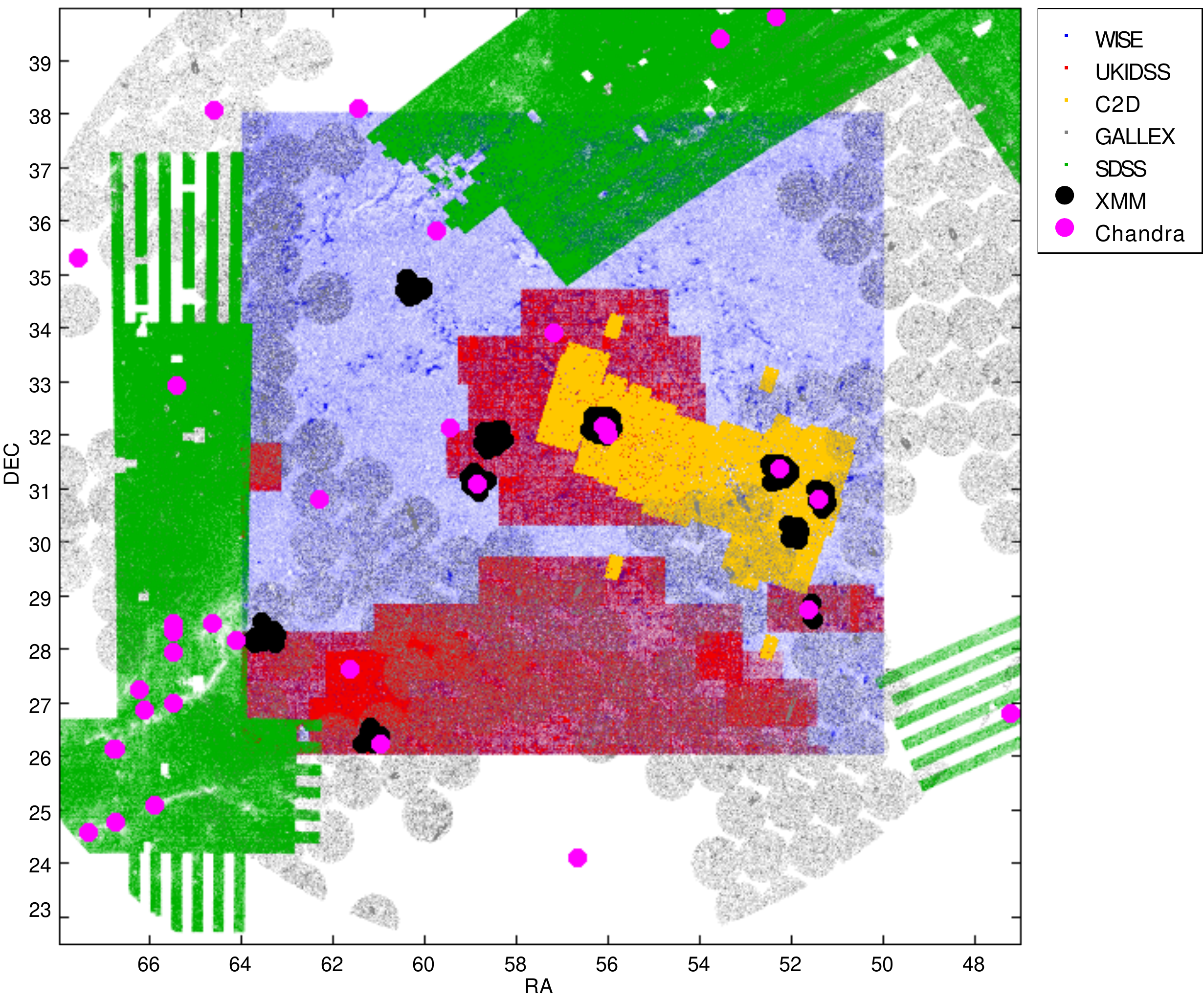}
\caption{\label{regions} Surveyed Region and its coverage by a sample of surveys: WISE (blue), Spitzer (c2d) (yellow), SDSS (green), UKIDSS (red), Gallex (grey), XMM (black) and  Chandra (magenta). }
\end{center}
\end{figure}

The Perseus star forming region is well covered by many large sky surveys (see Figure \ref{regions}) from  infrared to X-ray including: 2MASS, UKIRT Infrared Deep Sky Survey (UKIDSS),  {\it Spitzer} (e.g., c2d, \citet{c2d2003}), the InfraRed Imaging Surveyor (AKARI) \citep{akari2007}, Sloan Digital Sky Survey (SDSS, DR8+), the AAVSO Photometric All-Sky Survey (APASS, Release 7)\footnote{http://www.aavso.org/apass}, Chandra X-Ray Observatory and X-ray Multi-Mirror Mission (XMM-Newton) as well as catalogs based on older scanned photographic plates (US Naval Observatory: USNO-B1, UCAC3) and the USNO-B1/2MASS re-reduction, PPMXL \citep{ppmxl2010}.  

\begin{figure}
\begin{center}
\includegraphics[width=0.7\columnwidth]{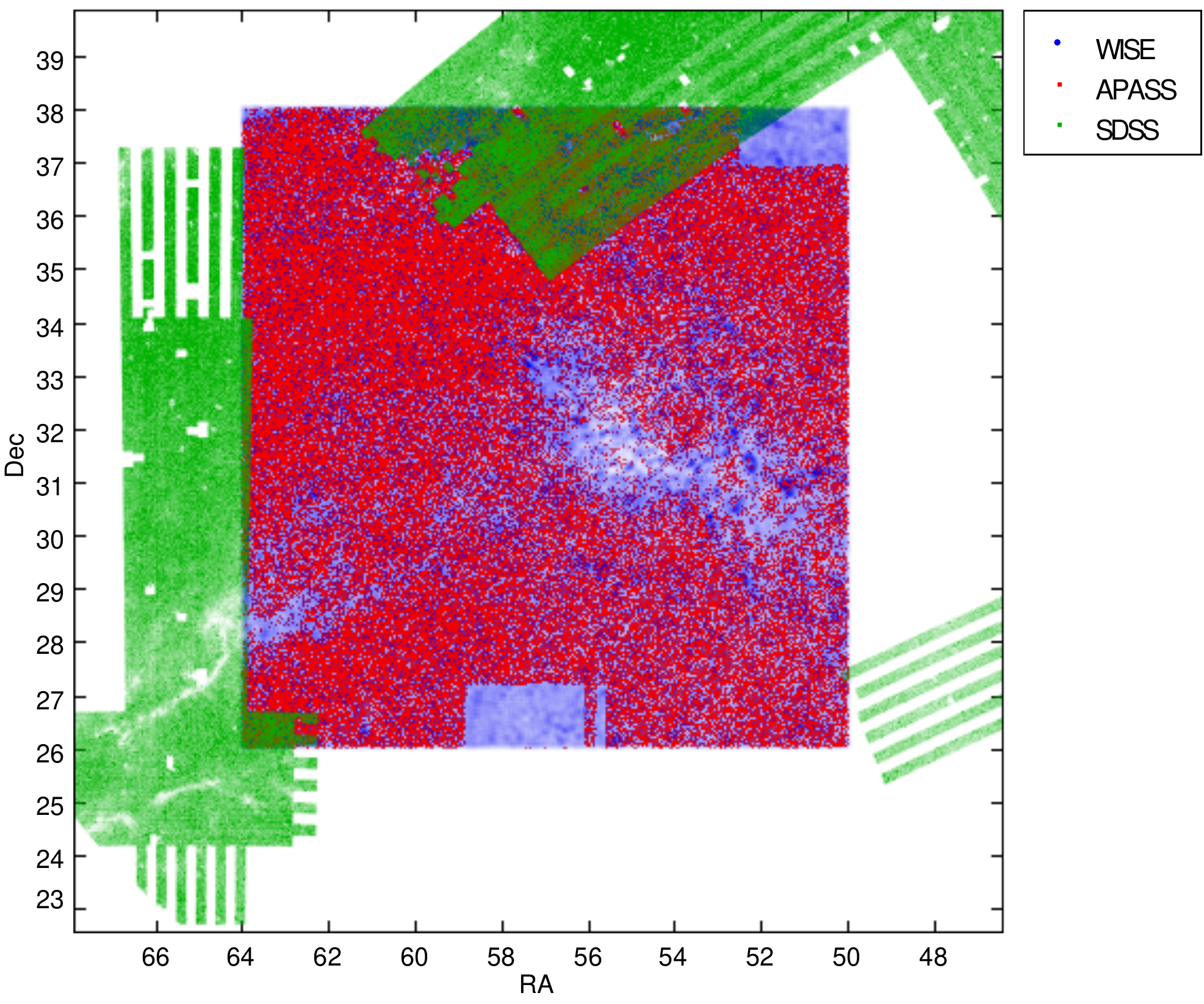}
\caption{\label{optical} Field coverage by SDSS, APASS and WISE}
\end{center}
\end{figure}

SDSS and APASS catalog measurements are adopted at optical wavelengths if available; otherwise, we have used PPMXL cataloged photometry for B and R bands.  Figure \ref{optical} presents APASS and SDSS coverage of our surveyed region, showing how APASS provided nearly complete, deep optical coverage. While PPMXL covers the entire area, it is intrinsically based on USNO-b1 and contains much larger errors than APASS. Because PPMXL contains corrected positions, proper motions and optical-infrared matches of stars, we used it to break degeneracies for multiple matches between different catalogs. X-ray observations were very concentrated toward the known star forming regions, and did not cover the entire region. We benefited from archival X-ray data in publications to identify known YSOs. Because of its limited spatial scope (Figure \ref{regions}) we neglected the UKIDSS catalog entirely. AKARI flux measurements were consistent with IRAC and MIPS observations, but only a few of our candidates had high quality AKARI data, therefore this catalog was not used in final SED fitting. 

There are several narrow field catalogs at different wavebands in Perseus region, but to keep a high degree of uniformity in source detection and flux calibration we do not account for such observations. We further downloaded all SIMBAD sources in the survey area to identify previously known YSOs or census contaminating sources such as planetary nebulae, background galaxies or other type of known evolved stars.

We employed the {\it CDS X-match} Service \footnote{http://cdsxmatch.u-strasbg.fr/xmatch,  Strasbourg Astronomical Data Center} to match different catalogs and investigate the astrometric uncertainties. This service employs the coordinate errors to  calculate the probability density of two sources from two different catalogs. The probability density  is given by convolution between two Gaussian distributions around each source \citet{pineau2011}. We  consider a completeness of 99.7 \% at 3$\sigma$ criterion (only 0.3\% of the counterparts could be missed).

\section{IDENTIFYING DISK CANDIDATES}
\label{selection}

We adopt the shape of the spectral energy distribution (SED) to identify sources with excess infrared emission as these are probable young stars with passively reradiating circumstellar disks \citep{lada87}. 
Following on the {\it Spitzer} based SED analysis of \citet{lada06} and  \citet{muench07}, we used a simple power-law, least-squares fit of the $3-12~\mu$m portion of the YSO's SEDs as traced by the W1, W2, and W3 $\w$ bands (3.4, 4.6 and 12~$\mu$m, respectively).

While \citet{muench07} showed that the {\it Spitzer} $3-8~\mu$m index was relatively insensitive to extinction, we acknowledge that including 12 microns into the SED slope fit will result in larger intrinsic dispersions in the slope distribution because of variations in the silicate  feature. Unreddened classical T-Tauri stars with $SiO_4$ in emission will have very large slopes. We do not anticipate silicate absorption due to A$_V$ to ``remove" a source from our sample, by lowering the measured $3-12$ micron slope. 

\begin{figure}
\begin{center}
\includegraphics[width=0.7\columnwidth]{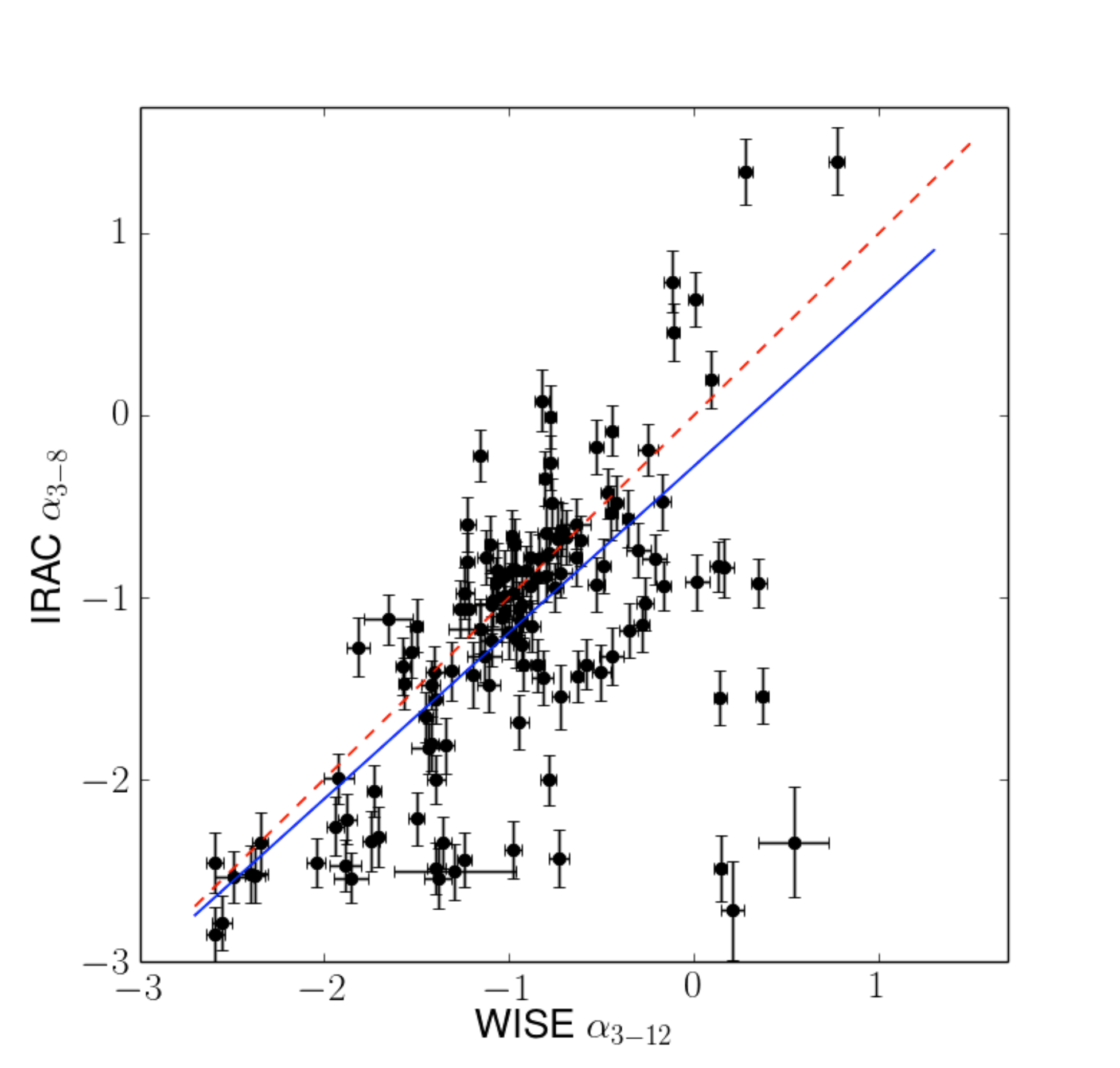}
\caption{\label{irac_wise} IRAC $\alpha_{3-8}$ vs. WISE $\alpha_{3-12}$.  Solid blue line presents a linear least square fit with slope of $0.92\pm0.2$ (excluding three  data points at right bottom corner with poor IRAC photometry) and red dashed line presents  equal values.}
\end{center}
\end{figure}

Figure \ref{irac_wise} presents a comparison between IRAC $\alpha_{3-8}$ and WISE $\alpha_{3-12}$ for 135 sources which have both data set in our sample. Solid line shows a least square fit (with slope of $0.92\pm0.2$, excluding three  data points at right bottom corner with poor IRAC photometry)  and dashed line presents equal values. There is a large scatter due to different wavelength,  band width for each wavelength, beam size and sensitivity for Spitzer and WISE. A selection bias is involved as well. In Figure \ref{irac_wise}  all IRAC data is included if available, but only $\w$  with  SNR$>$7 sources are selected in our initial sample. In general  for $\alpha_{3-8} <  -2$, we have found larger slopes for WISE     $\alpha_{3-12}$,  where anemic disks and stars are located (discussed further in section \ref{color-cuts}).  IRAC and $\w$ slopes are not expected  to  match completely, for this reason  we are not reporting classification for individual objects based on categories defined in \citet{muench07}.

The 22~$\mu$m was not considered in the initial slope-fit for two reasons: a) to keep detection uniformity for  sources which have not been detected or have very low SNR in W4  and b) to avoid the overestimation of slopes due to poor sensitivity of W4  for faint sources with W4~$> 8$. However 22~$\mu$m flux, when regarded as real, is considered  in final complete SED modeling to study the disk characteristics.

In the first attempt we filtered candidates based on $\w$ flags. 
We found a few strong YSO candidates which were located  nearby saturated stars or within nebulous regions, had been flagged as poor detection or photometry due to contamination. On the other hand in visual inspection we found sources with very poor detection   in cloudy areas which had high S/N fluxes reported with $\w$  high photometry quality flags but they  did not survive  our visual or point source profile  inspections. Finally  we decided to ignore the $\w$  flags at this stage and reconsider them after  manual inspection.

We considered all sources with S/N$>$7 in first 3 $\w$ bands  in our candidate pool and fitted a power-law to each source. Figure \ref{selection} shows the calculated slopes versus 2MASS {\it J}~magnitude for 48,692 sources with S/N$>$7 in first 3 $\w$ bands in grey dots. To select the sources with excess emission we looked at the slope distributions in bins of 0.5 $J$~magnitude and fitted a Gaussian to the  slopes distribution within each interval.  $J$ band  is the shortest  wavelength that we have uniform data for, and is thus the least contaminated wavelength by circumstellar disk emission, which makes it the best tracer of photospheric luminosity. 

The selection criteria can be considered to be a convolution of variation in the intrinsic (photospheric) color as a function of apparent magnitude and the $\w$ sensitivity function. 
We did not consider disentangling these two functions as a useful exercise, and instead posited all sources with an excess greater than 5 sigma from the typical color as evidence for infrared excess. 

The black locus in Figure \ref{selection} shows a 5$\sigma$ limit above the peak of distribution for each bin.  All sources above the locus have significant excess in slope (and hence in IR emission) and are considered  as  candidates with excess. On average the slope increases with the $J_{mag}$ after magnitude 12. Fainter objects in $J$ band are either embedded sources, or background galaxies. Hence, faint YSOs are missed in this selection, but in return the final sample is less contaminated with background galaxies.

The $J_{mag}$  error is not considered in calculating the locus, but it will not dramatically change the results. The maximum error in $J_{mag}$  in our entire data set   is only 1.7$\%$. The locus has been drawn based on $5\sigma$ above the mean population in each $J_{mag}$  bin which means a confidence of 99.999943$\%$ of IR excess. Considering the largest error for $J_{mag}$ around the locus, selected candidates  are still more than $4\sigma$  above the mean, which translates into a  99.994$\%$ confidence in IR excess.  Therefore by neglecting $J_{mag}$ error in our selection procedure we are not maintaining the $5\sigma$ confidence for all selected YSO candidates, but all of the selected candidates have statistically significant  IR excess.

669 sources met this criteria.  However we are aware that the sample may contain none-stellar objects, contamination  and false detections. In next step we filter candidate by manually inspecting them in optical and infrared images.

\subsection{Visual inspection}
\label{inspection}

$\w$ and 2MASS and if available SDSS images of each candidate which passed the locus selection were manually inspected. Visual inspection revealed that half of the selected sources cannot be considered resolved point sources. The first group were rejected as being identified as extended objects like galaxies and the second portion could not be identified as distinguished point sources  in nebulous regions in a combination of at least any three $\w$ bands. We performed the eye inspection in three rounds and categorized the sources as contaminated, saturated, extended and not resolved.  All extended objects were identified as galaxies and PNe in further investigations and were rejected. Candidates  that could not be resolved as point sources in at least any three $\w$ bands needed to be inspected for the luminosity profile. Contaminated, saturated and not resolved sources were accepted and back to the list if they were resolved in 2MASS and if luminosity profile presented a  point source in at least three $\w$ bands. 

354 out of 669 candidates survived the visual inspection. These final candidates are presented by open red circles (new candidates) and blue triangles (known candidates) in Figure \ref{selection}.
The known AGB stars are presented with black dots. The sources in our list which are less likely to be YSOs and closer to the AGB stars are noted by black crosses. They are separated from the YSO candidates employing magnitude diagrams and  will be discussed  in more details  in section \ref{candidate_analysis}. The known OB stars are also shown as yellow dots. The only  OB star which is above the locus is $BD +30 540$. \citet{rebull2007} identified this YSO as a class I source in their {\it Spitzer} c2d survey. It is a B8V star and  have an estimated mass of 2.6 M$_\odot$.

\begin{figure}
\begin{center}
\includegraphics[width=0.7\columnwidth]{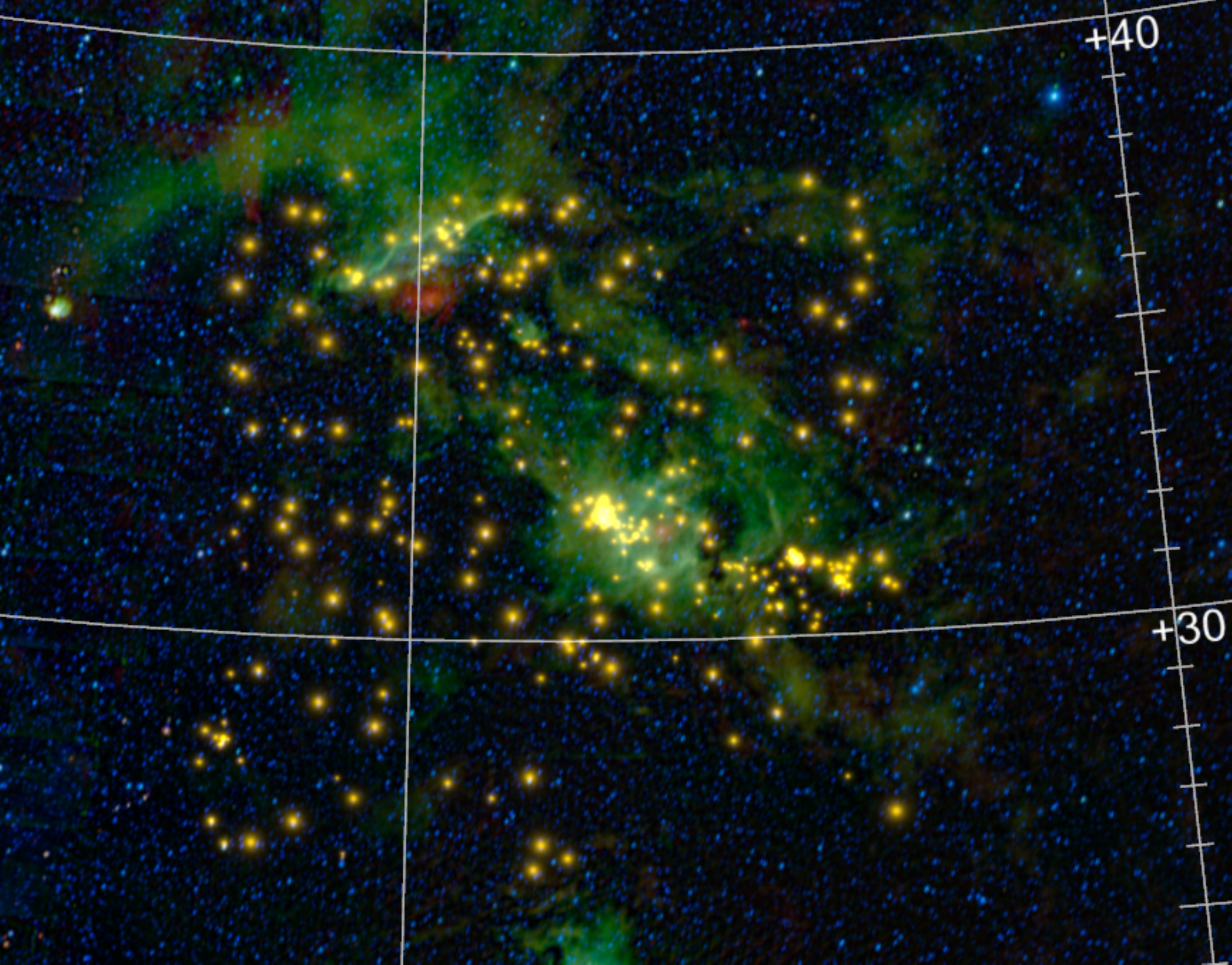}
\caption{\label{wwt}
YSO candidates overlaid on multi band WISE image of Perseus region ($4.6 \mu$m in blue, $12 \mu$m in green and $22 \mu$m in red).  The brightness of  each source presents its $\alpha_{3.4-12}$ slope. Larger slope presumably means stronger disks. Picture created using Microsoft Worldwide Telescope ($http://worldwidetelescope.org$)}
\end{center}
\end{figure}

Figure \ref{wwt} presents the spatial distribution of 354 final candidates on a $\w$ multi-color image. The brightness of each source is proportional to the the W1-3 slope; sources with larger slopes, which presumably have stronger disks, appear brighter in Figure \ref{wwt}.

\subsection{Comparing classification schemes}
\label{classification}

\begin{figure}
\begin{center}
\includegraphics[width=0.7\columnwidth]{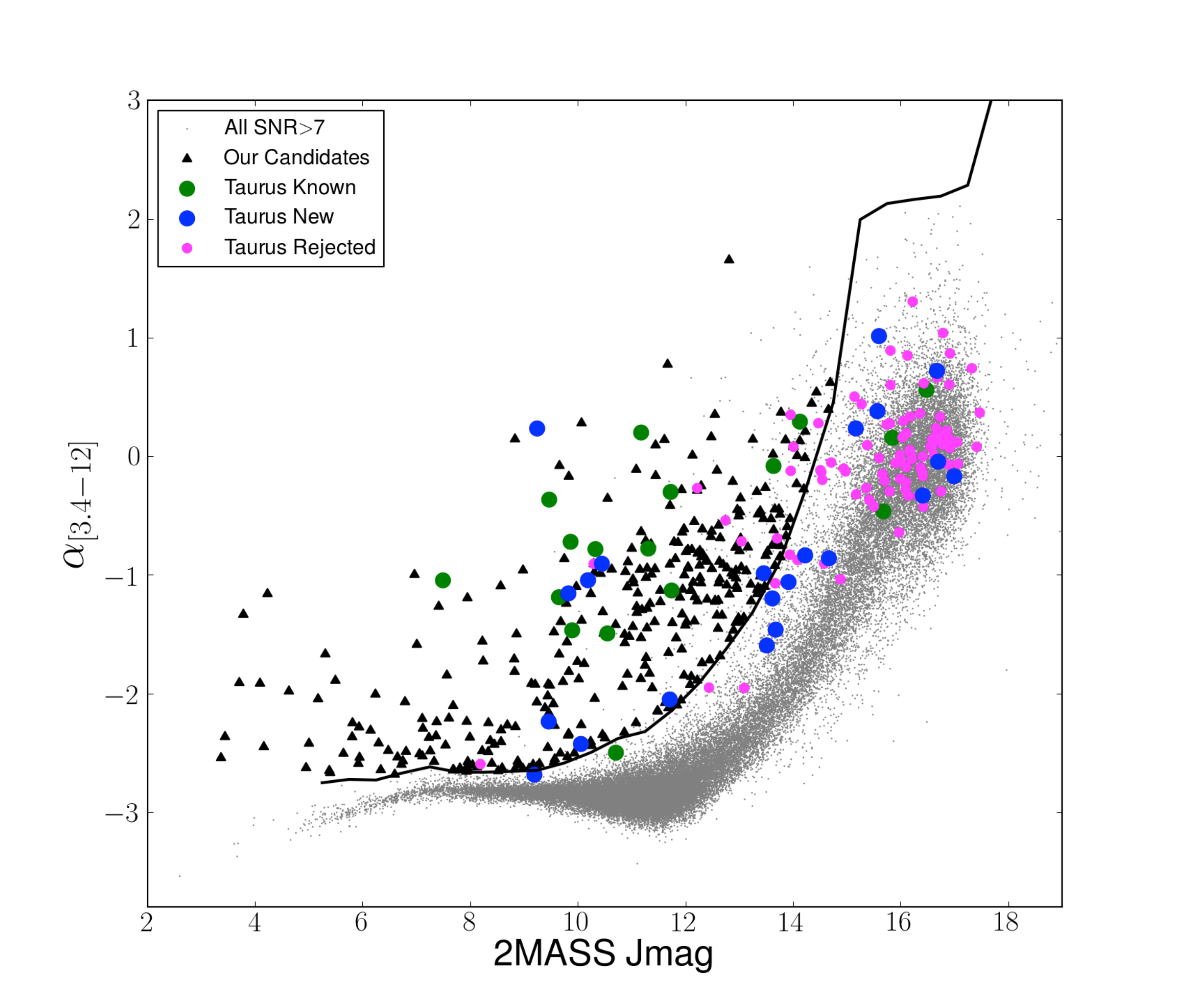}
\caption{\label{rebull} 
Comparing our results with \citet{rebull2011} study of Taurus-Auriga star forming region. Black triangles present our sample of 354 candidates in Perseus. Green, blue and pink dots presents known YSO candidates, new YSO candidates and rejected candidates respectively from Rebull et.al.
}
\end{center}
\end{figure}

\citet{rebull2011} identified new YSO candidates in Taurus-Auriga region using preliminary $\w$ data.  Nearly $25\%$ of our survey area is covered in their study, therefore we looked for common candidates to evaluate our detection method  accuracy. \citet{rebull2011} report 94 new YSO candidates in Taurus-Auriga region,  22 of which lie within our survey area. Astrometry and photometry of more than $25\%$ of sources have changed in $\w$ all-sky release, especially for bright sources in our list.  Therefore preliminary and all sky  release $\w$ catalogs do not contain identical IDs for all sources. Instead of matching IDs we performed a position match. Some photometries in the new $\w$ catalog also have sightly changed, but they are consistent enough with the preliminarily data  to help selecting  the right match. Rebull et al.  catalog is plotted over our selection diagram in Figure \ref{rebull}. Our complete catalog of 354 sources are shown in black triangles. Rebull et al. known YSOs are shown with green dots and their new YSO candidates are shown with blue dots. They have also rejected many sources (shown in magenta dots) after various inspections.  We have identified only 8 of Rebull's new YSO candidates in our field. 14 unidentified sources are almost fainter than 12 magnitude in 2MASS $J$ band and despite color excess they lie below our selection criteria.

We also have 17 of Rebull's {\it known} candidates in our field. 14 have been classified as YSO candidates in our sample, one does not have 2MASS match and therefore not selected in our sample and two  are below our selection criteria.

Rebull et al. also have 129 rejected candidates in our field which passed their color cuts in first place.
82 of their rejected sources (as background galaxies) match with our initial sample but  lie below our selection criteria. 35 do not have 2MASS point source match and therefore not selected in our list. 
We have 8 common sources in our visually rejected list which they have identified as confused, galaxy, PNe or foreground/background star. Noticeably 7 of their rejected sources do not have a match in new WISE within 1$''$, and 4 do not have a match within 2$''$ , therefore we cannot comment about these sources.

\subsection{Comparing with color-cuts}
\label{color-cuts}

We compared our candidates with the color-cut spaces described by  \citet{koenig2012}. The color cuts are defined in multiple color spaces based on known different types of YSOs and other contaminant objects  such as AGNs, galaxies with PAH-feature emissions and shock emission blobs. \citet{koenig2012}  selections are employing the results from  {\it Spitzer} surveys to identify various  objects with color excess \citep{guter2009,rebull2010} based on where they statistically lie on $\w+2MASS$ color-color spaces. 
Similar to $\alpha$ selection technique, color-cut  criteria in general can be  adjusted to extract weaker disks with smaller IR excess or exclude them.  In this section we compare our results with criteria defined by \citet{koenig2012} which is frequently adopted  in similar studies. 

Following \citet{koenig2012}, no AGNs were identified in our sample of candidates, but three sources (J03504369+3507088, J04022740+3057450, J03504333+3346024) might be considered as PAH emitting/star forming galaxies. These objects  have no match in the literature. 
They are out of the {\it Spitzer} field and not covered by SDSS, therefore their identity cannot be determined by the spectrum. 
It worths to mention that Koenig color-cuts are defined based on the $\w$ preliminary data, while photometry of $~25\%$ of sources have had slight changes in all sky release. That may partially affect their selection criteria at the borders.

\citet{koenig2012} also used color-cuts to categorize their YSOs into different class types.
\citet{muench07} (appendix A) explored the influence of dust extinction on $\alpha$ detection technique and presented that IRAC SED slope  for a diskless K0 star requires  A$_V > 100$ to inflect into a positive slope and even  A$_V > 200$ cannot inflect  $\alpha_{5.8-24}$ of background stars into positive.  Such large column densities within typical molecular clouds occur only in protosetllar envelop of  embedded YSOs. But smaller values of dust extinction also may affect $\alpha$ at the  classification of ClassI/ClassII candidates.  In contrast  mid-IR color cut techniques can be defined such that they minimize extinction-induced bias.  The goal of this study is only identifying YSOs and not classifying them. We present  the classification only to compare the slope-fitting and the color cut methods and will not report the class types for our final candidates.

According to \citet{muench07}, $\alpha_{IRAC3-8}=-2.66$ corresponds to the predicted slope of M0 star photosphere. Accepting  a typical precision of SED fits, they characterized sources with $-2.56<\alpha<-1.80$ as ``anemic disks". Transitions disks, disks with inner holes and heavily depleted optically thin disks may locate in this category. Objects with $\alpha_{IRAC3-8}<-2.56$ are considered stars, and objects with $\alpha_{IRAC3-8}>-1.80$ are accepted as IR excess objects ($-1.80<\alpha_{IRAC3-8}<-0.5$ as Class~II and $-0.5<\alpha_{IRAC3-8}$ as Class~I).

\begin{figure}
\begin{center}
\includegraphics[width=0.7\columnwidth]{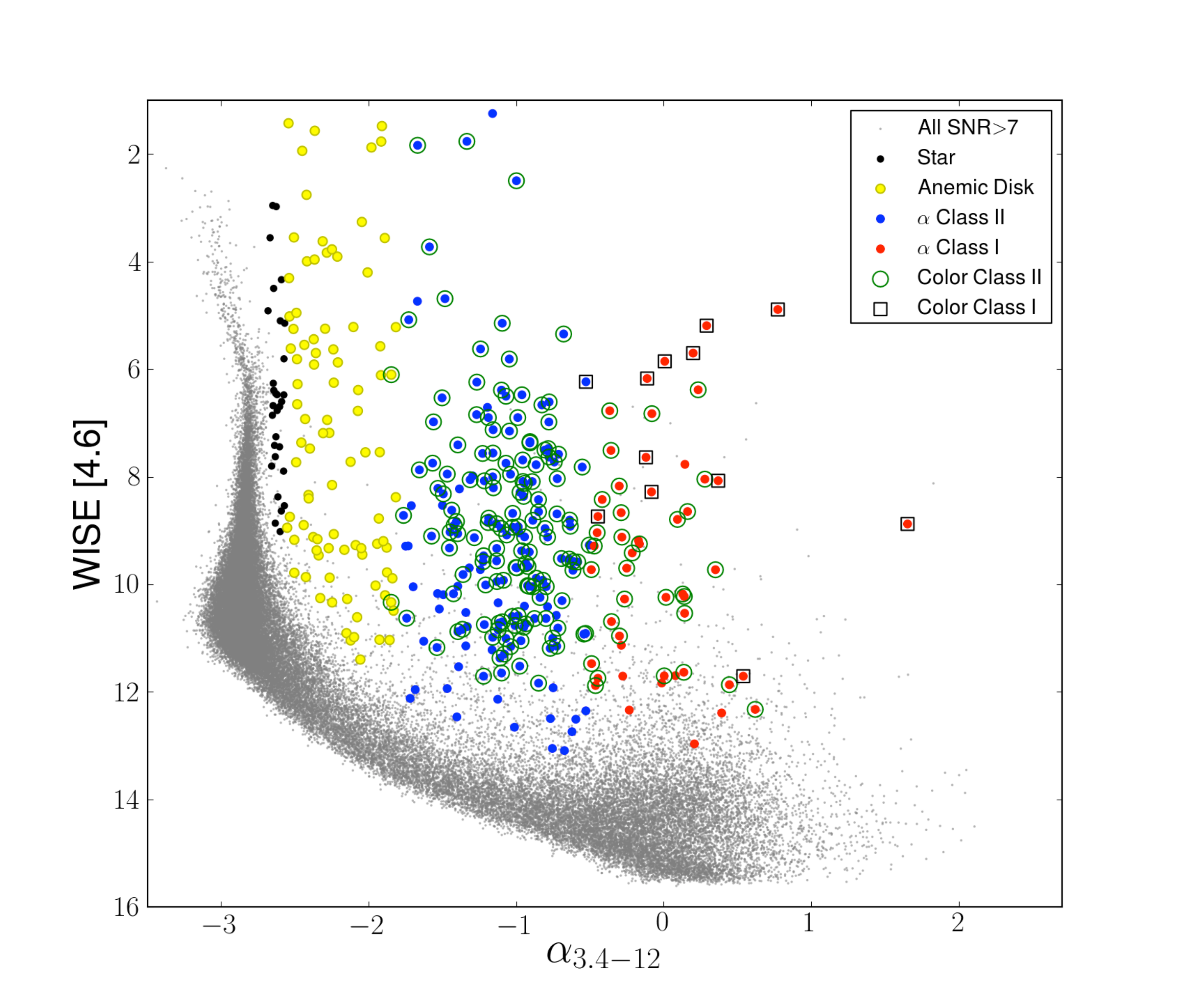}
\caption{\label{alpha_compare}
Comparing disk classification  based on the slopes ($\alpha$) and Koenig color cuts in one plot.   Black dots show $\alpha-$stars ($\alpha<-2.56$) and yellow dots show $\alpha-$anemic disks ($-2.56<\alpha<-1.8$). Blue dots show Class~II ($-1.8<\alpha<-0.5$) and red dots present Class~I $\alpha$-disks ($\alpha>-0.5$). Selected by $[3.4]-[4.6]$ and $[4.6]-[12]$ color-color diagrams, green open circles present Class~II and black open squares  present Class~I sources. }
\end{center}
\end{figure}

Figure \ref{alpha_compare} compares the disk classification by \citet{muench07} based on the slopes ($\alpha$) and Koenig color cuts in one plot.  
Grey dots show all 48,692 $\w$ sources in our field with SNR$>$7 for first three $\w$ bands. Black dots are $\alpha-$stars. Yellow dots show $\alpha-$anemic disks. Blue dots present  Class~II  and red dots present Class~I $\alpha$-disks.
Based on Koenig classifications, green open circles present Class~II and black open squares  present Class~I sources (selected in $[3.4]-[4.6]$ and $[4.6]-[12]$ color-color diagrams). There is a  good agreement in disk identification type between two methods.  However, color-cuts select less Class~I protostars and many $\alpha-$Class~I sources are identified as Class~II by color-cuts.

We should remind  that $\alpha$ selection fails to detect faint sources ($J_{mag} > 12$) even if they have noticeable color excess. On the other hand color-cut method misses the weak or evolved disks, even if they are bright sources. For example anemic disks have rarely been distinguished by color-cuts.  The comparison between the selected YSOs in \citet{rebull2011} and this work in Figure \ref{rebull} also presents how the faint sources even  with strong disks have been missed in $\alpha$ selection, and how the bright sources with weak disks have been missed by color-cut selection.
We suggest that to have a complete sample of faint to bright and weak to strong disks, both methods needed to be considered. 

To summarize, we establish our working sample of candidate young stellar objects as selected by the spectral energy distribution  power-law  slope in range of $3.4-12~\mu$m from $\w$ all sky point source catalog. Using this method we can identify the weak and more evolved circumstellar disks if they are hosted by a bright protostar. These type of disks are barely identified with color-color diagrams.  In contrast $\alpha_{3-12}$ selection  miss fainter sources, in particular those with  J$_{mag}>$12 sometimes even if they have strong disks. Color-color diagrams are capable to select many fainter sources but the results are  contaminated  ($\sim 67\%$ for example, in Rebull's  primary selected sources) with background galaxies and other fake detections compared to $\alpha$ selection method. For example  in pool of 1014 potential YSOs identified by \citet{rebull2011}, 686 were rejected as known galaxies or objects to be likely background galaxies in their final list.

\section{CANDIDATE ANALYSIS}
\label{candidate_analysis}
\subsection{Known YSOs}
\label{known_ysos}

We cross matched our candidates with the SIMBAD database to identify known YSOs and  possible contaminants.  Among 354 candidates passed our manual inspection as resolved point sources, J035523.11+310245.0 is a known massive X-ray binary and  was removed.  12 other sources  are known AGBs, Mira variables or other type of known evolved stars. 76 have matches with  previously known YSOs or T-tau stars  while 80 others are listed as YSO or T-tau candidates in SIMBAD. All these sources are categorized  as known YSOs in our final catalog and listed in Table \ref{tbl_known}.  185 sources remain in our list as  new YSO candidates, but  66 of them  identified to be more likely evolved dusty stars and were separated from final new YSO candidates list (Table \ref{tblAGBs}.  We will discuss them in section \ref{dusty_evolved_stars}. Finally 119 sources survived as new YSO candidates in our final catalog and they are listed in Table \ref{tbl_new}. The errors (range of Lower limit and upper limit for each parameters) are provided in online tables.

\subsection{Dusty evolved stars}
\label{dusty_evolved_stars}

\begin{figure}
\begin{center}
\plottwo{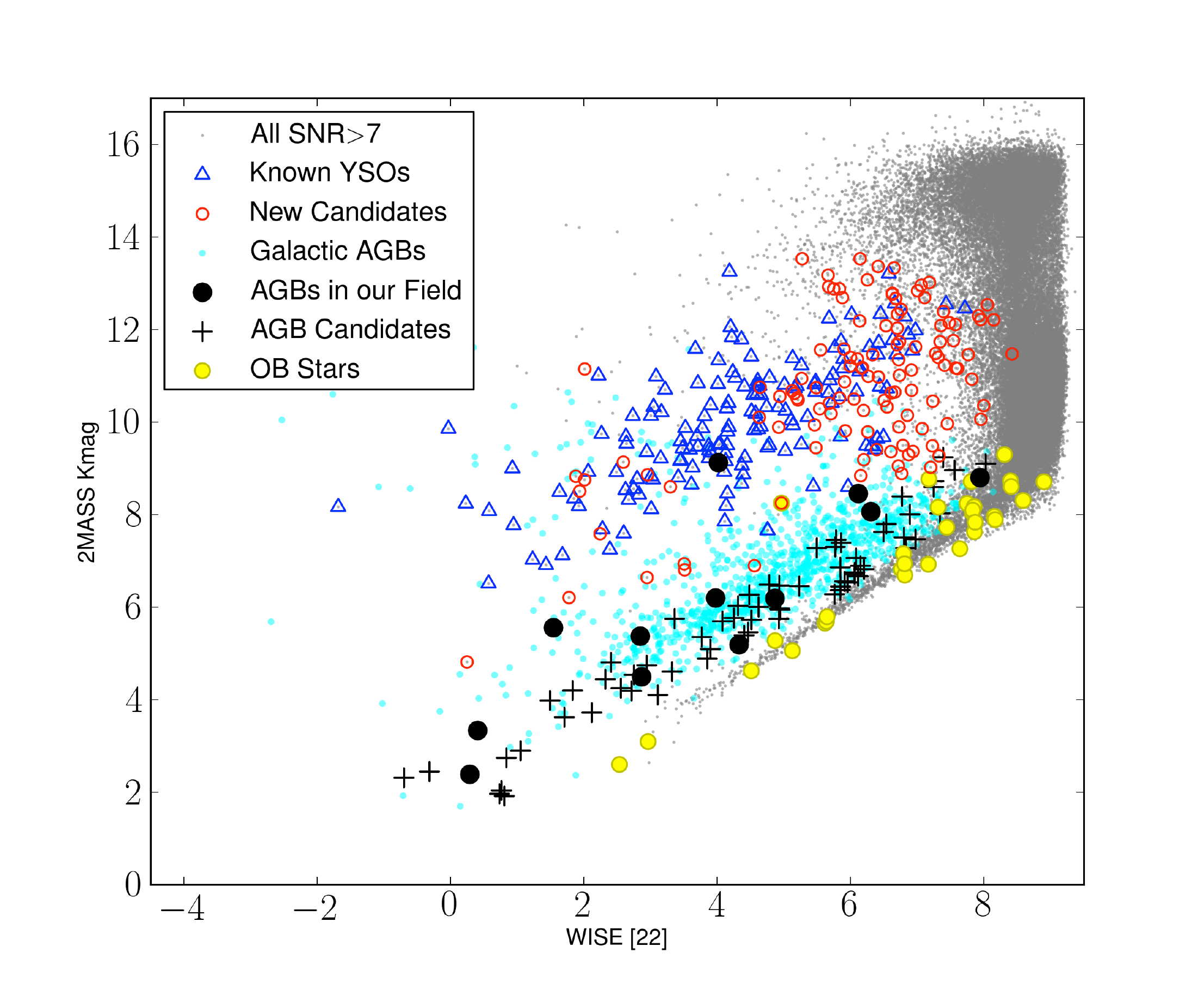}{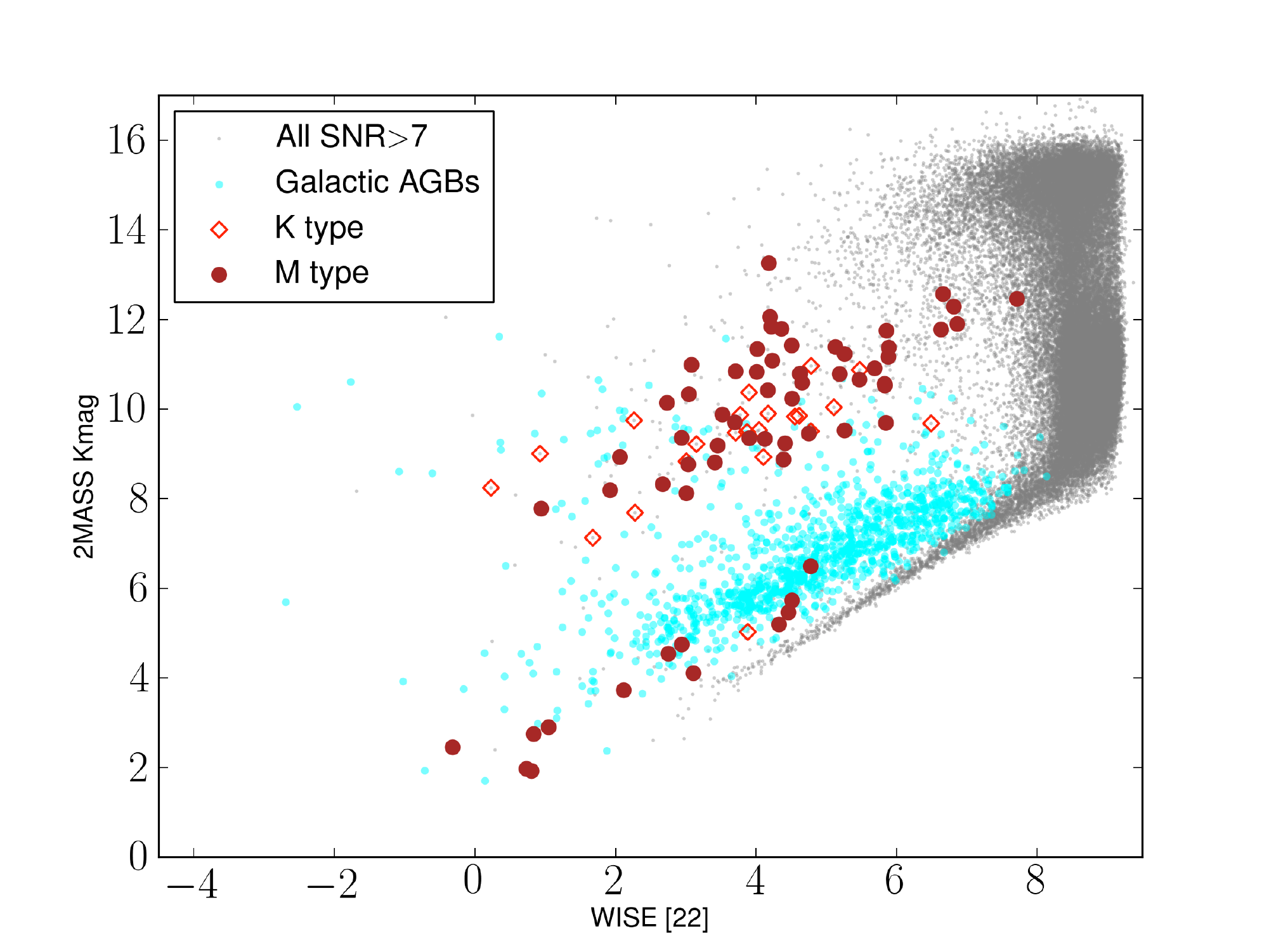}
\caption{\label{k_w4} Left: our candidate with infrared excess are divided into two groups. The brighter sources with brighter  K$_{mag}$ match with known Galactic AGB stars (cyan dots) in this plot and are likely AGB stars and not YSOs. They are noted with black crosses. Right: M and K type stars in our sample similarly follow the same pattern, empowering the suggestion that sources noted by cross in left panel are likely  M and K type evolved dusty stars. }
\end{center}
\end{figure}

We checked the distribution of our sample in various color and magnitude diagrams to look for possible category and grouping objects. In particular  22~$\mu$m is the best indicator to probe the embedded objects. Figure \ref{k_w4} shows 2MASS K$_{mag}$ plotted vs. $\w$  22~$\mu$m for all 353 point sources (J035523.11+310245.0, the known massive X-ray binary is removed).  It was noticeable that  sources are divided into two main populations. Sources with the same W4 magnitude are divided into  two brightness branches in  K$_{mag}$. The lower group  are usually bright stars with smaller slopes but they have statistically significant excess emission in longer wavelengths compared to main sequence stars. Known YSOs and YSO candidates  lie in the upper group, with larger K magnitudes. Extreme brightness in K$_{mag}$ of the lower population  suggests that they can be dusty evolved stars. The majority of them are  out of Perseus cloud and have low extinctions.

To characterize this  ``lower population" we compared them with the known AGB stars in our sample and in the literature. Grey dots in  Figure \ref{k_w4} present all $\w$ sources in our survey field with SNR$>$7 for all four bands which  mostly contain main sequence stars. Then we matched all known evolved stars in the Galaxy (including AGB stars, carbon stars and Mira variables) from SIMBAD with $\w$ catalog.  Cyan dots show evolved stars  which have a  $\w$ match with SNR$>$10 in all four bands and we call it AGB branch in our diagram. The AGB branch in this plot is above the main sequence and perfectly matches with the ``lower population". 10 of the known evolved stars in our sample also lie in this area. We selected  66 objects from the  ``lower population" in K$-$W4 diagram as   {\em dusty evolved star}  candidates. These objects are shown by  black crosses in Figure \ref{k_w4} left panel. Right panel in Figure \ref{k_w4} presents the K and M type stars in our sample which had known spectral type in the literature. While the majority of them locate within known and candidate YSOs, a smaller population lie within the AGB branch, indicating these objects are most likely evolved dusty stars rather than YSOs.

In addition we checked our AGB candidate with \citet{blum2006} color-magnitude diagrams  from their Large Magellanic Cloud {\em Spitzer} survey. In Figures $3-6$ of their paper they present the location of carbon stars and extreme AGB stars on [3.6] vs. J$-$[3.6],  [8] vs. J$-$[8],  [3.6] vs. [3.6]$-$[8] and [24] vs. [8]$-$[24].  We replaced [3.6] by W1 ($3.4~\mu$~m), [8] by W3 ($12~\mu$~m), and [24] by W4 (22 $\mu$~m) and applied their criteria.    
Although the wave bands are slightly different, our 66 AGB candidates reasonably matches the AGB stars criteria on their various plots. These candidates are listed in Table \ref{tblAGBs}.

We also did a statistical estimation of how many AGB stars we expect to identify in our field. A recent $\w$ study \citep{tu2013} estimated 470,000 of AGB stars in the Galaxy with contamination uncertainty of $~20\%$ of other sources including YSOs.  \citet{jackson2002} estimated a total of 200,000 AGB stars in the Galaxy. Using their volume density around solar neighborhood and integrating over 10~kpc (distance at which we can detect bright AGBs in our sample)  we expect to have 100-400 AGB stars in our field. Therefore 66  likely AGB candidates plus 12 known evolved stars in our field seems  a reasonable number.

\subsection{Spatial Distribution}

\begin{figure}
\begin{center}
\includegraphics[width=0.7\columnwidth]{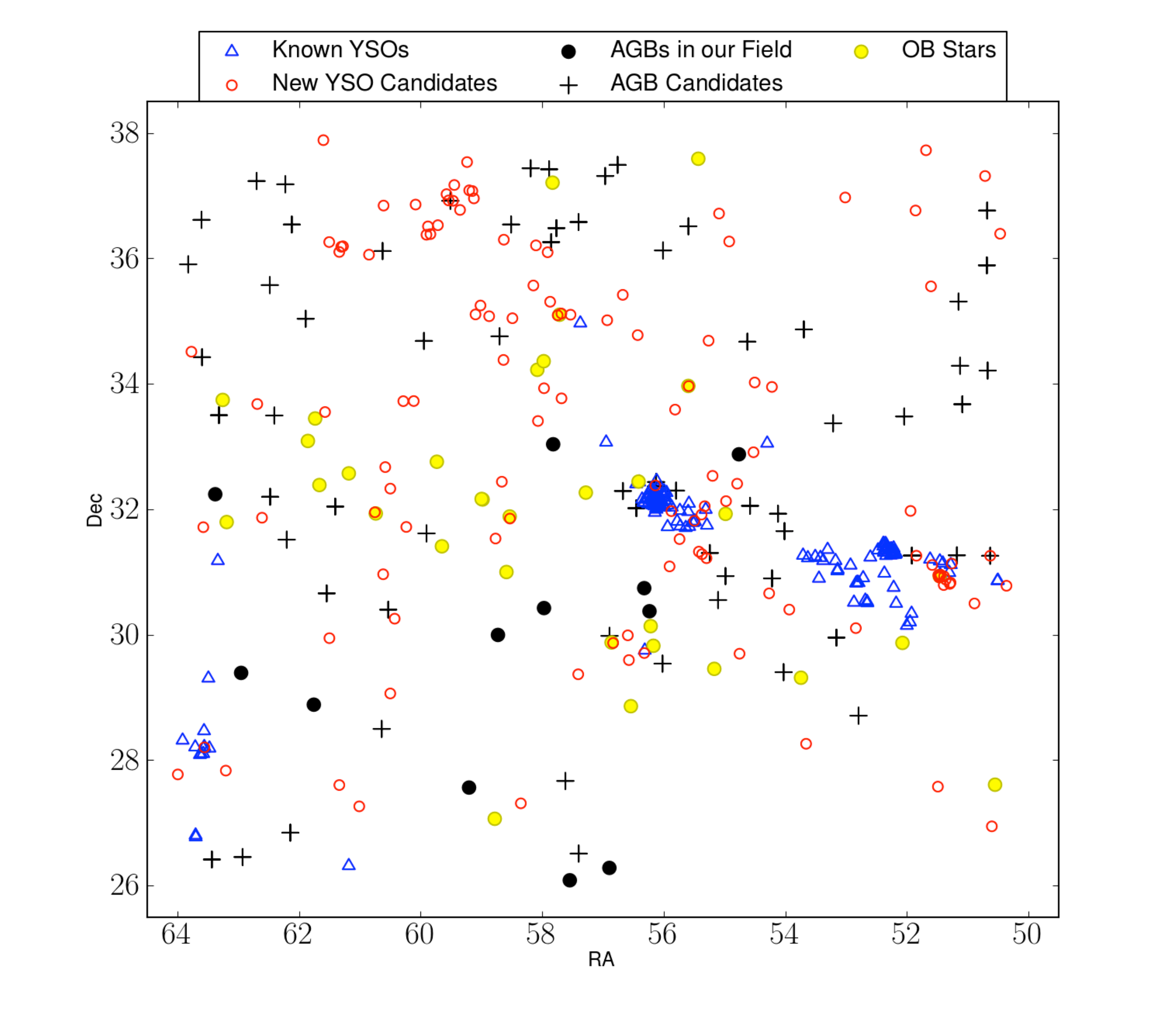}
\caption{\label{agb_dist}
Distribution of YSO, AGB and OB stars  in Per OB2.  As expiated YSOs are concentrated toward star forming clusters (IC348, NGC1333) and more numerous within the Perseus molecular cloud. AGB candidates are expected  to be randomly distributed in the field but they are following the cloud structure and also concentrated in North and  North-West of the field. }
\end{center}
\end{figure}

Figure \ref{agb_dist} presents the location of known and new YSO candidates,  known and new candidate AGB stars and OB stars  in $12^{\circ}\times12^{\circ}$ Per OB2 field. All the known YSOs (blue triangles)  and the majority of new YSO candidates (red circles) are located within the nebulous region, bright in green color in Figure \ref{wwt}. New YSO candidates in North-East corner of the field follow the shell like dust feature and a large number of new YSOs in North-West side, toward the California nebula,  are located where the remaining of the original Perseus cloud forms dense filamentary structures. 

OB stars and known AGBs are shown in yellow and black dots respectively. Black crosses present our AGB candidates discussed above.  These candidates are expected to be foreground/background sources, therefore we expect them to be randomly distributed in the field. Figure \ref{agb_dist} does not confirm a random distribution. AGB candidates shown with crosses are following the cloud structure and also concentrated in North-West.  Some of AGBs also might  be below our locus and not selected in our sample. Therefore we suggest that a fraction of these objects might be YSOs with inaccurate  photometry or at the border of contamination with AGBs.

\section{MODELING SPECTRAL ENERGY DISTRIBUTION} 
\label{modeling}

A common procedure to find the spectral energy distribution model that better reproduces  data is using $\chi^2$ minimization:  comparing  a grid of models to the data points, finding the $\chi^2$ value using the measurement errors, and picking  the model that minimizes the $\chi^2$ (e.g. \citet{robitaille07}). We call this solution the ``Best Fit" model. However, besides the Best Fit solution, there might be many other models that  have slightly larger $\chi^2$ but that are also a reasonable solution for our fitting problem. If we consider the distribution of various physical parameters from all models with  $\chi^2$ smaller than a certain threshold, the peak of such distribution may not occur at the value given by the Best Fit. In other words, the Best Fit solution might not be located in the region of the parameter space where more models are a good representation of our data. Alternatively, we can assume that the model parameters are random variables, then derive probability distribution functions (PDFs) for them given our data, and find the parameter values that maximize those PDFs. In this paper we call the latter solutions the ``Peak" values for the parameters, and we will use a Bayesian method to find them.

\subsection{Bayesian approach}

In Bayesian inference, the model parameters are considered as random variables for which posterior probability distributions can be derived from the likelihood of the solutions (that we obtain from the data and the measurement errors) and the parameter priors, where we encode the prior knowledge in these parameters before any data has been taken. Once we derive the posterior PDFs, we can draw samples from them in order to visualize the solution and find the absolute PDF maximum for each parameter. An efficient way to perform this sample is by using a Markov Chain Monte Carlo (MCMC) method that randomly steps across the parameter space and at each iteration decides whether to accept the step based on an acceptance probability that depends on the ratio of probabilities between the current position and the proposed new position. This approach is particularly useful for multivariate problems like the one at hand here.

Our Bayesian algorithm is very similar to the implementation of the \textsc{Chiburst} code for fitting SEDs of star-forming systems \citep{mg14}. Here we will describe only the aspects of the code that are relevant for the present work. As a model grid, we use the SED models of \citet{robitaille06}, that have calculated the radiative transfer for the emission from the YSO as it traverses the disk and/or envelope that surrounds it. Synthetic SEDs are produced for a broad range of stellar masses ($0.1-50 M_{\odot}$), ages ($10^3$-$10^7$ yr), as well as disk and envelope sizes and geometries. Given a set of photometry and assuming this model grid is a fair description of the actual distribution of physical parameters in YSOs, we attempt to find the probability that a particular model represents the properties of the observed YSO.

This allows us to explore the inherent degeneracies that arise when we attempt to fit a limited amount of photometric data with a multivariate model. By solving for the PDF rather than just finding solutions listed by increasing $\chi^2$, we are able to visualize all the likely solutions for a particular object photometry, given the observational uncertainties and assuming that the parameter space of the model grid is a fair universal sample of the population of YSOs in the Perseus cloud. We focus on the determination of three main model parameters, namely the YSO stellar mass ($m_*$), age ($t_*$), and $A_V$ in the line of sight towards our objects. To some extent, all other model parameters are determined by the selection of mass and age, or difficult to constrain without  unavailable  photometric bands.

\subsubsection{ The Probability Distribution Functions}
Suppose that we have obtained photometry data (D) of a YSO at different bands, with certain observational uncertainties associated. Given our data, we can calculate $P(\Theta|D)$, the posterior PDF for the set of parameters $\Theta$ as the product of the likelihood $P(D|\Theta)$ and the prior P(M). In the case of normally distributed measurement errors, the likelihood $P(D|\Theta)$ can be obtained from the distribution of reduced $\chi^2$ values:

\begin{equation}
P(D|\Theta) = \sum\exp\left(-1/2\, \chi_{\rm{red}}^2\right) 
\end{equation}

where the sum is marginalized for each model parameter over all possible models with a given value of the parameter. The \textit{prior} P(M) is a measure of any previous knowledge that we have on a particular parameter or set of parameters. For example, if we have reliable extinction measurements in the line of sight towards a particular YSO, then we can constrain the possible solutions to our  problem by constructing a prior on ($A_V$) that is compatible with those extinction measurements. Finally, we need to apply a normalization factor to our  posterior PDF to guarantee that the probability of at least one model being a representation for our YSO equals one.

\subsubsection{Priors}
We are mostly interested in obtaining a distribution of masses, evolutionary stages and optical extinctions for a sample of newly identified YSOs in the Perseus region. For the mass and age of the YSOs we have adopted the flat, uniform priors already set by the model grid, with boundaries set by $0.1\, \rm{M}_{\odot} < M_* < 50\, \rm{M}_{\odot}$ and $10^3\, \rm{yr} < t_* < 10^7\, \rm{yr} $. We also chose a flat, uniform prior for the disk inclination, since we expect most of the inclination effects to cancel out for a sample of many YSO models. We must emphasize here that these priors are discreet (i.e., not all possible values of mass, age and inclination are possible), but we consider that the sampling is such that any uncertainty due to sampling is smaller than the uncertainties imposed by the observational errors and the model degeneracies. Most of the other priors (disk and envelope mass, accretion rate, etc.) are set by the mass and age of a particular YSO, and have been thoroughly explained in \citep{robitaille06}. We adopt a distance of 320 kpc to the Perseus molecular cloud, but allow for a random scaling of the model YSOs to account for a distance uncertainty of 25$\%$. We also obscure the models and characterize this obscuration by an $A_V$ value, to account for any foreground extinction in the line of sight towards the Perseus cloud. We construct the prior on $A_V$ using an extinction map produced by \citet{lombardi2010}. The prior is simple: we restrict the possible values of $A_V$ to be within 0.4 dex of the value obtained from the extinction map. We will latter modify this prior to show the effect of the $A_V$ on determining the mass of the YSOs.

\subsubsection{Stepping across the parameter space}

Calculating the posterior PDF at every single point of the model grid and for all possible values of distance scaling and $A_V$ would be computationally time consuming, specially as more data points are added and additional model parameters (i.e. multiplicity of YSOs) are considered. Instead, we use a Monte Carlo Markov Chain (MCMC) approach to draw samples from the posterior PDF across the multidimensional space of parameters $\Theta$. We use the Metropolis algorithm  with a uniform transition probability for the Markov Chain  (i.e., given our current position in the parameter space, it is equally probable moving in any direction). After enough iterations of the MCMC, the histogram of model parameters in our chain should be a good representation of the posterior PDF. It is from this final histogram that we select the ``Peak" solutions for our parameters.

\subsection{Examples}

\begin{figure}
\begin{center}
\includegraphics[width=0.7\columnwidth]{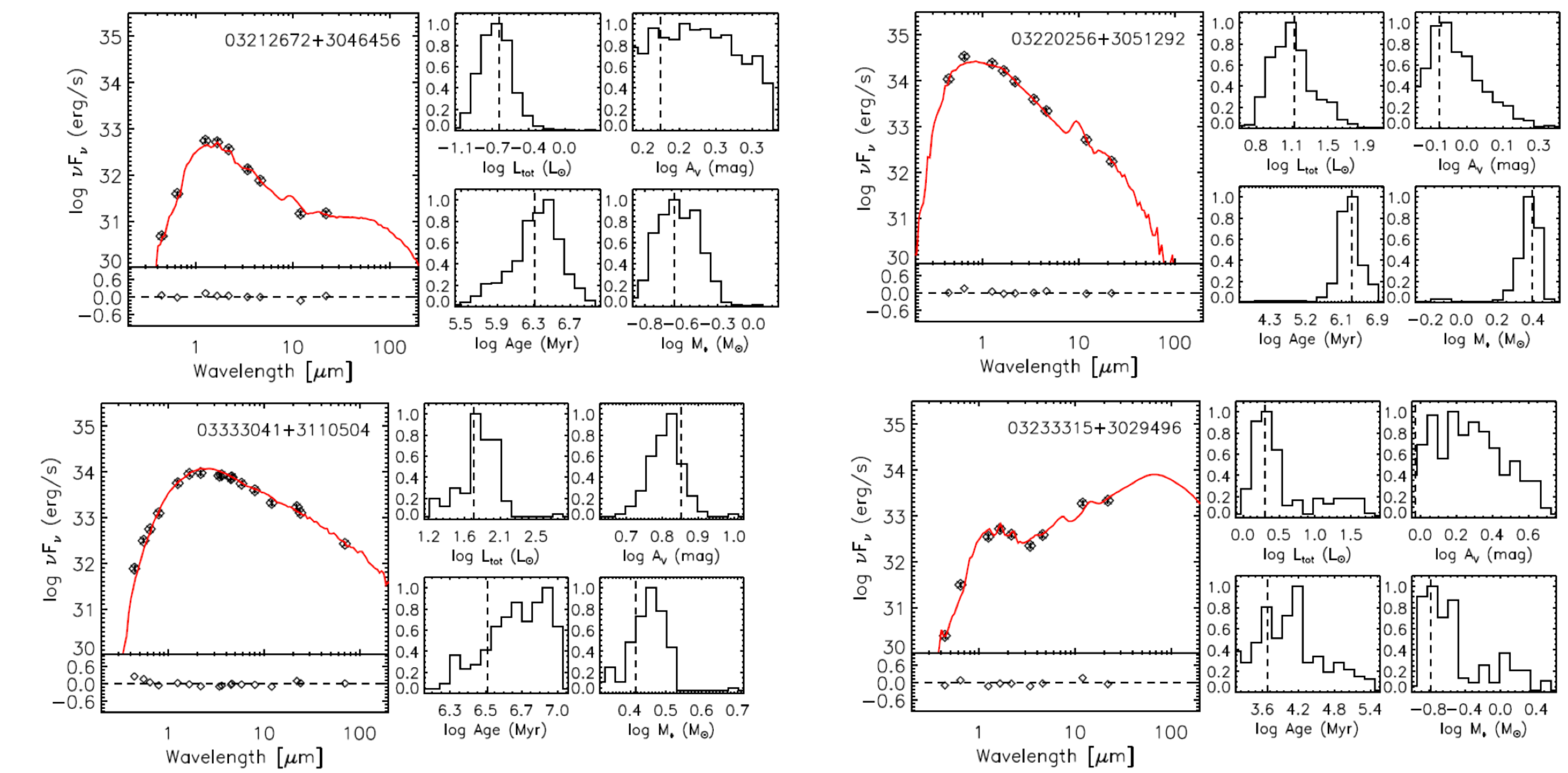}
\caption{\label{app1}
Multi-wavelength SEDs for a selection of objects in Perseus. The black diamonds are the extracted photometry, while the red line is the best fit (the one that minimizes the $\chi^2$). Also shown are the posterior PDFs for $m_*$, $t_*$, $L_{\rm{tot}}$ and $A_V$, with the best fit values represented by the dotted lines.}
\end{center}
\end{figure}

Figure \ref{app1} shows the best fits and associated posterior PDFs for several types of SEDs found in Perseus  sample. The method is quite successful at finding solutions for almost all types of SEDs. More importantly, the plotted PDFs show the usefulness of the method in determining the uniqueness of the solution.

It is important to realize the meaning of the best fit value as compared to the most likely solution as represented by the peak of the posterior PDF. They do not always coincide, which should not be surprising. The reason is that the best fit can be located in a region of the parameter space where not too many models agree with the data. The value obtained from the peak of the PDF, on the other hand, provides the most likely value for a given parameter given all possible combinations of a parameter value with all the other parameters, constrained by the information contained in the priors. In a sense, the posterior PDF is a full description of all possible solutions. This is a significant improvement over the listing of the ten best fitting models, since these can be all located very near the best fit value, but not necessarily include the peak of the PDF. Furthermore, as we will see later, bimodal solutions are also possible due to degeneracies between the parameters, and usually the 10 best fitting models all fall near one of the two possible solutions to the problem.

\subsubsection{Degeneracies}

Degeneracy is a common issue of all data-fitting problems that more than one combination of model parameters give a reasonable solution within the observational uncertainties. This problem of model degeneracies has been particularly neglected in the case of YSO SED fitting, partly because of the lack of enough data to break those degeneracies, but also due to a blind faith on the {\textquotedblleft}best fit{\textquotedblright} solution, even for objects with only a few photometric observations. Our method allows to clearly visualize the degeneracies between model parameters and hence it helps us consider other possible solutions that might be hidden in the complexity of our multi-parametrical model, and that might be in better agreement with independent determinations of the parameters. It also allows us to explore how those degeneracies behave as a function of the priors, i.e., how additional information or additional data points can break these degeneracies.

\begin{figure}
\begin{center}
\includegraphics[width=0.7\columnwidth]{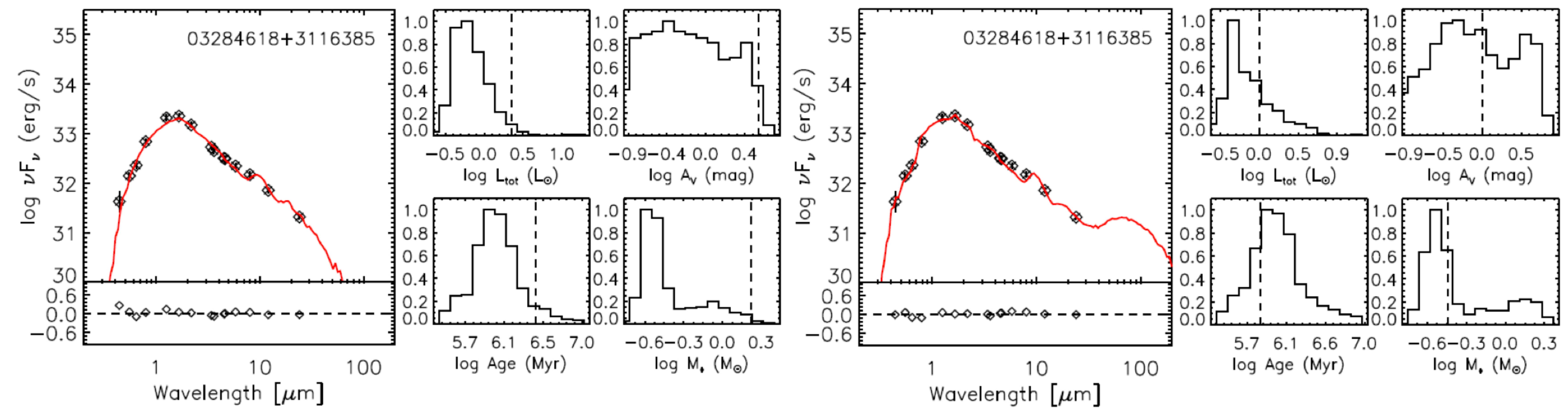}
\caption{\label{app2}
The $m_*$-$A_V$ degeneracy. Two possible solutions for a single set of photometry. The PDFs show the two possible solutions, while the best fit in each case tends to select only one.}
\end{center}
\end{figure}

\begin{figure}
\begin{center}
\includegraphics[width=0.7\columnwidth]{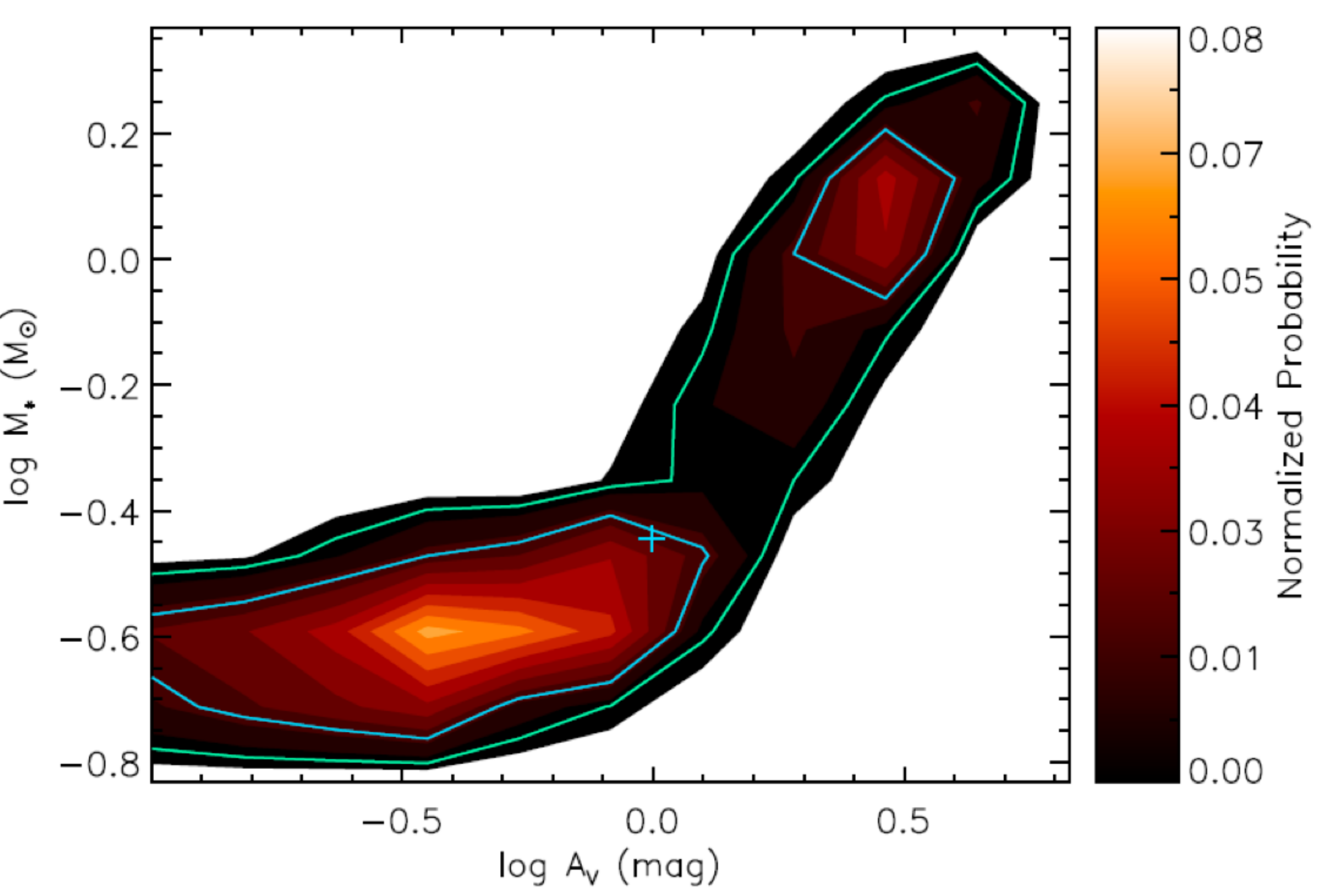}
\caption{\label{app3}
The $m_*$-$A_V$ plane with contours of normalized probability showing the degeneracy between these two model parameters. The blue cross indicates the best fit for this particular realization.}
\end{center}
\end{figure}

\begin{figure}
\begin{center}
\includegraphics[width=0.7\columnwidth]{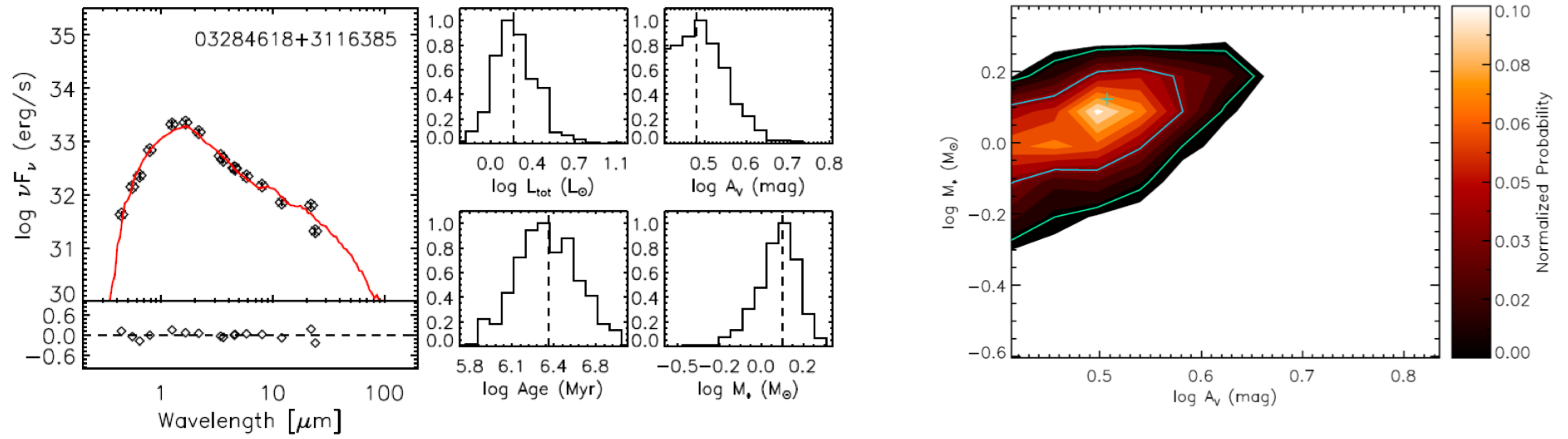}
\caption{\label{app4}
Only one solution is possible when the $A_V$ prior is modified to account for additional evidence on the extinction.}
\end{center}
\end{figure}

Figures  \ref{app2} and \ref{app3} shows an example of model degeneracies. Shown are two different realizations of the fit to the source J03284618+3116385. Two groups of solutions are possible, one with low foreground extinction ($A_V < 1$) and a low stellar mass, and one with higher $A_V$ and higher mass. Physically, this can be understood as a need for a higher stellar mass to produce the same UV and optical flux when more obscuration is at work. In the unconstrained case shown in the figure, both fits are equally satisfying, but the solution with small $A_V$ appears as more likely. Notice that the PDF for age remains unchanged. This implies that there is a strong degeneracy between stellar mass and extinction when fitting multi-wavelength SEDs of YSOs. Without a better prior for the $A_V$ (or the stellar mass), we might choose the low mass-solution, but additional evidence could make the other solution more likely. For example, if we know that the $A_V$ has to be greater than 1, then the low mass solution has to be neglected. This degeneracy is observed in many of the SEDs, in particular those of class IIa YSOs, as classified by \citep{guter2009}. Figure \ref{app4} shows the same effect much more clearly in the 2D $m_*$-$A_V$ plane. The choice of a particular $A_V$ determines the estimated mass of the YSO.

In fact, rather than thinking in terms of a {\textquotedblleft}best fit{\textquotedblright}, we should think in terms of likelihood of a solution given certain constrains on the values and all possible combinations of parameters.

It is possible to modify the prior on $A_V$ by considering independent determinations of the physical conditions. We have obtained extinction values from the maps produced by \citet{lombardi2010} and constrained the $A_V$ prior accordingly. Since the extinction map have their own uncertainties, we allow our $A_V$ values to vary within 0.4 dex of the value obtained from the maps. By doing so, we hope to break the degeneracy observed in Figure  \ref{app4} for at least some of the objects. It is important to notice here that the values of $A_V$ measured from the map represent in fact only an upper limit to the total extinction, since part of the measured $A_V$ can actually be coming from behind the source of interest. By selecting a range of values around the measured $A_V$, we are assuming that most of the extinction we see is due to foreground material between us and the source.
Figure \ref{app4} shows the solution and the resulting marginalized posterior PDFs after the prior has been modified as described, for the same object of Figures \ref{app2} and \ref{app3}. As expected, the additional information of the prior breaks the existing degeneracy and leaves only the most massive of the two solutions. This solution is in fact in better agreement with the expected mass for YSOs that we would be able to detect at the distance to the Perseus cloud. Of course, another way to break degeneracies might be by including additional data-points at longer wavelengths that are compatible with only one of the two possible solutions. Photometry from the PACS instrument onboard the Herschel Space Observatory will be instrumental for this task.

\subsection{Results and discussion}
\label{discussion}

\subsubsection{SED slope analysis}

\begin{figure}
\begin{center}
\includegraphics[width=0.7\columnwidth]{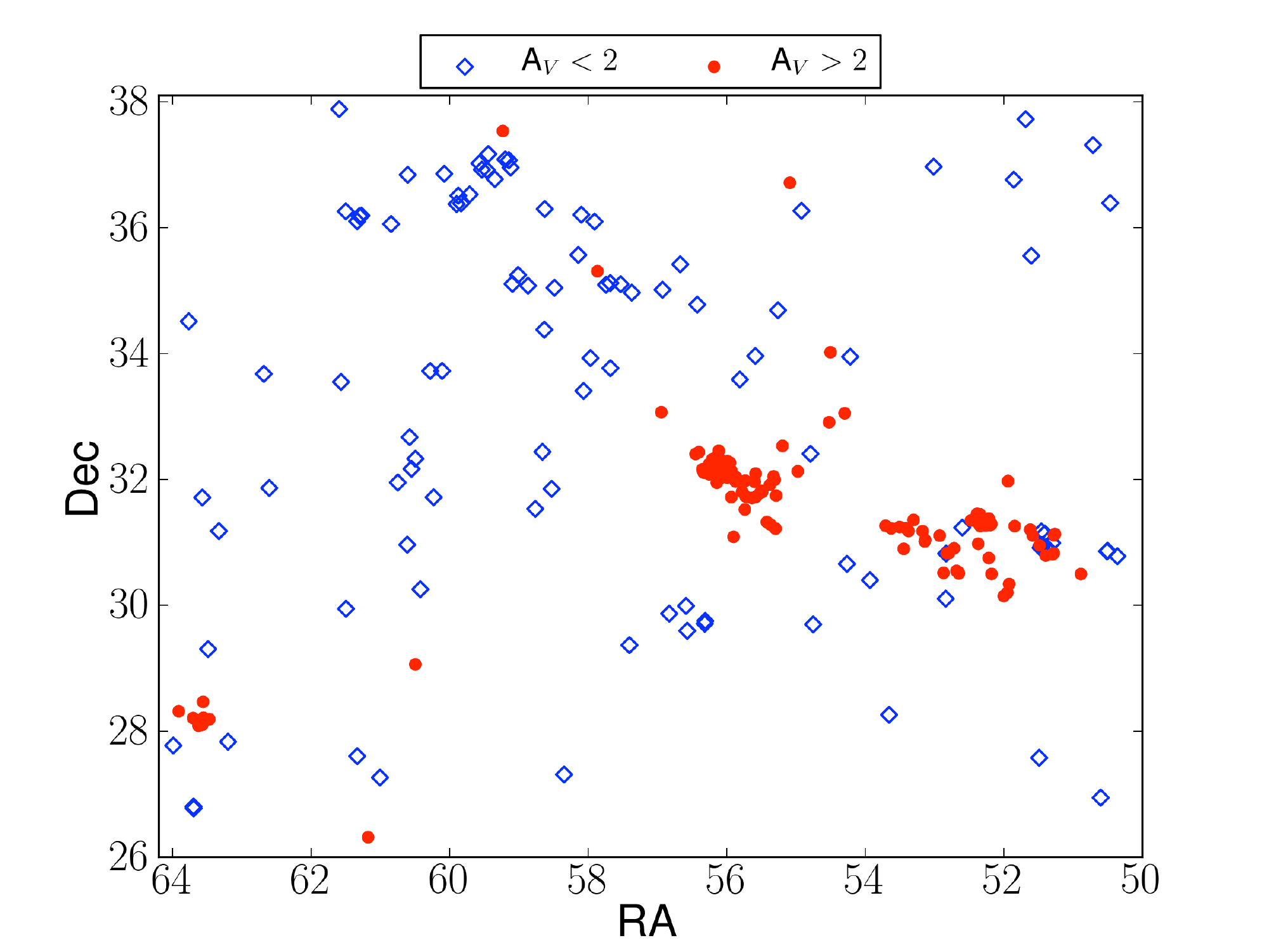}
\caption{\label{av_dist}
Spatial distribution of extinction for 275 YSO candidates. A$_V$s are calculated based on 2MASS K band extinction map by Lombardi et al. }
\end{center}
\end{figure}

We have applied our SED fitting method to 275 identified YSOs (total of known and new candidates) in order to perform a census of masses and evolutionary stages. The calculated parameters are highly dependent on the extinction, and therefore we had to estimate the dust absorptions in order to constrain A$_V$  before the SED fitting was applied.    The extinction values are calculated based on 2MASS K band extinction map provided by \citet{lombardi2010}.  Figure \ref{av_dist}  presents the location of high (A$_V > 2$) and low (A$_V < 2$) extinction regions. Currently active star forming regions such as NGC~1333, IC~348, L1448, L1455 and Taurus all are embedded within  Perseus molecular clouds and have higher extinctions.

\begin{figure}
\begin{center}
\plottwo{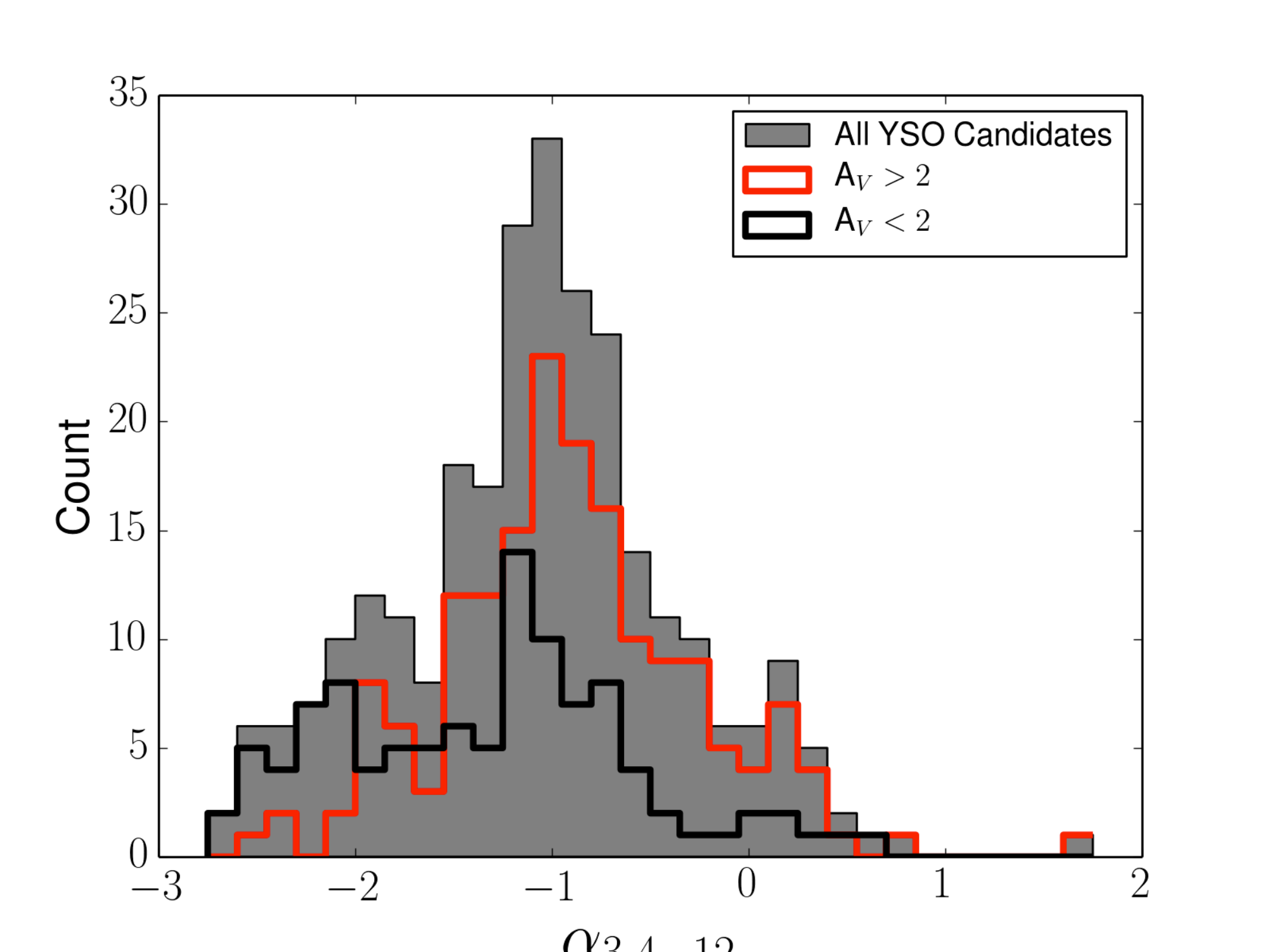}{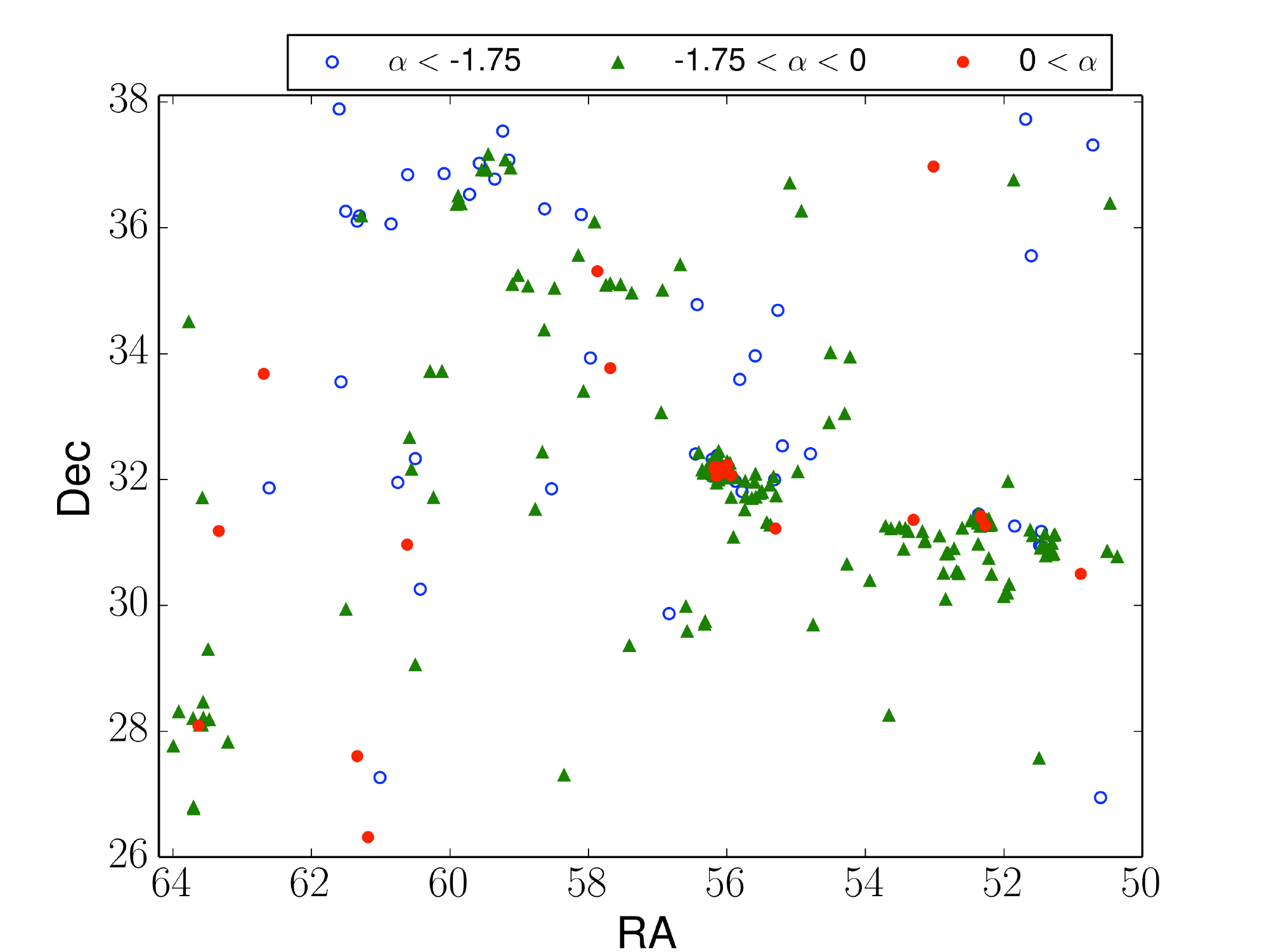}
\caption{\label{slope_hist}
Distribution of $\alpha_{3.4-12}$ for 275 YSO candidates. The location of members of three noticeable groups are shown in right panel.  $\alpha<-1.75$ in blue circles, $-1.75<\alpha<0$ in green triangles and $0<\alpha$  in red dots.}
\end{center}
\end{figure}

Figures \ref{slope_hist} presents the distribution of calculated $\alpha_{3.4-12}$  and position of candidates in  three distinct  groups.  For the first group,  $\alpha_{3.4-12}$ peaks at $\approx -2$  and these sources  are noted by blue circles in right panel for objects with $\alpha _{3.4-12} < -1.75$.  
This group have more evolved disks and are spread over the field with a concentration at North-West of the field, above the California nebula.

The main peak at $\alpha_{3.4-12} \approx -1$ contains many YSOs with stronger disks.  Green triangles present this group, selected as $ -1.75< \alpha_{3.4-12} < 0$ in the right panel. They are more concentrated within the Perseus and Taurus  star forming regions and molecular cloud. 

Candidates with very strong disks are presented with red dots and are selected as $ 0 < \alpha_{3.4-12}$. They are highly concentrated within active star forming clusters, NGC1333, IC348, and Taurus with a few spread over the field. It is not clear if they are formed individually or escaped from their original birthplace.

Distinct  population in Perseus region has been noted and discussed in several studies.  
For example \citet{herbig98} found evidence of different generation of stars within and around IC348 cluster. Also suggested by a hipparcos study  (\citet{dezeeuw99} and references therein), and later confirmed by \citet{belikov2002} Per OB2 contains two kinematically separated subgroups.  \citet{belikov2002}  identified more than 800 members 
for Per OB2 within a 50~pc diameter. 
We also have identified three separated evolutionary stages which will be discussed more in next section.

\subsubsection{Parameters from SED fitting}

Using the Bayesian MCMC method described, we have obtained best fit and peak values for all the YSOs in our sample. We focus on the peak values for stellar mass, age, total luminosity and extinction. Given the limited amount of information at wavelengths longer than $22~\mu$m, we do not attempt to constrain the disk/envelope properties with accurate  details. Instead, we are interested in studying the distribution of masses and ages in our wide select field  that includes NGC~1333 and IC~348, as well as another less dense  concentration of YSOs  located near California nebula.  

We should note that the derived parameters using \citet{robitaille07} models  are ``model-dependent-fit parameters" and not  physical values.  The YSO grid of models are randomly sampled in mass and age, and then use evolutionary tracks from \citet{siess2000} to calculate all the other stellar properties. This means that the ``age" derived here is not measured with respect to any physical stellar process, such as the start of deuterium burning. Rather, the reported ages are  indications of the relative evolutionary stages of the members which are dependent  on model assumptions and priors.

 \citet{willis2013} found a noticeable difference in total mass of their large complete sample of YSOs in NGC6334, derived from the same set of models used in this paper, compared to the expected total mass for a complete Kroupa IMF. Robitaille models employed in this study are simulated based on observation and physical parameters, therefor results are valid as a relative ratio of the evolutionary stage of all sources in the survey that satisfies the goal of this study.

\begin{figure}
\begin{center}
\plottwo{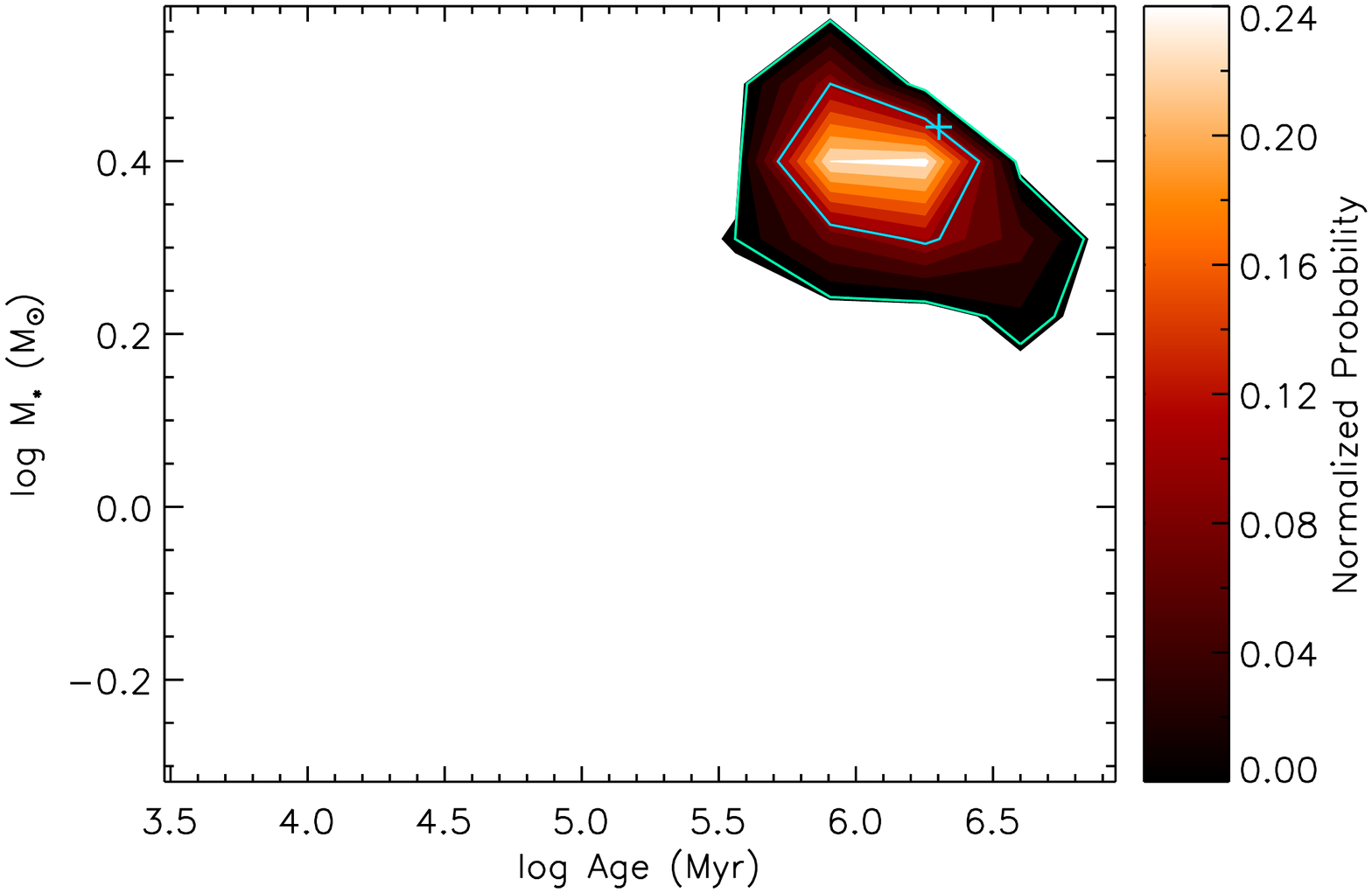}{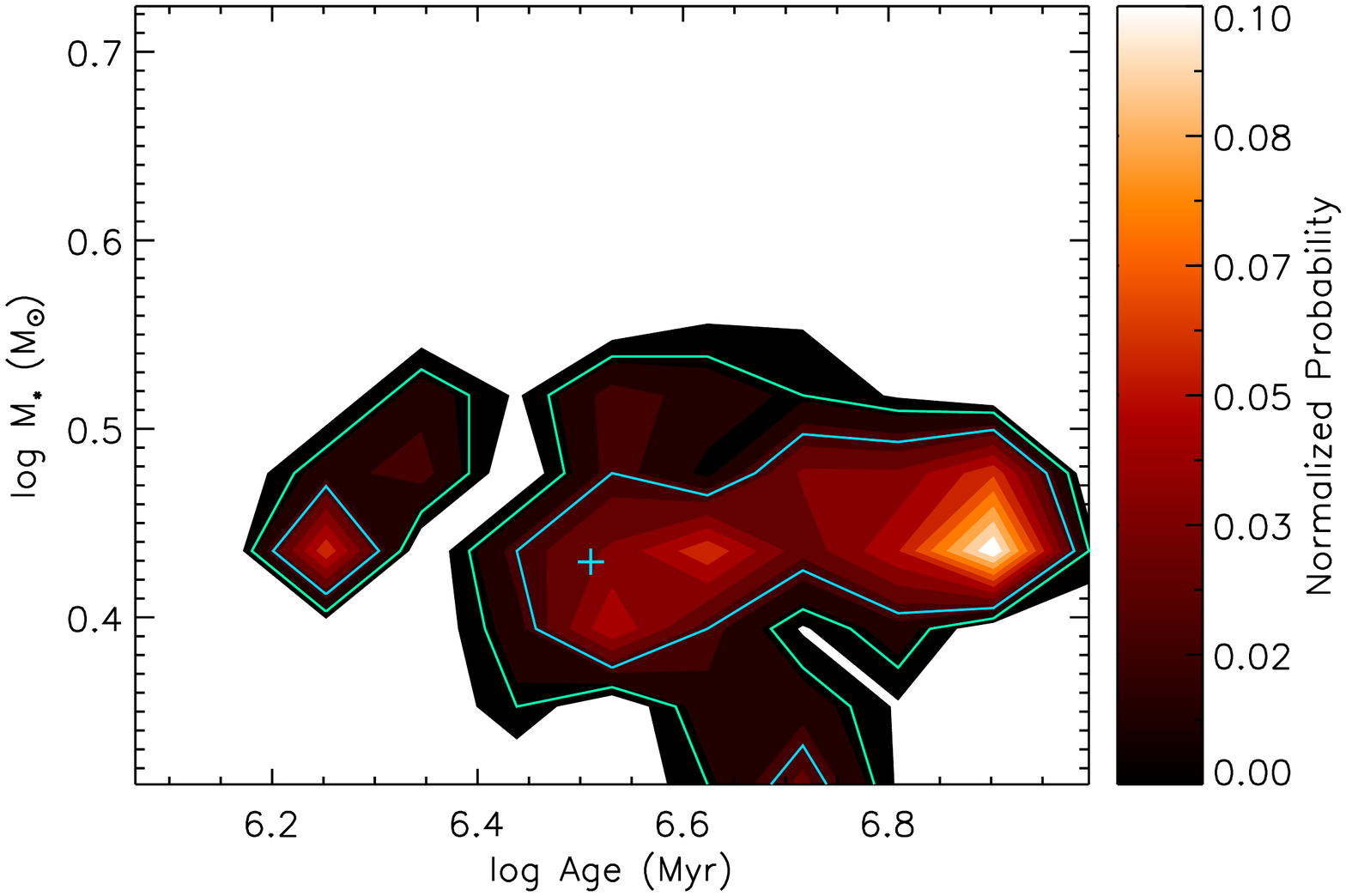}
\caption{\label{age_mass}
Mass-Age probability distribution for all fitted SED models for two different YSO candidates. The brightest region presents the most probable values from different fitted SED models. In the right panel the best-fit values (blue cross)is close  to the most probable values but in the right panel there is a large difference.}
\end{center}
\end{figure}

Figure \ref{age_mass} shows the age probability vs. mass probability for 2MASS J03311069+3049405. The blue cross which presents the best fit is very close to the most probable models for both mass and age in left panel. In some cases, however, the best fit model is very different from the probability peak. 2MASS 03253790+3108207 (right panel) is a good example where $Best~Fit$ and $Peak~Value$ do not match well.
There are two main reasons for this mismatch: first, the PDFs plotted here are marginalized versions of the PDFs, and are therefore integrated over all other parameters not shown in the plot. It is therefore likely that the absolute maximum of the multi-dimensional PDF does not always coincide with the maximum of a marginal distribution. Also, degeneracies in the models, specially given the lack of far-IR data points, might produce multiple maxima in the probability distributions.

We also realized that the correct estimation of the optical extinction is very important to constrain the mass and other physical parameters. The fitting tool lets $A_V$ to accept the values in a wide range, to find the best SED models. But not constraining $A_V$  will dramatically affect the final results.  We fixed this problem by constraining   $A_V$ from observed values of extinctions toward each individual candidate using the A$_K$ value from the extinction map provided by \citet{lombardi2010} toward Taurus and Perseus star forming regions, based on 2MASS counts. We have recorded the best fit values for the parameters  for each source, as well as the most likely solution or the peak value  as obtained from the peak of the posterior PDF. 

\begin{figure}
\begin{center}
\plottwo{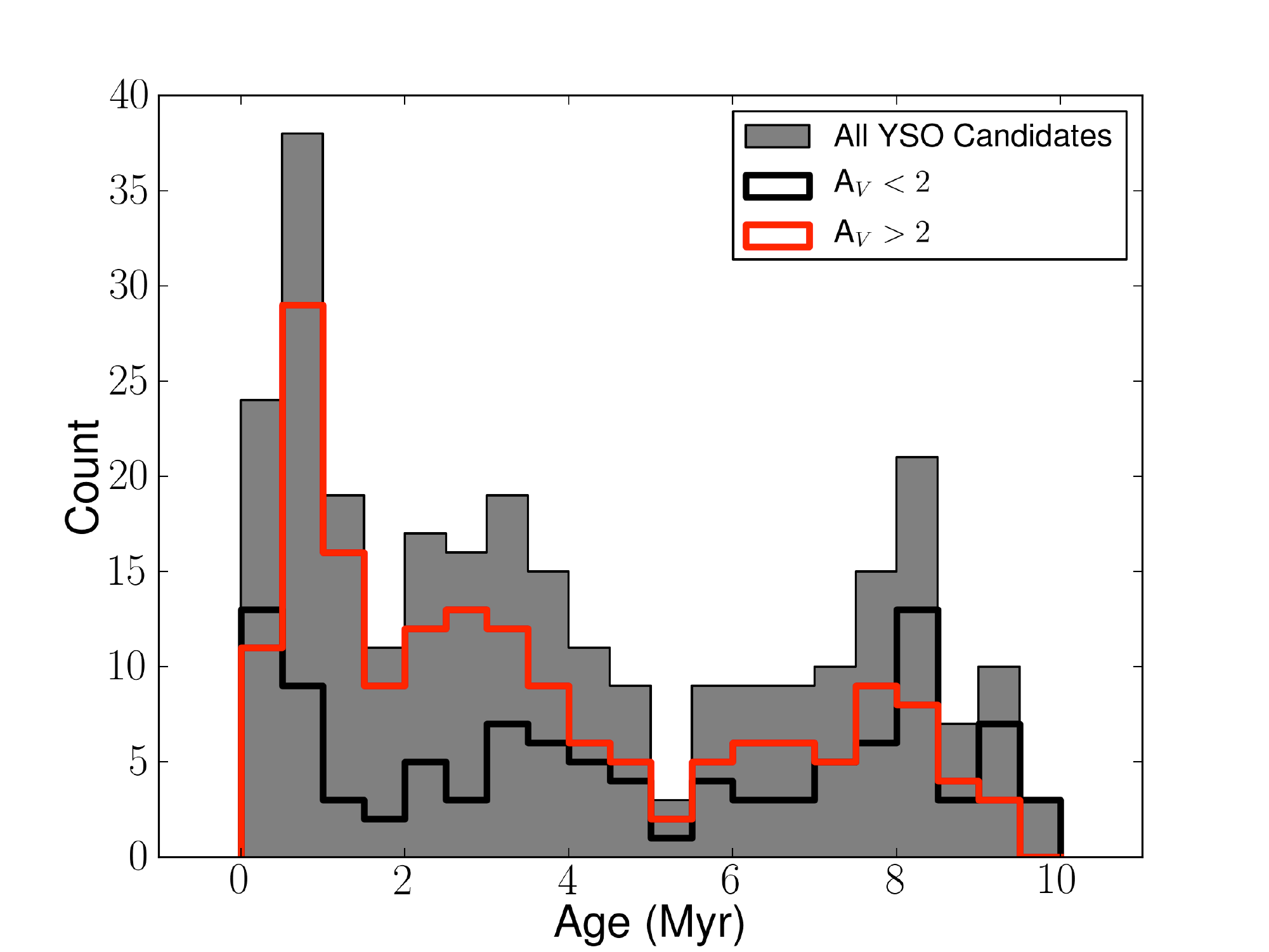}{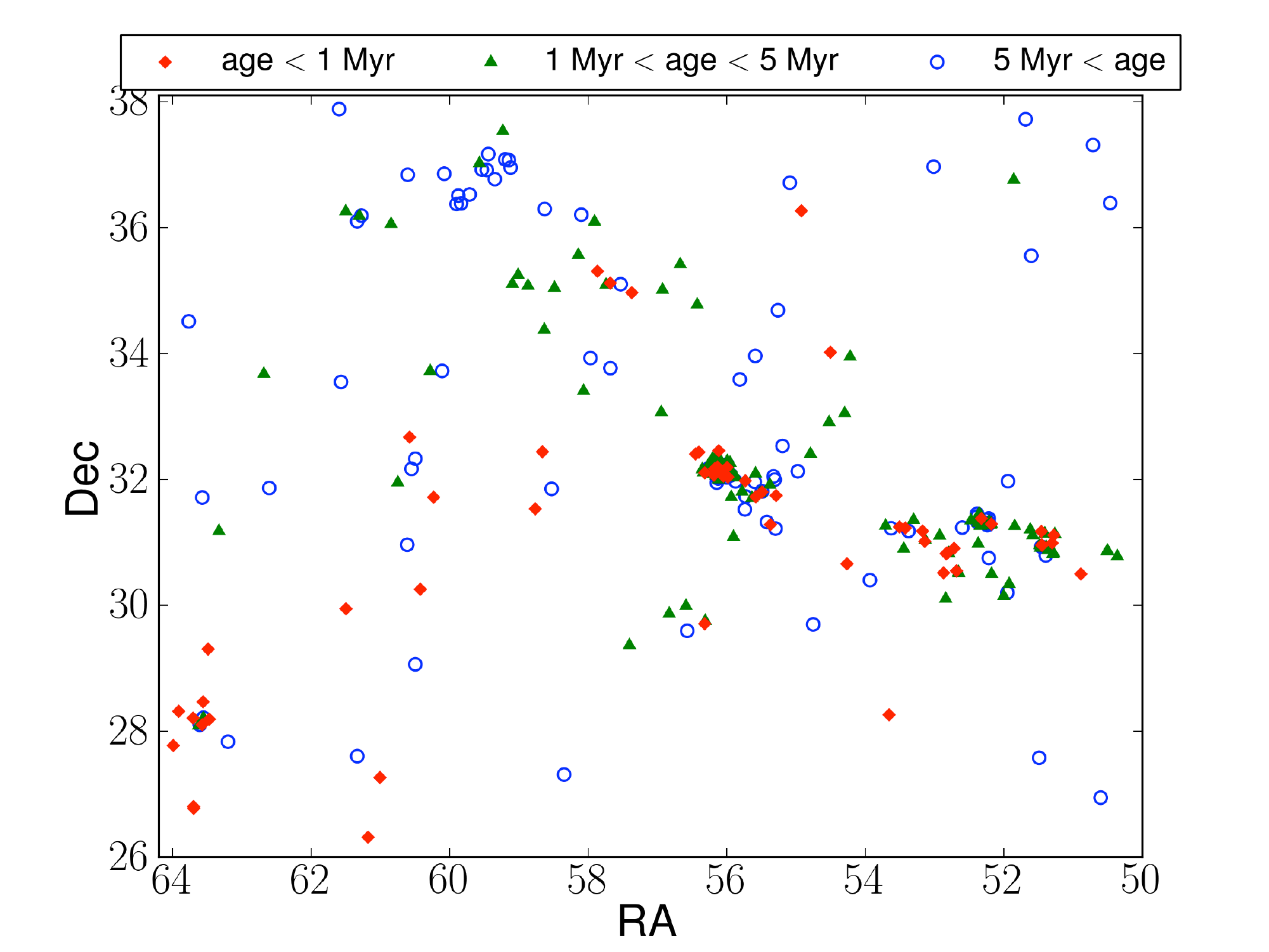}
\caption{\label{age_hist}
Age distribution for 275 YSO candidates. The location of members of three noticeable groups are shown in right panel.  $age<1$~Myr  in red dots, $1<age<5$~Myr  in green triangles and $5<age$  in blue circles.}
\end{center}
\end{figure}

Figure \ref{age_hist} presents the distribution of estimated ages for our candidates and the age spatial distribution in the field. Three distinct  age groups are noticeable in left histogram:  age$<$1~Myr, 1$<$age$<$5~Myr and 5~Myr$<$age. As expected, the  younger YSOs are having stronger disks and locate  within the high extinction inner parts of the Perseus cloud and NGC1333 and IC348 clusters. Noticeably the North-Western association do not contain any  young YSOs and only a few mid-age ones. 

\begin{figure}
\begin{center}
\plottwo{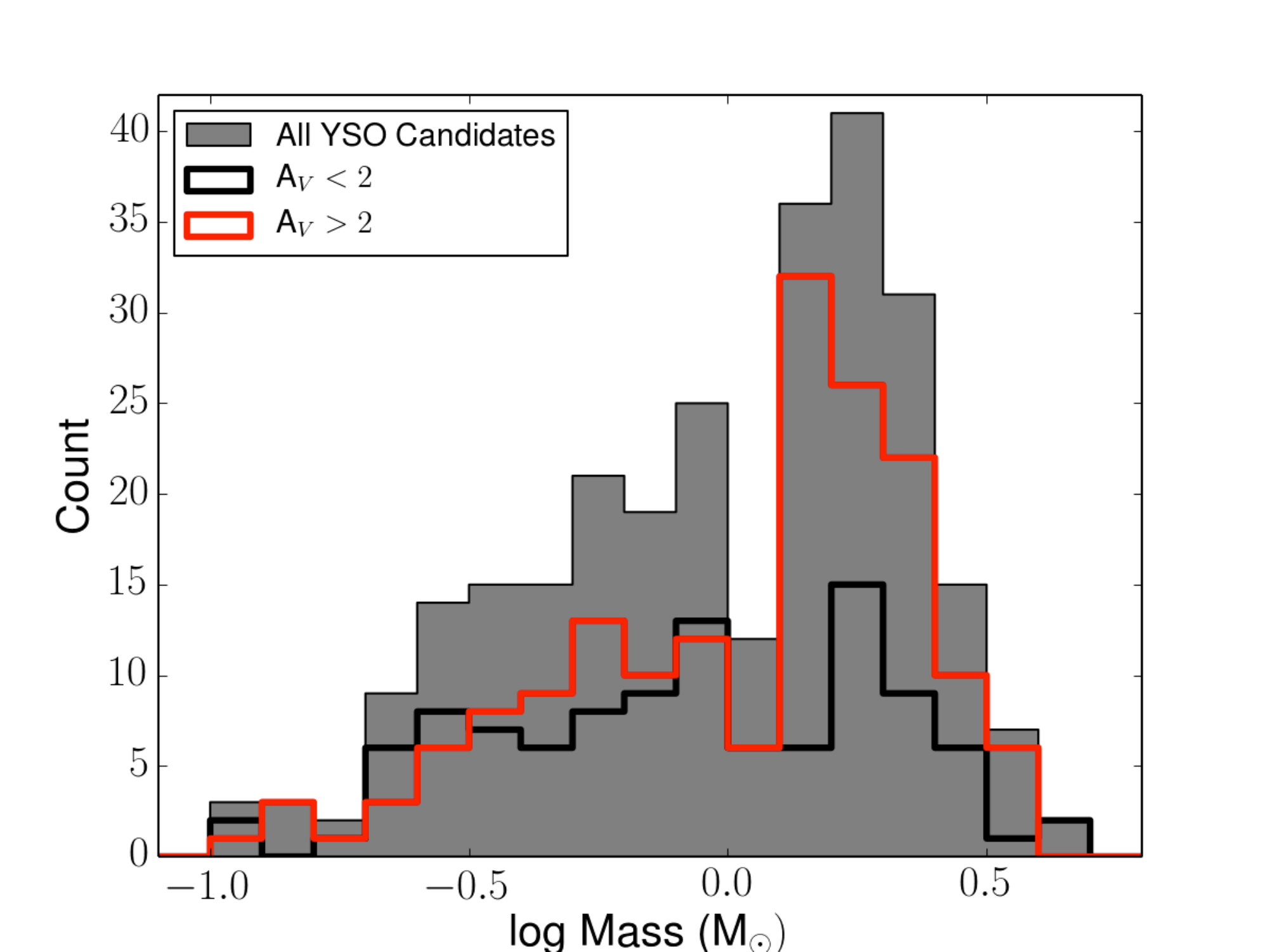}{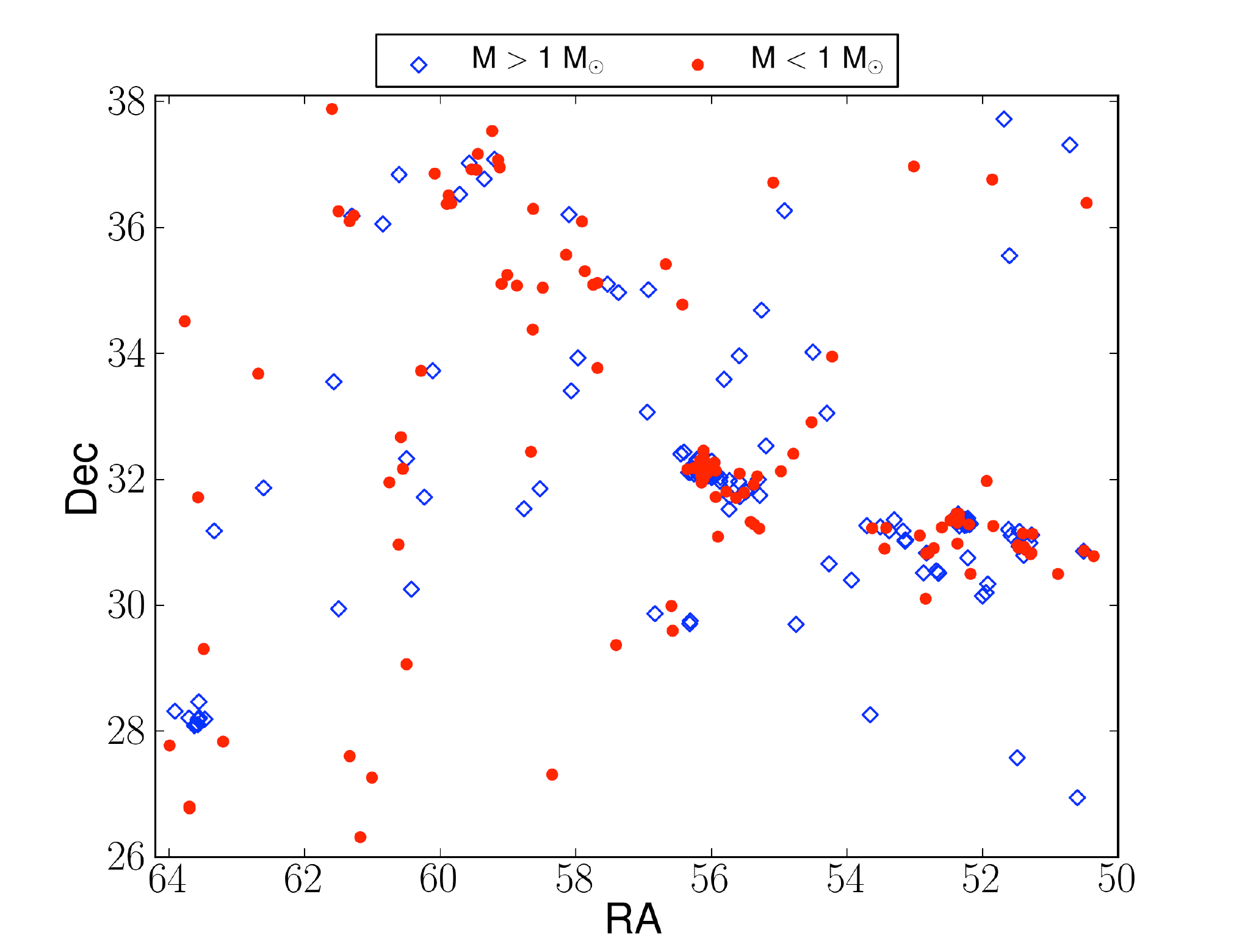}
\caption{\label{mass_hist}
Mass distribution for 275 YSO candidates.  The location of  two noticeable groups are shown in right panel. M$<$ 1~M$_\odot$ in red dots  and M$>$ 1~M$_\odot$ in blue squares.}
\end{center}
\end{figure}

The estimated masses for our YSO candidates varies in a wide range from 0.1 to 5 M$_\odot$. Figure \ref{mass_hist} presents the mass  distribution and the location of low mass and high mass candidates.  The majority of YSOs with masses larger than 1~M$_\odot$ are located within areas with larger A$_V$, i.e. within Perseus molecular cloud and particularly star-forming clusters. 
The massive YSOs from older populations spread over the field are probably too evolved to have strong, detectable disks.

\section{summary}
\label{summary}
We have performed a census of  YSO candidates in Perseus OB2 association covering $\sim$144 deg$^2$ using  $\w$ catalog. We derived physical characteristics of all  identified YSOs  within the region by employing other optical and infrared data including 2MASS, {\it Spitzer}, SDSS, PPMXL and APASS.  Following  \citet{lada06}  \citet{muench07} we calculated the SED slope in range of $3-12~\mu$m for 48,692 sources which had a S/N$\leq$7 in first three $\w$ bands. 669 point sources survived to have slopes larger than the 5$\sigma$ above the peak in bins of 0.5 2MASS J magnitude. Removing known non-stellar extended objects such as galaxies and PNe  we still had some sources remained in our sample that could not be identified as point sources in optical or infrared images or did not show point source profiles in at least three out of four $\w$ bands. Among 354 remained point sources one was identified as a massive X-ray binary and 12 as evolved stars such as carbon stars or Mira variables. 156 of remained sources were previously identified as YSO or YSO candidates. In this work we present 119 new YSO candidate toward the Perseus  region. We also identified 66 new point sources with infrared excess emission but brighter than normal YSOs in the region. We separated them as likely AGB and evolved star candidates.

The majority of known candidates are concentrated toward active star forming regions such as  Taurus, IC348 and NGC1333 clusters. We add more YSO candidates in these regions which had poor photometry in c2d or other previous studies to be confirmed as  YSOs. New candidates also follow the remaining gas of the original Perseus cloud with a concentration  toward North-West near California nebula. In total, new candidates have weaker disks and are more scattered within the field, while previously known sources locate within high A$_V$ regions in Perseus molecular cloud.

We employed the SED fitting models described by \citet{robitaille07}, but used a Markov Chain Monte Carlo method to explore the large parameter space of the model grid. Instead of selecting the model with the  lowest $\chi^2$ solution as the best fit, we accept the physical parameters from the most probable fitted models. Similar to the slope ($\alpha_{3.4-12}$) distribution, derived mass and ages also present separated populations. $\alpha_{3.4-12}$ histogram shows two separated peaks at $\approx -2$ and $\approx -1$. Candidates with slopes larger than zero (i.e. very strong disks) are mainly found within IC348 and NGC1333 clusters with a few scattered in the field. The mid-slope population which peaks around $-1$ are also located within the Perseus cloud and nebulous region.

As expected, younger sources (age$<$1~Myr) are located within active star forming clusters while older population (age$>$5~Myr)  are scattered within the field with a noticeable concentration toward North-West. 
In contrast both low mass (M$<$1~M$_\odot$) and higher mass (M$>$1~M$_\odot$) candidates are found equally scattered within the filed or located within the Perseus cloud. 

Finally, we compared our method of selecting circumstellar disks with other YSO selecting methods based on optical and infrared color and magnitudes. While our $\alpha$ method is very successful to identify bright sources with weak disks that will be filtered in other methods, our method misses YSOs with strong disks if they are faint and in particular fainter than J$_{mag}=12$. 
A combination of both  $\alpha$ and color-cut selection would be a complete method to identify both strong and weak disks.

\noindent
{\bf Acknowledgment}

We thank Matthew Templeton for providing us with APASS data and Marco Lombardi and Joao Alves for sharing Taurus-Perseus A$_K$ map based on their 2MASS study. We  also would like to  thank Luisa Rebull for helpful discussion and comments and the anonymous referee for detailed comments and
suggestions that helped to improve this work. 
This publication makes use of data products from the Wide-field Infrared Survey Explorer, which is a joint project of the University of California, Los Angeles, and the JPL/California Institute of Technology, funded by NASA. This publication makes use of data products from the Two Micron All Sky Survey, which is a joint project of the University of Massachusetts and the Infrared Processing and Analysis Center/California Institute of Technology, funded by NASA and  NSF.
This research has made use of the NASA/ IPAC Infrared Science Archive, which is operated by JPL, California Institute of Technology, under contract with NASA. This research was made possible through the use of the AAVSO Photometric All-Sky Survey (APASS), funded by the Robert Martin Ayers Sciences Fund. This research has made use of the SIMBAD database and X-Match tool, operated at CDS, Strasbourg, France.
This research funded by the National Aeronautics and Space Administration Grants NNX08AJ66G, NNX10AD68G, NNX12AI55G, and JPL RSA 717352 to the Smithsonian Astrophysical Observatory.

\begin{figure}
\begin{center}
\includegraphics[width=0.7\columnwidth]{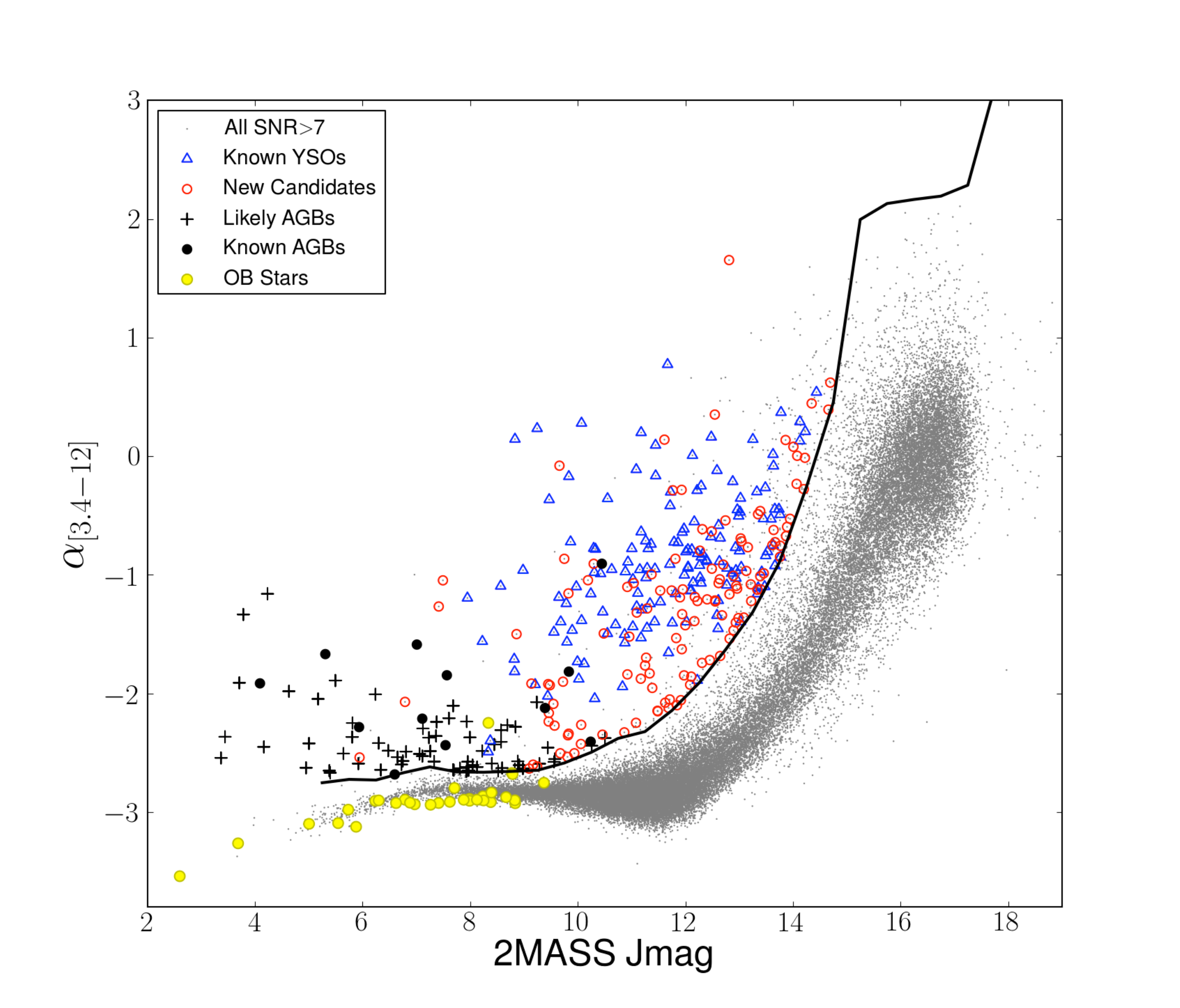}
\caption{\label{selection}
$\alpha_{3.4-12}$ versus 2MASS J magnitude were used to identify the YSO candidates. Grey dots show all $\w$ point sources with SNR$>$7 in first three bands. Black locus separates sources with $\alpha_{3.4-12}$ which are 5$\sigma$ above the Gaussian peak of slope distribution in each  $J_{mag}$ 0.5 magnitude bins. New identified YSO candidates are presented with open red circles while known candidates from literature are shown in blue triangles. Known AGB stars are presented in black bold dots in this plot while AGB  candidates identified in this study are presented in black crosses. Yellow dots present OB stars in Per OB2.  }
\end{center}
\end{figure}

\clearpage

\begingroup
\let\clearpage\relax 

\rotate
\begin{deluxetable}{ccccc}
\tabletypesize{\scriptsize}
\tablecaption{Physical parameters  measured for the entire sample \label{tbl1}}
\tablewidth{0pt}
\tablehead{
\colhead{Catalog}  & \colhead{Coverage Type} & \colhead{Coverage of $12^\circ\times12^\circ$}&
\colhead{$ Waveband$} & \colhead{Number of Sources}
}
\startdata
2MASS&all sky& 100$\%$ & J, H, K  & 887,729\\
WISE&all sky& 100$\%$ &3.4, 4.6, 12, 22  $\mu$m&1,657,265\\
USNO-UCAC3&all sky&100$\%$ &579-642 nm&205,279\\
USNO-B1&all sky&100$\%$ & B, R, I &2,072,601\\
PPMXL&all sky&100$\%$ &B, R, I &833,291\\
NOMAD&all sky&100$\%$ &B, R, I&815,242 \\
GSC&all sky&100$\%$ &J, V, N, U B&850,673\\
AKARI&all sky&100$\%$ &9, 18 $\mu$m& 1399\\
SDSS DR8&selected regions& $\sim10\%$ &u, g, i, r, z &59,4967\\
UKIDSS DR7&selected regions& $\sim50\%$& J,H,K,Y,Z &1,842,326\\
C2D-cloud&selected regions&$\sim15\%$&3.6, 4.5, 5.8, 8.0, 24,70,160 $\mu$m&777484\\
C2D-off&selected regions&$\sim2\%$&3.6, 4.5, 5.8, 8.0, 24,70,160 $\mu$m&55,193\\
Chandra &selected regions&$< 2\%$&0.3-3 \AA&887\\
XMM-Newton&selected regions&$< 2\%$&0.6-6 \AA&819\\

\enddata
\end{deluxetable}

\rotate
\begin{deluxetable}{cccccccccc}
\tabletypesize{\scriptsize}
\tablecaption{Physical parameters  measured for the 119 new candidates \label{tbl_new}}
\tablewidth{0pt}
\tablehead{
\colhead{Order} & \colhead{Catalog Number}   &\colhead{RA} & \colhead{Dec}& \colhead{Name} &\colhead{$\alpha$} &\colhead{ Mass$_{BF}$} & \colhead{Mass$_{peak}$} & \colhead{Age$_{BF}$} &
\colhead{Age$_{peak}$} \\
\colhead{} &\colhead{}   &\colhead{(deg)} & \colhead{(deg)}&
\colhead{} &\colhead{} &\colhead{(M$_\odot$)} &
\colhead{(M$_\odot$)} & \colhead{(Myr)} & \colhead{(Myr)}
}

\startdata
  1 & J032126.73+304645.5 & 50.3613824 & 30.7793282 &  & -1.74 $\pm$ 0.05 & 0.2 & 0.35 $\pm$ 0.07 & 0.56 & 3.06 $\pm$ 0.23\\
  2 & J032151.66+362327.7 & 50.4652808 & 36.3910322 &  & -1.39 $\pm$ 0.05 & 1.42 & 0.93 $\pm$ 0.22 & 9.67 & 9.32 $\pm$ 0.13\\
  3 & J032224.49+265637.9 & 50.602025 & 26.943878 & BD+26   545 & -2.63 $\pm$ 0.04 & 1.97 & 1.86 $\pm$ 0.06 & 9.79 & 8.35 $\pm$ 0.1\\
  4 & J032231.00+311527.2 & 50.629184 & 31.25754 &  & 2.62 $\pm$ 0.04 & 2.54 & 2.8 $\pm$ 0.0 & 1.04 & 1.12 $\pm$ 0.0\\
  5 & J032251.21+371842.7 & 50.713394 & 37.3118834 &  & -2.25 $\pm$ 0.06 & 1.16 & 1.15 $\pm$ 0.22 & 4.53 & 5.85 $\pm$ 0.23\\
  6 & J032333.14+302949.8 & 50.8881155 & 30.4971848 & IRAS 03204+3019 & 1.65 $\pm$ 0.03 & 0.52 & 0.64 $\pm$ 0.32 & 0.03 & 0.03 $\pm$ 0.6\\
  7 & J032503.08+310755.9 & 51.2628556 & 31.1322201 &  & -1.54 $\pm$ 0.05 & 0.23 & 0.38 $\pm$ 0.18 & 3.37 & 3.41 $\pm$ 0.46\\
  8 & J032507.48+304941.9 & 51.2811739 & 30.8283286 &  & -1.07 $\pm$ 0.05 & 0.33 & 0.46 $\pm$ 0.16 & 2.42 & 3.82 $\pm$ 0.26\\
  9 & J032510.59+304834.7 & 51.2941413 & 30.8096523 &  & -1.08 $\pm$ 0.05 & 0.24 & 0.39 $\pm$ 0.33 & 4.18 & 3.53 $\pm$ 0.26\\
  10 & J032526.49+305237.4 & 51.3604161 & 30.877062 &  & -1.04 $\pm$ 0.04 & 0.66 & 0.35 $\pm$ 0.61 & 8.02 & 3.55 $\pm$ 0.25\\
  11 & J032533.16+305544.1 & 51.3881758 & 30.9289289 &  & -1.19 $\pm$ 0.04 & 0.2 & 0.3 $\pm$ 0.09 & 0.38 & 1.12 $\pm$ 0.31\\
  12 & J032534.52+304728.0 & 51.3938412 & 30.7911121 &  & -0.69 $\pm$ 0.04 & 1.29 & 1.46 $\pm$ 0.21 & 3.3 & 8.81 $\pm$ 0.25\\
  13 & J032546.87+305720.4 & 51.4453023 & 30.9556944 & 2MASS J03254686+3057204 & -0.61 $\pm$ 0.04 & 1.38 & 1.39 $\pm$ 0.37 & 0.05 & 0.04 $\pm$ 0.61\\
  14 & J032550.10+305554.1 & 51.4587901 & 30.9317144 & BD+30   540 & -2.25 $\pm$ 0.05 & 2.49 & 2.63 $\pm$ 0.17 & 4.34 & 6.15 $\pm$ 0.19\\
  15 & J032552.76+305449.0 & 51.469849 & 30.9136219 & 2MASS J03255275+3054490 & -1.53 $\pm$ 0.05 & 0.65 & 0.5 $\pm$ 0.11 & 2.63 & 2.8 $\pm$ 0.2\\
  16 & J032555.72+305705.4 & 51.4822053 & 30.9515063 &  & -1.88 $\pm$ 0.05 & 1.06 & 0.89 $\pm$ 0.1 & 4.61 & 3.31 $\pm$ 0.22\\
  17 & J032557.49+273436.3 & 51.4895635 & 27.5767754 &  & -0.29 $\pm$ 0.04 & 1.5 & 1.19 $\pm$ 0.3 & 6.3 & 9.16 $\pm$ 0.16\\
  18 & J032619.81+310637.1 & 51.5825609 & 31.1103143 &  & -1.19 $\pm$ 0.04 & 1.62 & 1.36 $\pm$ 0.06 & 2.84 & 1.03 $\pm$ 0.34\\
  19 & J032624.73+353310.1 & 51.603061 & 35.552788 & HD 278643 & -2.62 $\pm$ 0.04 & 2.07 & 2.03 $\pm$ 0.15 & 4.23 & 5.58 $\pm$ 0.35\\
  20 & J032644.45+374319.9 & 51.6852328 & 37.7221956 & HD 275417 & -2.5 $\pm$ 0.04 & 1.97 & 1.99 $\pm$ 0.29 & 9.79 & 9.03 $\pm$ 0.19\\
  21 & J032722.41+311528.7 & 51.8434013 & 31.2579774 &  & -1.85 $\pm$ 0.05 & 0.42 & 0.56 $\pm$ 0.08 & 2.35 & 2.6 $\pm$ 0.28\\
  22 & J032725.66+364542.0 & 51.8569303 & 36.7616939 &  & -1.22 $\pm$ 0.06 & 0.1 & 0.11 $\pm$ 0.3 & 2.44 & 4.81 $\pm$ 0.32\\
  23 & J032745.45+315821.0 & 51.9394058 & 31.9725183 &  & -0.91 $\pm$ 0.04 & 0.85 & 0.95 $\pm$ 0.08 & 8.03 & 6.43 $\pm$ 0.16\\
  24 & J033121.45+300614.3 & 52.8393947 & 30.1039948 &  & -1.12 $\pm$ 0.05 & 0.12 & 0.13 $\pm$ 0.11 & 2.3 & 2.87 $\pm$ 0.27\\
  25 & J033203.36+365810.0 & 53.0140256 & 36.9694462 &  & 0.45 $\pm$ 0.06 & 0.61 & 0.74 $\pm$ 0.38 & 5.27 & 9.59 $\pm$ 0.12\\
  26 & J033437.78+281536.8 & 53.6574328 & 28.2602328 &  & -1.67 $\pm$ 0.06 & 1.46 & 1.5 $\pm$ 0.0 & 0.51 & 0.19 $\pm$ 0.0\\
  27 & J033544.16+302357.5 & 53.9340301 & 30.3993128 &  & -0.01 $\pm$ 0.05 & 1.49 & 1.47 $\pm$ 0.11 & 3.03 & 5.7 $\pm$ 0.17\\
  28 & J033652.68+335658.0 & 54.2195016 & 33.9494472 &  & -1.7 $\pm$ 0.05 & 0.65 & 0.79 $\pm$ 0.13 & 1.64 & 3.02 $\pm$ 0.24\\
  29 & J033703.64+303929.0 & 54.2651805 & 30.6580796 & TYC 2355-740-1 & -0.86 $\pm$ 0.03 & 1.27 & 1.94 $\pm$ 0.1 & 0.46 & 0.57 $\pm$ 0.11\\
  30 & J033800.72+340112.6 & 54.5030092 & 34.020184 & 2MASS J03380072+3401126 & -1.73 $\pm$ 0.06 & 1.07 & 1.54 $\pm$ 0.0 & 0.27 & 0.29 $\pm$ 0.0\\
  31 & J033805.39+325428.5 & 54.5224599 & 32.9079195 &  & -1.16 $\pm$ 0.07 & 0.62 & 0.47 $\pm$ 0.07 & 6.58 & 2.91 $\pm$ 0.3\\
  32 & J033900.55+294145.7 & 54.7523308 & 29.6960465 & V* V1185 Tau & -0.08 $\pm$ 0.04 & 3.3 & 2.17 $\pm$ 0.1 & 4.35 & 7.95 $\pm$ 0.13\\
  33 & J033909.97+322421.8 & 54.7915742 & 32.4060745 &  & -2.15 $\pm$ 0.08 & 0.76 & 0.85 $\pm$ 0.26 & 3.51 & 4.86 $\pm$ 0.23\\
  34 & J033941.31+361604.3 & 54.9221469 & 36.2678697 & GSC 02367-01706 & -1.0 $\pm$ 0.15 & 2.76 & 2.8 $\pm$ 0.0 & 0.1 & 0.11 $\pm$ 0.0\\
  35 & J033953.88+320742.6 & 54.9745076 & 32.1285048 &  & -0.53 $\pm$ 0.07 & 1.25 & 0.3 $\pm$ 0.16 & 4.39 & 8.39 $\pm$ 0.22\\
  36 & J034021.42+364248.4 & 55.0892783 & 36.7134484 &  & -0.28 $\pm$ 0.04 & 0.3 & 0.31 $\pm$ 0.11 & 8.02 & 7.9 $\pm$ 0.12\\
  37 & J034046.96+323153.7 & 55.1956827 & 32.5315911 & V* IP Per & -1.92 $\pm$ 0.05 & 2.44 & 2.19 $\pm$ 0.04 & 6.73 & 5.39 $\pm$ 0.17\\
  38 & J034102.75+344111.8 & 55.2614645 & 34.686616 & HD 278919 & -2.27 $\pm$ 0.04 & 1.91 & 1.85 $\pm$ 0.3 & 8.85 & 7.29 $\pm$ 0.0\\
  39 & J034110.99+311308.0 & 55.2958295 & 31.2188936 &  & 0.0 $\pm$ 0.06 & 0.3 & 0.34 $\pm$ 0.03 & 8.02 & 7.83 $\pm$ 0.15\\
  40 & J034117.95+320250.9 & 55.3247937 & 32.0474798 &  & -0.75 $\pm$ 0.06 & 0.21 & 0.29 $\pm$ 0.13 & 3.6 & 8.28 $\pm$ 0.26\\
  41 & J034128.70+311658.1 & 55.3695924 & 31.2828254 &  & -0.23 $\pm$ 0.07 & 0.47 & 0.35 $\pm$ 0.21 & 0.22 & 0.4 $\pm$ 0.32\\
  42 & J034130.52+315452.3 & 55.3771731 & 31.9145545 &  & -1.36 $\pm$ 0.05 & 0.18 & 0.72 $\pm$ 0.12 & 2.6 & 3.01 $\pm$ 0.3\\
  43 & J034141.12+311924.7 & 55.4213386 & 31.3235488 &  & -1.34 $\pm$ 0.05 & 0.44 & 0.59 $\pm$ 0.06 & 6.21 & 6.88 $\pm$ 0.16\\
  44 & J034158.52+314855.7 & 55.4938374 & 31.8154967 & SSTc2d J034158.6+314855 & -1.07 $\pm$ 0.04 & 2.22 & 2.12 $\pm$ 0.22 & 9.89 & 8.46 $\pm$ 0.3\\
  45 & J034221.27+335743.5 & 55.5886603 & 33.9620971 & BD+33   698B & -2.16 $\pm$ 0.05 & 1.91 & 1.89 $\pm$ 0.05 & 8.85 & 7.49 $\pm$ 0.0\\
  46 & J034257.36+313124.5 & 55.7390175 & 31.5234811 &  & -1.33 $\pm$ 0.07 & 1.22 & 1.1 $\pm$ 0.21 & 6.66 & 5.42 $\pm$ 0.17\\
  47 & J034314.92+333516.1 & 55.8122066 & 33.5878064 & HD 278934 & -1.9 $\pm$ 0.04 & 2.29 & 2.21 $\pm$ 0.13 & 9.02 & 9.12 $\pm$ 0.1\\
  48 & J034329.67+315808.9 & 55.8736578 & 31.9691642 &  & -1.86 $\pm$ 0.09 & 1.22 & 1.13 $\pm$ 0.23 & 9.86 & 7.41 $\pm$ 0.18\\
  49 & J034336.78+310516.5 & 55.9032525 & 31.0879277 &  & -1.01 $\pm$ 0.04 & 0.54 & 0.45 $\pm$ 0.16 & 7.16 & 2.81 $\pm$ 0.38\\
  50 & J034432.72+322243.0 & 56.1363454 & 32.3786267 &  & -2.35 $\pm$ 0.04 & 1.75 & 1.1 $\pm$ 0.12 & 1.48 & 1.69 $\pm$ 0.24\\
  51 & J034516.57+294224.4 & 56.319061 & 29.7067829 &  & -1.28 $\pm$ 0.04 & 3.62 & 3.39 $\pm$ 0.12 & 0.0 & 0.0 $\pm$ 0.33\\
  52 & J034542.58+344631.9 & 56.42743 & 34.775547 & NSV  1272 & -1.76 $\pm$ 0.04 & 2.0 & 0.95 $\pm$ 0.11 & 0.7 & 2.31 $\pm$ 0.25\\
  53 & J034617.26+293540.6 & 56.5719328 & 29.5946356 &  & -0.86 $\pm$ 0.04 & 0.99 & 0.79 $\pm$ 0.11 & 7.21 & 6.23 $\pm$ 0.21\\
  54 & J034621.58+295920.8 & 56.5899559 & 29.9891256 &  & -1.36 $\pm$ 0.05 & 0.37 & 0.35 $\pm$ 0.13 & 6.07 & 4.13 $\pm$ 0.22\\
  55 & J034641.48+352502.0 & 56.6728344 & 35.4172376 &  & -1.0 $\pm$ 0.03 & 1.01 & 0.81 $\pm$ 0.04 & 2.38 & 3.31 $\pm$ 0.27\\
  56 & J034719.71+295200.0 & 56.8321391 & 29.8666734 & HD 281258 & -1.93 $\pm$ 0.04 & 1.78 & 1.91 $\pm$ 0.08 & 7.63 & 3.5 $\pm$ 0.26\\
  57 & J034742.94+350044.1 & 56.9289537 & 35.0122731 & NSV  1302 & -1.1 $\pm$ 0.04 & 1.55 & 1.4 $\pm$ 0.15 & 5.9 & 3.59 $\pm$ 0.25\\
  58 & J034937.39+292205.2 & 57.4058169 & 29.368137 &  & -1.12 $\pm$ 0.04 & 0.37 & 0.34 $\pm$ 0.09 & 3.61 & 3.65 $\pm$ 0.2\\
  59 & J035007.75+350603.3 & 57.5322953 & 35.1009271 & GSC 02364-00805 & -1.72 $\pm$ 0.07 & 1.42 & 1.01 $\pm$ 0.14 & 9.67 & 9.5 $\pm$ 0.11\\
  60 & J035043.33+334602.5 & 57.6805755 & 33.7673872 &  & 0.62 $\pm$ 0.04 & 0.24 & 0.24 $\pm$ 0.19 & 9.42 & 8.08 $\pm$ 0.3\\
  61 & J035043.70+350708.9 & 57.6820945 & 35.1191647 &  & -0.6 $\pm$ 0.06 & 0.53 & 0.53 $\pm$ 0.17 & 0.04 & 0.04 $\pm$ 0.14\\
  62 & J035058.53+350528.5 & 57.7438986 & 35.0912715 &  & -0.97 $\pm$ 0.05 & 0.13 & 0.25 $\pm$ 0.17 & 2.6 & 4.04 $\pm$ 0.22\\
  63 & J035128.19+351825.8 & 57.8674971 & 35.3071818 &  & 0.14 $\pm$ 0.04 & 0.46 & 0.5 $\pm$ 0.24 & 0.36 & 0.44 $\pm$ 0.36\\
  64 & J035138.16+360546.2 & 57.9090149 & 36.0961775 &  & -0.49 $\pm$ 0.05 & 1.25 & 0.19 $\pm$ 0.0 & 4.39 & 4.0 $\pm$ 0.22\\
  65 & J035152.40+335541.1 & 57.9683514 & 33.9280997 & HD 279119 & -2.09 $\pm$ 0.04 & 1.91 & 1.88 $\pm$ 0.04 & 8.85 & 7.62 $\pm$ 0.0\\
  66 & J035216.28+332422.1 & 58.0678615 & 33.4061622 & HD 279128 & -1.27 $\pm$ 0.05 & 2.59 & 2.89 $\pm$ 0.04 & 2.71 & 2.1 $\pm$ 0.28\\
  67 & J035224.19+361221.4 & 58.1008192 & 36.2059517 & HD 279075 & -2.5 $\pm$ 0.05 & 1.91 & 1.77 $\pm$ 0.14 & 8.85 & 8.28 $\pm$ 0.25\\
  68 & J035234.42+353358.9 & 58.1434471 & 35.5663787 &  & -1.12 $\pm$ 0.04 & 0.36 & 0.35 $\pm$ 0.24 & 4.49 & 2.27 $\pm$ 0.2\\
  69 & J035323.82+271838.3 & 58.3492848 & 27.310649 &  & -0.99 $\pm$ 0.04 & 0.11 & 0.38 $\pm$ 0.14 & 7.59 & 6.91 $\pm$ 0.18\\
  70 & J035357.09+350239.2 & 58.4878805 & 35.0442238 &  & -1.13 $\pm$ 0.04 & 1.87 & 0.69 $\pm$ 0.03 & 3.61 & 4.17 $\pm$ 0.3\\
  71 & J035407.02+315101.8 & 58.5292786 & 31.850522 & BD+31   666E & -2.6 $\pm$ 0.04 & 1.91 & 1.97 $\pm$ 0.07 & 8.85 & 8.94 $\pm$ 0.13\\
  72 & J035431.06+361747.7 & 58.629428 & 36.2966037 &  & -2.06 $\pm$ 0.07 & 1.17 & 0.91 $\pm$ 0.14 & 9.81 & 7.49 $\pm$ 0.17\\
  73 & J035432.45+342245.7 & 58.6352227 & 34.3793737 &  & -1.22 $\pm$ 0.05 & 0.52 & 0.44 $\pm$ 0.1 & 5.17 & 4.55 $\pm$ 0.19\\
  74 & J035438.91+322616.8 & 58.6621312 & 32.4380016 &  & -0.85 $\pm$ 0.05 & 0.25 & 0.27 $\pm$ 0.21 & 0.38 & 0.38 $\pm$ 0.31\\
  75 & J035503.75+313157.4 & 58.7656359 & 31.5326213 &  & -0.91 $\pm$ 0.03 & 2.15 & 2.13 $\pm$ 0.09 & 7.32 & 0.91 $\pm$ 0.32\\
  76 & J035528.93+350435.5 & 58.8705711 & 35.0765338 &  & -1.32 $\pm$ 0.04 & 0.75 & 0.72 $\pm$ 0.16 & 3.19 & 3.06 $\pm$ 0.24\\
  77 & J035603.22+351450.5 & 59.0134429 & 35.2473748 &  & -1.22 $\pm$ 0.04 & 1.03 & 0.62 $\pm$ 0.18 & 0.77 & 3.02 $\pm$ 0.29\\
  78 & J035622.85+350618.5 & 59.0952433 & 35.1051579 &  & -0.46 $\pm$ 0.05 & 0.12 & 0.21 $\pm$ 0.19 & 3.62 & 3.67 $\pm$ 0.24\\
  79 & J035629.21+365717.2 & 59.1217283 & 36.9547837 &  & -0.75 $\pm$ 0.09 & 0.3 & 0.3 $\pm$ 0.1 & 8.02 & 9.6 $\pm$ 0.12\\
  80 & J035635.23+370421.2 & 59.1468169 & 37.0725814 &  & -1.93 $\pm$ 0.07 & 0.79 & 0.75 $\pm$ 0.05 & 5.67 & 7.57 $\pm$ 0.13\\
  81 & J035647.69+370500.3 & 59.1987252 & 37.0834308 &  & -1.52 $\pm$ 0.05 & 1.54 & 1.42 $\pm$ 0.11 & 7.1 & 7.9 $\pm$ 0.17\\
  82 & J035656.17+373202.1 & 59.2340637 & 37.5339431 &  & -1.95 $\pm$ 0.04 & 0.57 & 0.91 $\pm$ 0.08 & 2.99 & 2.73 $\pm$ 0.22\\
  83 & J035723.92+364615.4 & 59.3496905 & 36.7709473 &  & -2.15 $\pm$ 0.07 & 1.17 & 1.23 $\pm$ 0.13 & 6.97 & 8.79 $\pm$ 0.24\\
  84 & J035746.60+371006.7 & 59.4441938 & 37.1685303 &  & -1.12 $\pm$ 0.07 & 0.46 & 0.31 $\pm$ 0.1 & 8.87 & 8.4 $\pm$ 0.15\\
  85 & J035751.71+365501.2 & 59.4654968 & 36.917001 &  & -1.68 $\pm$ 0.09 & 0.86 & 0.61 $\pm$ 0.1 & 9.46 & 8.35 $\pm$ 0.11\\
  86 & J035808.98+365520.7 & 59.5374326 & 36.9224253 &  & -0.62 $\pm$ 0.07 & 0.3 & 0.3 $\pm$ 0.12 & 8.02 & 7.72 $\pm$ 0.08\\
  87 & J035817.28+370120.5 & 59.5720174 & 37.0223663 &  & -1.84 $\pm$ 0.05 & 0.62 & 1.07 $\pm$ 0.04 & 2.75 & 3.06 $\pm$ 0.24\\
  88 & J035851.36+363140.0 & 59.7140014 & 36.5277881 & HD 279222 & -2.34 $\pm$ 0.05 & 1.91 & 2.03 $\pm$ 0.22 & 8.85 & 8.3 $\pm$ 0.1\\
  89 & J035920.81+362308.0 & 59.836732 & 36.385564 &  & -0.77 $\pm$ 0.07 & 0.38 & 0.82 $\pm$ 0.11 & 6.72 & 8.31 $\pm$ 0.11\\
  90 & J035930.57+363031.4 & 59.8773908 & 36.5087443 &  & -0.67 $\pm$ 0.08 & 0.3 & 0.23 $\pm$ 0.19 & 8.02 & 8.53 $\pm$ 0.18\\
  91 & J035936.50+362231.5 & 59.9021105 & 36.3754322 &  & -1.01 $\pm$ 0.07 & 0.77 & 0.67 $\pm$ 0.07 & 5.56 & 9.32 $\pm$ 0.17\\
  92 & J040019.35+365123.6 & 60.0806382 & 36.8565579 &  & -2.12 $\pm$ 0.08 & 1.22 & 0.95 $\pm$ 0.04 & 9.86 & 8.25 $\pm$ 0.12\\
  93 & J040026.64+334326.5 & 60.1110131 & 33.7240537 & HD 281352 & -1.49 $\pm$ 0.04 & 1.54 & 1.59 $\pm$ 0.15 & 7.1 & 8.22 $\pm$ 0.11\\
  94 & J040056.13+314301.3 & 60.2338993 & 31.7170481 & TYC 2357-1345-1 & -1.16 $\pm$ 0.04 & 3.27 & 2.2 $\pm$ 0.28 & 0.44 & 0.49 $\pm$ 0.44\\
  95 & J040107.93+334320.9 & 60.2830825 & 33.7224846 &  & -1.21 $\pm$ 0.04 & 1.13 & 0.55 $\pm$ 0.05 & 1.14 & 1.88 $\pm$ 0.33\\
  96 & J040141.98+301516.3 & 60.424928 & 30.2545303 & V* WW Tau & -2.54 $\pm$ 0.09 & 4.24 & 4.1 $\pm$ 0.06 & 0.49 & 0.62 $\pm$ 0.2\\
  97 & J040159.15+321941.2 & 60.4964818 & 32.328137 & HD 281479 & -2.24 $\pm$ 0.04 & 1.81 & 1.92 $\pm$ 0.07 & 8.29 & 8.31 $\pm$ 0.35\\
  98 & J040159.27+290344.2 & 60.4969797 & 29.062297 &  & -1.47 $\pm$ 0.06 & 1.29 & 0.76 $\pm$ 0.12 & 6.92 & 7.51 $\pm$ 0.0\\
  99 & J040219.09+324015.2 & 60.5795592 & 32.6709063 &  & -0.8 $\pm$ 0.04 & 0.2 & 0.23 $\pm$ 0.05 & 0.92 & 0.63 $\pm$ 0.4\\
  100 & J040225.72+365025.4 & 60.6071825 & 36.8404101 & HD 279280 & -2.35 $\pm$ 0.04 & 1.96 & 2.02 $\pm$ 0.08 & 5.06 & 8.36 $\pm$ 0.34\\
  101 & J040227.39+305745.3 & 60.6141306 & 30.9625879 &  & 0.39 $\pm$ 0.04 & 0.24 & 0.3 $\pm$ 0.13 & 9.42 & 9.22 $\pm$ 0.05\\
  102 & J040259.96+315703.9 & 60.7498591 & 31.9510947 &  & -2.05 $\pm$ 0.04 & 0.6 & 0.51 $\pm$ 0.11 & 0.31 & 2.23 $\pm$ 0.27\\
  103 & J040323.26+360327.2 & 60.8469493 & 36.0575797 &  & -2.33 $\pm$ 0.06 & 1.36 & 1.36 $\pm$ 0.08 & 9.49 & 4.97 $\pm$ 0.2\\
  104 & J040401.78+271545.4 & 61.0074512 & 27.2626172 & IRAS 04010+2707 & -1.92 $\pm$ 0.05 & 0.81 & 0.78 $\pm$ 0.3 & 0.17 & 0.28 $\pm$ 0.21\\
  105 & J040506.06+361129.2 & 61.2752807 & 36.1914563 &  & -1.63 $\pm$ 0.07 & 0.95 & 0.89 $\pm$ 0.13 & 7.92 & 6.9 $\pm$ 0.13\\
  106 & J040512.95+361102.9 & 61.3039635 & 36.1841551 &  & -2.53 $\pm$ 0.06 & 1.28 & 1.04 $\pm$ 0.1 & 1.33 & 1.2 $\pm$ 0.25\\
  107 & J040520.46+273608.2 & 61.3352698 & 27.6022827 & IRAS 04023+2728 & 0.08 $\pm$ 0.04 & 0.27 & 0.26 $\pm$ 0.1 & 6.18 & 6.04 $\pm$ 0.13\\
  108 & J040520.61+360605.0 & 61.3358901 & 36.1014137 &  & -2.08 $\pm$ 0.06 & 0.65 & 0.81 $\pm$ 0.17 & 5.11 & 5.36 $\pm$ 0.2\\
  109 & J040559.62+295638.2 & 61.4984288 & 29.9439538 &  & -1.05 $\pm$ 0.03 & 2.68 & 2.1 $\pm$ 0.07 & 0.37 & 0.52 $\pm$ 0.46\\
  110 & J040600.41+361531.6 & 61.5017086 & 36.2587885 &  & -1.83 $\pm$ 0.05 & 1.22 & 0.87 $\pm$ 0.11 & 9.86 & 4.38 $\pm$ 0.17\\
  111 & J040616.84+333256.3 & 61.5701758 & 33.5489916 & HD 281534 & -2.26 $\pm$ 0.05 & 1.91 & 2.1 $\pm$ 0.08 & 8.85 & 7.58 $\pm$ 0.24\\
  112 & J040623.36+375259.9 & 61.597374 & 37.883377 &  & -2.1 $\pm$ 0.06 & 1.04 & 0.92 $\pm$ 0.05 & 5.55 & 9.26 $\pm$ 0.18\\
  113 & J041025.61+315150.7 & 62.6067479 & 31.8640941 & HD 281664 & -2.43 $\pm$ 0.06 & 1.91 & 1.83 $\pm$ 0.15 & 8.85 & 7.21 $\pm$ 0.2\\
  114 & J041044.36+334036.1 & 62.6848703 & 33.6767131 &  & 0.14 $\pm$ 0.04 & 0.98 & 0.55 $\pm$ 0.13 & 2.17 & 3.04 $\pm$ 0.24\\
  115 & J041248.58+274956.2 & 63.2024526 & 27.832299 & 2MASS J04124858+2749563 & -0.54 $\pm$ 0.04 & 0.33 & 0.46 $\pm$ 0.16 & 3.86 & 6.73 $\pm$ 0.11\\
  116 & J041412.92+281212.2 & 63.5538407 & 28.2033957 & [CGI2005]  4 & -1.05 $\pm$ 0.05 & 3.68 & 3.78 $\pm$ 0.09 & 1.47 & 1.21 $\pm$ 0.62\\
  117 & J041417.39+314247.7 & 63.5724876 & 31.7132503 &  & -0.95 $\pm$ 0.04 & 0.51 & 0.41 $\pm$ 0.47 & 8.07 & 7.37 $\pm$ 0.22\\
  118 & J041504.40+343040.1 & 63.7683363 & 34.5111655 &  & -1.4 $\pm$ 0.07 & 1.29 & 0.77 $\pm$ 0.34 & 6.92 & 8.29 $\pm$ 0.11\\
  119 & J041558.00+274617.1 & 63.9916924 & 27.7714435 & 2MASS J04155799+2746175 & -1.13 $\pm$ 0.04 & 0.23 & 0.22 $\pm$ 0.0 & 0.39 & 0.39 $\pm$ 0.31\\

\enddata
\end{deluxetable}

\rotate
\begin{deluxetable}{cccccccccc}
\tabletypesize{\scriptsize}
\tablecaption{Physical parameters  measured for the 156 known candidates \label{tbl_known}}
\tablewidth{0pt}
\tablehead{
\colhead{Order} &\colhead{Catalog Number}   &\colhead{RA} & \colhead{Dec}& \colhead{Name} &\colhead{$\alpha$} &\colhead{ Mass$_{BF}$} & \colhead{Mass$_{peak}$} & \colhead{Age$_{BF}$} &
\colhead{Age$_{peak}$} \\
\colhead{} &\colhead{}   &\colhead{(deg)} & \colhead{(deg)}&
\colhead{} &\colhead{} &\colhead{(M$_\odot$)} &
\colhead{(M$_\odot$)} & \colhead{(Myr)} & \colhead{(Myr)}
}
\startdata
 1 & J032200.52+305153.1 & 50.5022079 & 30.8647526 & 2MASS J03220052+3051531 & -1.45 $\pm$ 0.04 & 0.8 & 0.95 $\pm$ 0.05 & 1.53 & 1.3 $\pm$ 0.34\\
  2 & J032202.57+305129.3 & 50.5107177 & 30.8581628 & 2MASS J03220256+3051292 & -1.56 $\pm$ 0.04 & 2.75 & 2.62 $\pm$ 0.04 & 2.01 & 2.13 $\pm$ 0.25\\
  3 & J032506.71+310652.9 & 51.2779972 & 31.1147102 & 2MASS J03250672+3106528 & -1.4 $\pm$ 0.04 & 1.82 & 1.43 $\pm$ 0.23 & 0.72 & 0.52 $\pm$ 0.24\\
  4 & J032512.59+305921.9 & 51.3024683 & 30.9894241 & 2MASS J03251260+3059215 & -1.16 $\pm$ 0.04 & 1.65 & 1.6 $\pm$ 0.21 & 1.06 & 0.63 $\pm$ 0.28\\
  5 & J032537.91+310820.9 & 51.407985 & 31.1391488 & 2MASS J03253790+3108207 & -0.92 $\pm$ 0.04 & 0.55 & 0.95 $\pm$ 0.32 & 6.41 & 2.8 $\pm$ 0.26\\
  6 & J032548.90+305725.8 & 51.4537885 & 30.9571679 & 2MASS J03254886+3057258 & -0.96 $\pm$ 0.04 & 1.49 & 1.61 $\pm$ 0.08 & 9.08 & 0.68 $\pm$ 0.4\\
  7 & J032549.83+311023.8 & 51.4576476 & 31.1732939 & 2MASS J03254982+3110237 & -2.02 $\pm$ 0.04 & 3.24 & 1.79 $\pm$ 0.11 & 0.84 & 0.81 $\pm$ 0.36\\
  8 & J032628.22+311207.8 & 51.6175851 & 31.2021747 & 2MASS J03262821+3112078 & -0.95 $\pm$ 0.03 & 2.04 & 1.77 $\pm$ 0.05 & 0.44 & 1.92 $\pm$ 0.4\\
  9 & J032741.48+302016.8 & 51.9228373 & 30.3380137 & SSTc2d J032741.5+302017 & -1.02 $\pm$ 0.03 & 1.22 & 1.32 $\pm$ 0.15 & 2.21 & 3.21 $\pm$ 0.29\\
  10 & J032747.68+301204.4 & 51.9486767 & 30.2012397 & NAME LDN 1455 IRS 2 & -0.11 $\pm$ 0.04 & 2.68 & 3.26 $\pm$ 0.22 & 2.88 & 6.92 $\pm$ 0.39\\
  11 & J032800.09+300846.9 & 52.0004148 & 30.1463655 & 2MASS J03280010+3008469 & -1.4 $\pm$ 0.04 & 1.34 & 1.38 $\pm$ 0.27 & 3.47 & 3.47 $\pm$ 0.39\\
  12 & J032842.43+302953.1 & 52.1768142 & 30.4980955 & [EDJ2009] 146 & -0.91 $\pm$ 0.04 & 0.39 & 0.63 $\pm$ 0.14 & 1.91 & 1.9 $\pm$ 0.33\\
  13 & J032843.26+311732.9 & 52.1802544 & 31.2924824 & HBC 340 & -0.12 $\pm$ 0.04 & 1.48 & 2.12 $\pm$ 0.12 & 1.89 & 0.51 $\pm$ 0.42\\
  14 & J032843.55+311736.6 & 52.1814862 & 31.2935018 & HBC 341 & -0.29 $\pm$ 0.06 & 3.17 & 1.47 $\pm$ 0.11 & 0.87 & 2.09 $\pm$ 0.42\\
  15 & J032846.19+311638.6 & 52.192479 & 31.2773971 & EM* LkHA  351 & -1.57 $\pm$ 0.04 & 1.33 & 1.37 $\pm$ 0.45 & 2.51 & 2.3 $\pm$ 0.25\\
  16 & J032847.83+311655.2 & 52.1992954 & 31.2820006 & 2MASS J03284782+3116552 & -0.96 $\pm$ 0.04 & 1.29 & 0.58 $\pm$ 0.47 & 3.3 & 1.39 $\pm$ 0.53\\
  17 & J032851.02+311818.4 & 52.212614 & 31.3051357 & EM* LkHA  352A & -0.74 $\pm$ 0.04 & 2.61 & 2.19 $\pm$ 0.13 & 8.35 & 6.61 $\pm$ 0.36\\
  18 & J032851.19+311954.8 & 52.2133275 & 31.3319087 & 2MASS J03285119+3119548 & -0.42 $\pm$ 0.04 & 2.78 & 1.52 $\pm$ 0.07 & 0.92 & 1.35 $\pm$ 0.3\\
  19 & J032852.16+312245.3 & 52.2173627 & 31.3792507 & 2MASS J03285216+3122453 & -1.0 $\pm$ 0.04 & 1.47 & 1.4 $\pm$ 0.07 & 9.36 & 7.52 $\pm$ 0.21\\
  20 & J032852.18+304505.5 & 52.217418 & 30.7515404 & 2MASS J03285217+3045055 & -0.78 $\pm$ 0.04 & 1.99 & 2.14 $\pm$ 0.17 & 9.84 & 7.13 $\pm$ 0.09\\
  21 & J032854.62+311651.3 & 52.2275887 & 31.2809407 & 2MASS J03285461+3116512 & -0.99 $\pm$ 0.04 & 1.95 & 1.79 $\pm$ 0.43 & 7.29 & 8.18 $\pm$ 0.37\\
  22 & J032856.64+311835.6 & 52.2360079 & 31.30989 & 2MASS J03285663+3118356 & -0.85 $\pm$ 0.04 & 2.64 & 1.83 $\pm$ 0.19 & 6.25 & 5.75 $\pm$ 0.24\\
  23 & J032856.96+311622.3 & 52.2373374 & 31.27287 & 2MASS J03285694+3116222 & -0.49 $\pm$ 0.04 & 1.58 & 1.6 $\pm$ 0.06 & 8.48 & 7.48 $\pm$ 0.45\\
  24 & J032859.55+312146.7 & 52.2481447 & 31.3629832 & EM* LkHA  353 & -1.22 $\pm$ 0.04 & 2.28 & 2.32 $\pm$ 1.12 & 7.65 & 6.67 $\pm$ 0.11\\
  25 & J032903.77+311603.8 & 52.2657417 & 31.2677457 & V* V512 Per & 0.77 $\pm$ 0.04 & 1.22 & 25.69 $\pm$ 0.09 & 1.73 & 3.73 $\pm$ 1.09\\
  26 & J032903.86+312148.7 & 52.2661055 & 31.3635495 & 2MASS J03290386+3121487 & -1.23 $\pm$ 0.04 & 1.98 & 2.37 $\pm$ 0.16 & 8.22 & 7.29 $\pm$ 0.09\\
  27 & J032917.67+312244.9 & 52.3236529 & 31.3791443 & NAME NGC 1333 IRS 2 & -1.1 $\pm$ 0.04 & 2.17 & 2.49 $\pm$ 0.16 & 4.94 & 0.54 $\pm$ 0.56\\
  28 & J032918.72+312325.5 & 52.3280363 & 31.3904245 & 2MASS J03291872+3123254 & 0.09 $\pm$ 0.04 & 1.82 & 1.34 $\pm$ 0.11 & 4.75 & 1.94 $\pm$ 0.33\\
  29 & J032919.77+312457.1 & 52.3324158 & 31.4158781 & 2MASS J03291977+3124572 & 0.14 $\pm$ 0.04 & 2.56 & 2.55 $\pm$ 0.1 & 3.95 & 3.78 $\pm$ 0.23\\
  30 & J032921.55+312110.3 & 52.3398199 & 31.3528862 & 2MASS J03292155+3121104 & -0.88 $\pm$ 0.05 & 1.17 & 0.95 $\pm$ 0.13 & 6.03 & 6.88 $\pm$ 0.16\\
  31 & J032921.87+311536.2 & 52.3411571 & 31.2600729 & EM* LkHA  271 & -1.53 $\pm$ 0.04 & 1.08 & 1.82 $\pm$ 0.09 & 7.46 & 2.46 $\pm$ 0.28\\
  32 & J032923.15+312030.3 & 52.3464585 & 31.3417503 & EM* LkHA  355 & -0.89 $\pm$ 0.04 & 1.48 & 0.96 $\pm$ 0.32 & 8.59 & 4.59 $\pm$ 0.19\\
  33 & J032923.23+312653.0 & 52.3468186 & 31.4480694 & 2MASS J03292322+3126531 & -0.53 $\pm$ 0.04 & 0.39 & 0.39 $\pm$ 0.1 & 5.14 & 3.8 $\pm$ 0.38\\
  34 & J032925.92+312640.1 & 52.3580264 & 31.4444756 & 2MASS J03292591+3126401 & -0.78 $\pm$ 0.03 & 1.48 & 1.87 $\pm$ 0.09 & 2.24 & 2.38 $\pm$ 0.4\\
  35 & J032926.79+312647.4 & 52.3616596 & 31.44652 & 2MASS J03292681+3126475 & -1.94 $\pm$ 0.05 & 1.6 & 1.39 $\pm$ 0.16 & 5.2 & 3.44 $\pm$ 0.25\\
  36 & J032928.89+305841.8 & 52.3703977 & 30.978281 & [EDJ2009] 248 & -0.77 $\pm$ 0.04 & 0.23 & 0.33 $\pm$ 0.29 & 5.25 & 4.26 $\pm$ 0.21\\
  37 & J032929.26+311834.7 & 52.371941 & 31.309651 & 2MASS J03292925+3118347 & -1.34 $\pm$ 0.04 & 0.36 & 0.51 $\pm$ 0.19 & 2.17 & 2.42 $\pm$ 0.41\\
  38 & J032929.78+312102.6 & 52.3741039 & 31.3507368 & 2MASS J03292978+3121027 & -0.69 $\pm$ 0.04 & 0.43 & 0.73 $\pm$ 0.12 & 2.68 & 2.67 $\pm$ 0.26\\
  39 & J032930.39+311903.3 & 52.3766345 & 31.3176101 & EM* LkHA  356 & -1.13 $\pm$ 0.04 & 2.01 & 1.82 $\pm$ 0.15 & 8.07 & 7.42 $\pm$ 0.65\\
  40 & J032932.56+312436.9 & 52.38569 & 31.4102637 & EM* LkHA  357 & -0.92 $\pm$ 0.04 & 0.55 & 0.43 $\pm$ 0.11 & 6.42 & 6.05 $\pm$ 0.22\\
  41 & J032932.87+312712.5 & 52.3869611 & 31.4534793 & 2MASS J03293286+3127126 & -1.11 $\pm$ 0.06 & 0.22 & 0.41 $\pm$ 0.23 & 4.8 & 9.09 $\pm$ 0.18\\
  42 & J032954.03+312053.0 & 52.475129 & 31.3480571 & 2MASS J03295403+3120529 & -0.64 $\pm$ 0.04 & 0.43 & 0.56 $\pm$ 0.17 & 2.68 & 2.88 $\pm$ 0.34\\
  43 & J033024.09+311404.3 & 52.6004078 & 31.2345343 & 2MASS J03302409+3114043 & -0.98 $\pm$ 0.05 & 0.22 & 0.27 $\pm$ 0.07 & 6.61 & 8.04 $\pm$ 0.13\\
  44 & J033035.94+303024.5 & 52.6497626 & 30.5068066 & 2MASS J03303593+3030244 & -0.96 $\pm$ 0.04 & 3.29 & 2.99 $\pm$ 0.12 & 1.09 & 1.17 $\pm$ 0.36\\
  45 & J033036.97+303127.7 & 52.6540781 & 30.5243642 & SSTc2d J033037.0+303128 & -1.16 $\pm$ 0.04 & 2.87 & 2.38 $\pm$ 0.1 & 7.94 & 1.09 $\pm$ 0.52\\
  46 & J033043.99+303246.9 & 52.683332 & 30.5463736 & [EDJ2009] 269 & -0.99 $\pm$ 0.04 & 2.44 & 2.22 $\pm$ 0.16 & 6.73 & 0.78 $\pm$ 0.45\\
  47 & J033052.52+305417.7 & 52.7188596 & 30.9049298 & 2MASS J03305252+3054177 & -0.47 $\pm$ 0.04 & 0.12 & 0.12 $\pm$ 0.25 & 0.1 & 0.18 $\pm$ 0.29\\
  48 & J033110.70+304940.5 & 52.7945859 & 30.8279314 & SSTc2d J033110.7+304941 & -0.61 $\pm$ 0.04 & 0.57 & 0.32 $\pm$ 0.25 & 2.11 & 1.15 $\pm$ 0.36\\
  49 & J033114.71+304955.3 & 52.811322 & 30.8320395 & [EDJ2009] 272 & -0.45 $\pm$ 0.04 & 0.11 & 0.13 $\pm$ 0.12 & 0.62 & 0.09 $\pm$ 0.56\\
  50 & J033118.31+304939.4 & 52.8263068 & 30.8276353 & SSTc2d J033118.3+304940 & -0.71 $\pm$ 0.04 & 1.48 & 2.13 $\pm$ 0.14 & 2.24 & 0.78 $\pm$ 0.3\\
  51 & J033120.11+304917.5 & 52.8338214 & 30.8215501 & [EDJ2009] 274 & -0.75 $\pm$ 0.04 & 0.37 & 0.36 $\pm$ 0.17 & 0.04 & 0.04 $\pm$ 0.37\\
  52 & J033128.87+303053.2 & 52.8703166 & 30.5147946 & 2MASS J03312887+3030531 & -1.57 $\pm$ 0.04 & 2.25 & 1.69 $\pm$ 0.35 & 2.13 & 0.87 $\pm$ 0.25\\
  53 & J033142.41+310624.7 & 52.9267459 & 31.106877 & [EDJ2009] 278 & -1.45 $\pm$ 0.04 & 0.48 & 0.64 $\pm$ 0.13 & 2.14 & 1.83 $\pm$ 0.46\\
  54 & J033233.00+310221.6 & 53.1375316 & 31.0393575 & [EDJ2009] 281 & -1.02 $\pm$ 0.04 & 1.29 & 1.4 $\pm$ 0.12 & 3.3 & 1.26 $\pm$ 0.27\\
  55 & J033234.06+310055.7 & 53.1419191 & 31.0154864 & IRAS 03295+3050 & -0.94 $\pm$ 0.04 & 2.17 & 1.64 $\pm$ 0.52 & 1.69 & 0.95 $\pm$ 0.33\\
  56 & J033241.70+311045.7 & 53.1737635 & 31.1793821 & 2MASS J03324171+3110461 & -0.45 $\pm$ 0.03 & 0.27 & 1.44 $\pm$ 0.09 & 1.58 & 0.94 $\pm$ 0.48\\
  57 & J033312.84+312124.1 & 53.3035245 & 31.3566981 & SSTc2d J033312.8+312124 & 0.01 $\pm$ 0.04 & 2.37 & 3.14 $\pm$ 0.05 & 7.2 & 3.01 $\pm$ 0.37\\
  58 & J033330.42+311050.4 & 53.3767691 & 31.180692 & EM* LkHA  327 & -1.24 $\pm$ 0.05 & 2.69 & 2.98 $\pm$ 0.32 & 3.24 & 7.86 $\pm$ 0.21\\
  59 & J033341.30+311340.9 & 53.4220887 & 31.2280551 & 2MASS J03334129+3113410 & -0.85 $\pm$ 0.04 & 1.08 & 0.45 $\pm$ 0.23 & 7.46 & 0.68 $\pm$ 1.0\\
  60 & J033346.93+305350.1 & 53.4455792 & 30.8972686 & [EDJ2009] 299 & -1.1 $\pm$ 0.06 & 0.16 & 0.28 $\pm$ 0.24 & 8.87 & 4.15 $\pm$ 0.18\\
  61 & J033401.67+311439.6 & 53.5069625 & 31.2443577 & EM* LkHA  328 & -0.97 $\pm$ 0.04 & 1.47 & 1.57 $\pm$ 0.18 & 1.19 & 0.95 $\pm$ 0.24\\
  62 & J033430.79+311324.3 & 53.6283159 & 31.2234194 & [EDJ2009] 302 & -1.03 $\pm$ 0.05 & 0.18 & 0.57 $\pm$ 0.16 & 3.06 & 5.64 $\pm$ 0.21\\
  63 & J033449.86+311550.1 & 53.7077643 & 31.263929 & 2MASS J03344987+3115498 & -1.42 $\pm$ 0.04 & 0.73 & 1.47 $\pm$ 0.12 & 0.97 & 1.81 $\pm$ 0.21\\
  64 & J033711.38+330303.0 & 54.2974288 & 33.0508484 & 2MASS J03371138+3303032 & -0.95 $\pm$ 0.03 & 1.25 & 1.69 $\pm$ 0.19 & 1.97 & 2.56 $\pm$ 0.32\\
  65 & J034109.14+314437.9 & 55.2880837 & 31.7438668 & IRAS 03380+3135 & -1.5 $\pm$ 0.04 & 0.63 & 2.59 $\pm$ 0.12 & 0.47 & 0.52 $\pm$ 0.31\\
  66 & J034114.12+315946.1 & 55.308873 & 31.996148 & 2MASS J03411412+3159462 & -1.88 $\pm$ 0.05 & 1.97 & 2.04 $\pm$ 0.14 & 9.79 & 8.31 $\pm$ 0.1\\
  67 & J034157.45+314836.6 & 55.4894069 & 31.810175 & [EDJ2009] 314 & -1.27 $\pm$ 0.04 & 2.28 & 2.21 $\pm$ 0.36 & 7.82 & 8.39 $\pm$ 0.62\\
  68 & J034157.76+314800.7 & 55.490705 & 31.8002143 & [EDJ2009] 315 & -0.89 $\pm$ 0.04 & 0.42 & 1.73 $\pm$ 0.1 & 0.45 & 0.87 $\pm$ 0.37\\
  69 & J034204.35+314711.4 & 55.5181477 & 31.7865217 & [EDJ2009] 319 & -0.84 $\pm$ 0.04 & 0.43 & 0.21 $\pm$ 0.39 & 1.81 & 1.83 $\pm$ 0.44\\
  70 & J034219.28+314326.8 & 55.5803535 & 31.7241352 & 2MASS J03421927+3143269 & -0.8 $\pm$ 0.04 & 1.08 & 1.17 $\pm$ 0.26 & 7.46 & 0.98 $\pm$ 0.34\\
  71 & J034220.34+320530.9 & 55.584758 & 32.0919442 & 2MASS J03422033+3205310 & -0.94 $\pm$ 0.05 & 0.42 & 0.37 $\pm$ 0.04 & 7.36 & 2.79 $\pm$ 0.31\\
  72 & J034223.33+315742.6 & 55.5972411 & 31.9618447 & [C93]  46 & -1.73 $\pm$ 0.04 & 1.97 & 1.87 $\pm$ 0.18 & 9.79 & 8.62 $\pm$ 0.08\\
  73 & J034232.92+314220.5 & 55.6371724 & 31.7057131 & [EDJ2009] 326 & -0.58 $\pm$ 0.04 & 0.45 & 0.57 $\pm$ 0.06 & 3.15 & 1.83 $\pm$ 0.51\\
  74 & J034254.70+314345.1 & 55.7279365 & 31.7292217 & [EDJ2009] 332 & -1.4 $\pm$ 0.05 & 1.41 & 1.43 $\pm$ 0.19 & 3.38 & 5.77 $\pm$ 0.19\\
  75 & J034255.96+315841.9 & 55.7331887 & 31.9783205 & 2MASS J03425596+3158419 & -0.36 $\pm$ 0.03 & 2.14 & 1.97 $\pm$ 0.08 & 0.65 & 0.6 $\pm$ 0.17\\
  76 & J034306.80+314820.4 & 55.7783399 & 31.8056847 & [EDJ2009] 336 & -1.89 $\pm$ 0.08 & 0.26 & 0.8 $\pm$ 0.07 & 1.57 & 3.02 $\pm$ 0.28\\
  77 & J034328.20+320159.1 & 55.8675276 & 32.0331078 & V* V338 Per & -1.4 $\pm$ 0.05 & 0.75 & 1.33 $\pm$ 0.08 & 2.11 & 2.13 $\pm$ 0.27\\
  78 & J034344.49+314309.3 & 55.9353919 & 31.7192681 & 2MASS J03434449+3143092 & -0.83 $\pm$ 0.04 & 0.8 & 0.83 $\pm$ 0.14 & 1.05 & 1.18 $\pm$ 0.67\\
  79 & J034344.61+320817.7 & 55.9359083 & 32.1382646 & 2MASS J03434461+3208177 & -0.94 $\pm$ 0.05 & 0.76 & 0.71 $\pm$ 0.08 & 2.77 & 2.34 $\pm$ 0.28\\
  80 & J034345.16+320358.6 & 55.9381899 & 32.0663019 & 2MASS J03434517+3203585 & 0.37 $\pm$ 0.03 & 2.44 & 2.05 $\pm$ 0.25 & 6.73 & 6.45 $\pm$ 0.64\\
  81 & J034348.81+321551.4 & 55.9534082 & 32.2643026 & 2MASS J03434881+3215515 & -0.8 $\pm$ 0.05 & 0.39 & 0.39 $\pm$ 0.32 & 5.14 & 4.91 $\pm$ 0.32\\
  82 & J034356.03+320213.2 & 55.9834625 & 32.0370212 & 2MASS J03435602+3202132 & -1.07 $\pm$ 0.04 & 0.69 & 1.62 $\pm$ 0.14 & 0.37 & 0.67 $\pm$ 0.47\\
  83 & J034358.12+321357.0 & 55.9922053 & 32.2325024 & Cl* IC  348    LRL     323 & 0.21 $\pm$ 0.07 & 0.3 & 0.23 $\pm$ 0.16 & 8.02 & 8.86 $\pm$ 0.3\\
  84 & J034358.56+321727.5 & 55.9940193 & 32.2909808 & 2MASS J03435856+3217275 & -0.93 $\pm$ 0.04 & 1.84 & 1.0 $\pm$ 0.29 & 3.06 & 3.21 $\pm$ 0.27\\
  85 & J034358.90+321127.1 & 55.9954458 & 32.1908736 & 2MASS J03435890+3211270 & -1.06 $\pm$ 0.06 & 0.18 & 0.59 $\pm$ 0.38 & 0.18 & 0.95 $\pm$ 0.5\\
  86 & J034359.09+321421.3 & 55.9962085 & 32.2392547 & 2MASS J03435907+3214213 & -0.26 $\pm$ 0.05 & 0.55 & 0.79 $\pm$ 0.05 & 6.41 & 6.45 $\pm$ 0.67\\
  87 & J034359.64+320154.0 & 55.9985229 & 32.0316944 & 2MASS J03435964+3201539 & -0.53 $\pm$ 0.04 & 3.03 & 2.46 $\pm$ 0.31 & 2.7 & 8.51 $\pm$ 0.2\\
  88 & J034406.78+320754.0 & 56.028277 & 32.1316908 & 2MASS J03440678+3207540 & -0.35 $\pm$ 0.05 & 0.14 & 0.45 $\pm$ 0.07 & 1.03 & 6.33 $\pm$ 0.41\\
  89 & J034408.47+320716.5 & 56.0353011 & 32.1212639 & HD 281160 & -1.71 $\pm$ 0.04 & 1.97 & 1.92 $\pm$ 0.08 & 9.79 & 7.81 $\pm$ 0.44\\
  90 & J034409.14+320709.3 & 56.0381115 & 32.1192589 & BD+31   641A & -1.5 $\pm$ 0.04 & 1.97 & 1.95 $\pm$ 0.41 & 9.79 & 9.48 $\pm$ 0.1\\
  91 & J034411.62+320313.1 & 56.048419 & 32.0536554 & 2MASS J03441162+3203131 & -0.94 $\pm$ 0.04 & 1.19 & 1.48 $\pm$ 0.24 & 1.33 & 0.98 $\pm$ 0.46\\
  92 & J034418.16+320456.9 & 56.0756684 & 32.0824977 & 2MASS J03441816+3204570 & -1.13 $\pm$ 0.09 & 0.29 & 0.45 $\pm$ 0.33 & 0.17 & 0.48 $\pm$ 0.39\\
  93 & J034421.61+321037.7 & 56.0900646 & 32.1771427 &  & -0.63 $\pm$ 0.04 & 1.08 & 0.97 $\pm$ 0.26 & 7.46 & 1.01 $\pm$ 0.45\\
  94 & J034422.29+320542.6 & 56.0928827 & 32.0951789 &  & 0.35 $\pm$ 0.05 & 1.63 & 1.46 $\pm$ 0.08 & 4.83 & 3.3 $\pm$ 0.58\\
  95 & J034422.34+321200.4 & 56.0930856 & 32.200113 &  & -1.22 $\pm$ 0.04 & 1.87 & 1.41 $\pm$ 0.14 & 6.19 & 9.21 $\pm$ 0.16\\
  96 & J034422.57+320153.6 & 56.0940739 & 32.0315756 & 2MASS J03442257+3201536 & -0.79 $\pm$ 0.04 & 2.55 & 2.43 $\pm$ 0.37 & 5.91 & 5.96 $\pm$ 0.44\\
  97 & J034426.68+320820.5 & 56.1111899 & 32.1390304 & 2MASS J03442668+3208203 & -0.5 $\pm$ 0.05 & 1.34 & 1.61 $\pm$ 0.3 & 5.24 & 2.25 $\pm$ 0.47\\
  98 & J034427.21+322028.7 & 56.1133914 & 32.3413181 & 2MASS J03442721+3220288 & -0.45 $\pm$ 0.07 & 0.25 & 0.28 $\pm$ 0.16 & 4.28 & 4.03 $\pm$ 0.22\\
  99 & J034427.24+321420.9 & 56.1135363 & 32.2391618 &  & -1.09 $\pm$ 0.1 & 0.6 & 0.54 $\pm$ 0.1 & 6.02 & 4.57 $\pm$ 0.23\\
  100 & J034427.89+322718.9 & 56.1162134 & 32.4552638 & 2MASS J03442789+3227189 & -1.09 $\pm$ 0.04 & 0.17 & 0.16 $\pm$ 0.11 & 1.23 & 0.51 $\pm$ 0.34\\
  101 & J034428.50+315954.0 & 56.1187879 & 31.9983367 &  & -1.42 $\pm$ 0.05 & 0.88 & 0.93 $\pm$ 0.1 & 8.15 & 4.95 $\pm$ 0.2\\
  102 & J034429.73+321039.6 & 56.1238824 & 32.1776775 &  & -0.28 $\pm$ 0.04 & 1.58 & 1.41 $\pm$ 0.14 & 0.94 & 3.32 $\pm$ 0.27\\
  103 & J034429.79+320054.6 & 56.1241648 & 32.0151765 &  & -0.72 $\pm$ 0.06 & 0.66 & 0.77 $\pm$ 0.12 & 8.28 & 6.47 $\pm$ 0.38\\
  104 & J034430.82+320955.8 & 56.1284416 & 32.1655078 & 2MASS J03443081+3209558 & -0.17 $\pm$ 0.05 & 2.65 & 2.33 $\pm$ 0.06 & 6.09 & 6.74 $\pm$ 0.17\\
  105 & J034431.20+320622.1 & 56.1300119 & 32.1061402 & V* V705 Per & -2.49 $\pm$ 0.11 & 2.09 & 2.56 $\pm$ 0.23 & 5.35 & 2.6 $\pm$ 0.25\\
  106 & J034432.03+321143.7 & 56.1334627 & 32.1954928 & Cl* IC  348      H     140 & 0.28 $\pm$ 0.04 & 2.23 & 2.22 $\pm$ 0.13 & 6.52 & 0.46 $\pm$ 0.61\\
  107 & J034434.20+320946.4 & 56.1425188 & 32.1629105 & BD+31   643A & -2.07 $\pm$ 0.05 & 3.13 & 3.31 $\pm$ 0.16 & 2.07 & 2.42 $\pm$ 0.62\\
  108 & J034434.80+315655.2 & 56.1450193 & 31.9486743 & 2MASS J03443481+3156552 & -1.4 $\pm$ 0.05 & 0.98 & 0.63 $\pm$ 0.28 & 4.09 & 7.89 $\pm$ 0.32\\
  109 & J034434.97+321531.1 & 56.145737 & 32.2586537 & 2MASS J03443498+3215311 & -0.3 $\pm$ 0.06 & 0.57 & 0.57 $\pm$ 0.07 & 2.83 & 4.64 $\pm$ 0.22\\
  110 & J034435.35+321004.5 & 56.1473265 & 32.1679432 & 2MASS J03443536+3210045 & -1.19 $\pm$ 0.04 & 2.55 & 3.22 $\pm$ 0.24 & 3.65 & 2.39 $\pm$ 0.29\\
  111 & J034435.67+320303.6 & 56.1486569 & 32.0510013 & 2MASS J03443568+3203035 & 0.13 $\pm$ 0.04 & 0.47 & 0.95 $\pm$ 0.1 & 9.51 & 7.89 $\pm$ 0.46\\
  112 & J034436.94+320645.3 & 56.1539579 & 32.1125835 & V* V918 Per & -1.92 $\pm$ 0.08 & 1.97 & 2.1 $\pm$ 0.11 & 1.55 & 0.79 $\pm$ 0.17\\
  113 & J034437.40+320901.0 & 56.1558428 & 32.1502995 & V* V710 Per & 0.16 $\pm$ 0.06 & 2.22 & 1.51 $\pm$ 0.16 & 2.9 & 3.65 $\pm$ 0.25\\
  114 & J034437.88+320804.0 & 56.1578576 & 32.1344452 & V* V920 Per & -1.66 $\pm$ 0.13 & 1.73 & 1.55 $\pm$ 0.14 & 7.67 & 1.01 $\pm$ 0.3\\
  115 & J034437.96+320329.7 & 56.1582056 & 32.0582586 & V* V712 Per & -0.16 $\pm$ 0.04 & 1.26 & 1.65 $\pm$ 0.11 & 2.38 & 3.48 $\pm$ 0.33\\
  116 & J034438.00+321137.0 & 56.1583528 & 32.193614 & V* V713 Per & -1.16 $\pm$ 0.17 & 0.77 & 0.16 $\pm$ 0.99 & 5.56 & 5.98 $\pm$ 0.42\\
  117 & J034438.17+321021.2 & 56.1590704 & 32.1725691 & 2MASS J03443814+3210215 & 0.54 $\pm$ 0.19 & 5.13 & 25.69 $\pm$ 0.28 & 0.04 & 0.52 $\pm$ 0.86\\
  118 & J034438.44+320735.7 & 56.1601957 & 32.1266101 & V* V715 Per & -0.64 $\pm$ 0.08 & 1.64 & 1.53 $\pm$ 0.77 & 4.5 & 0.98 $\pm$ 0.34\\
  119 & J034438.72+320843.0 & 56.1613728 & 32.1452907 & V* V717 Per & -1.29 $\pm$ 0.33 & 0.57 & 1.79 $\pm$ 0.14 & 0.01 & 3.72 $\pm$ 1.09\\
  120 & J034439.16+320918.4 & 56.1631908 & 32.1551189 & V* V921 Per & -1.38 $\pm$ 0.07 & 2.68 & 1.41 $\pm$ 0.32 & 1.18 & 1.28 $\pm$ 0.21\\
  121 & J034439.78+321804.0 & 56.1657808 & 32.3011331 & 2MASS J03443979+3218041 & -0.82 $\pm$ 0.06 & 0.82 & 0.81 $\pm$ 0.34 & 4.22 & 1.71 $\pm$ 0.35\\
  122 & J034441.73+321202.3 & 56.1738994 & 32.2006505 & 2MASS J03444173+3212022 & 0.14 $\pm$ 0.04 & 0.49 & 0.22 $\pm$ 0.2 & 1.37 & 1.46 $\pm$ 0.78\\
  123 & J034442.09+320901.2 & 56.1753791 & 32.1503355 & Cl* IC  348     CB     119 & -0.72 $\pm$ 0.05 & 0.55 & 0.54 $\pm$ 0.2 & 0.67 & 1.09 $\pm$ 0.55\\
  124 & J034442.56+321002.4 & 56.1773661 & 32.1673597 & 2MASS J03444256+3210025 & 0.02 $\pm$ 0.07 & 0.18 & 0.14 $\pm$ 0.51 & 0.27 & 0.44 $\pm$ 0.49\\
  125 & J034443.76+321030.3 & 56.1823352 & 32.1751 & V* V719 Per & -0.25 $\pm$ 0.05 & 1.14 & 1.35 $\pm$ 0.33 & 9.65 & 3.75 $\pm$ 0.58\\
  126 & J034444.59+320812.5 & 56.1857922 & 32.136829 & V* V925 Per & -0.21 $\pm$ 0.07 & 0.57 & 1.35 $\pm$ 0.15 & 3.76 & 4.23 $\pm$ 0.38\\
  127 & J034444.70+320402.5 & 56.18629 & 32.0673732 & V* V926 Per & -1.04 $\pm$ 0.04 & 1.85 & 1.61 $\pm$ 0.08 & 6.92 & 0.8 $\pm$ 0.39\\
  128 & J034447.71+321911.8 & 56.1988298 & 32.3199551 & [C93]  80 & -0.98 $\pm$ 0.05 & 1.78 & 1.58 $\pm$ 0.06 & 7.63 & 2.96 $\pm$ 0.24\\
  129 & J034450.65+321906.4 & 56.2110465 & 32.3184715 & V* V927 Per & -2.4 $\pm$ 0.09 & 2.24 & 2.58 $\pm$ 0.37 & 4.19 & 2.73 $\pm$ 0.18\\
  130 & J034456.13+320915.0 & 56.2338955 & 32.1541925 & 2MASS J03445614+3209152 & -1.44 $\pm$ 0.1 & 1.98 & 1.64 $\pm$ 0.05 & 3.3 & 2.35 $\pm$ 0.43\\
  131 & J034501.41+320501.7 & 56.2558937 & 32.0838185 & 2MASS J03450142+3205017 & -2.04 $\pm$ 0.05 & 1.56 & 1.81 $\pm$ 0.07 & 8.36 & 4.09 $\pm$ 0.26\\
  132 & J034501.51+321051.3 & 56.2563233 & 32.1809357 & 2MASS J03450151+3210512 & -0.73 $\pm$ 0.06 & 0.94 & 0.93 $\pm$ 0.14 & 3.65 & 8.05 $\pm$ 0.13\\
  133 & J034501.74+321427.6 & 56.2572626 & 32.2410271 & 2MASS J03450174+3214276 & -1.24 $\pm$ 0.04 & 1.41 & 1.17 $\pm$ 0.08 & 3.38 & 2.84 $\pm$ 0.25\\
  134 & J034507.62+321027.9 & 56.2817792 & 32.1744282 & 2MASS J03450762+3210279 & -1.75 $\pm$ 0.05 & 1.64 & 1.56 $\pm$ 0.05 & 7.44 & 2.82 $\pm$ 0.33\\
  135 & J034514.71+294503.1 & 56.3113186 & 29.7508854 & HD 281192 & -1.09 $\pm$ 0.06 & 2.87 & 2.85 $\pm$ 0.26 & 4.67 & 4.32 $\pm$ 0.23\\
  136 & J034516.34+320619.9 & 56.3180861 & 32.1055416 & EM* LkHA   99 & -0.8 $\pm$ 0.04 & 2.29 & 1.5 $\pm$ 0.15 & 7.17 & 0.59 $\pm$ 0.17\\
  137 & J034520.45+320634.4 & 56.3352286 & 32.1095804 & 2MASS J03452046+3206344 & -1.15 $\pm$ 0.04 & 2.16 & 2.05 $\pm$ 0.24 & 2.02 & 2.44 $\pm$ 0.39\\
  138 & J034525.14+320930.3 & 56.3547711 & 32.1584269 & 2MASS J03452514+3209301 & -0.79 $\pm$ 0.04 & 0.57 & 0.53 $\pm$ 0.27 & 3.11 & 1.11 $\pm$ 0.3\\
  139 & J034536.85+322556.8 & 56.4035733 & 32.4324705 & EM* LkHA  329 & -1.31 $\pm$ 0.04 & 1.25 & 1.43 $\pm$ 0.14 & 0.65 & 0.67 $\pm$ 0.28\\
  140 & J034548.28+322411.8 & 56.4512063 & 32.4032995 & IRAS 03426+3214 & -1.82 $\pm$ 0.06 & 2.46 & 3.11 $\pm$ 0.1 & 3.92 & 0.5 $\pm$ 0.31\\
  141 & J034747.14+330403.2 & 56.9464261 & 33.0675598 & [OH83] B5  3 & -0.55 $\pm$ 0.03 & 0.56 & 1.81 $\pm$ 0.16 & 1.41 & 1.41 $\pm$ 0.23\\
  142 & J034929.05+345800.6 & 57.3710713 & 34.9668454 & 2MASS J03492906+3458006 & -0.89 $\pm$ 0.04 & 2.36 & 3.13 $\pm$ 0.16 & 3.15 & 0.07 $\pm$ 0.93\\
  143 & J040443.06+261856.3 & 61.1794518 & 26.3156406 & IRAS 04016+2610 & 0.29 $\pm$ 0.04 & 0.77 & 0.63 $\pm$ 0.1 & 0.01 & 0.01 $\pm$ 0.52\\
  144 & J041320.01+311047.2 & 63.3334141 & 31.1797938 & HD 281789 & 0.23 $\pm$ 0.04 & 4.31 & 4.24 $\pm$ 0.43 & 0.7 & 1.96 $\pm$ 0.75\\
  145 & J041353.28+281123.1 & 63.4720167 & 28.1897541 & IRAS 04108+2803A & -0.08 $\pm$ 0.04 & 0.49 & 1.65 $\pm$ 0.08 & 0.25 & 0.59 $\pm$ 0.48\\
  146 & J041357.38+291819.1 & 63.4890872 & 29.305319 & IRAS 04108+2910 & -0.78 $\pm$ 0.03 & 0.6 & 0.62 $\pm$ 0.06 & 0.42 & 0.16 $\pm$ 0.31\\
  147 & J041413.58+281249.0 & 63.5566088 & 28.2136261 & V* FM Tau & -0.78 $\pm$ 0.03 & 1.86 & 2.3 $\pm$ 0.39 & 8.41 & 8.21 $\pm$ 0.58\\
  148 & J041414.59+282757.9 & 63.5608221 & 28.4661111 & V* FN Tau & -0.37 $\pm$ 0.04 & 3.97 & 2.79 $\pm$ 0.07 & 3.9 & 0.41 $\pm$ 0.65\\
  149 & J041417.00+281057.6 & 63.5708561 & 28.1826907 & V* CW Tau & -1.48 $\pm$ 0.07 & 3.74 & 2.91 $\pm$ 0.12 & 9.87 & 4.01 $\pm$ 0.18\\
  150 & J041417.61+280609.5 & 63.5733926 & 28.1026472 & [BCG93]  1 & -0.3 $\pm$ 0.03 & 0.29 & 1.64 $\pm$ 0.1 & 0.22 & 0.64 $\pm$ 0.72\\
  151 & J041426.27+280603.1 & 63.6094781 & 28.1008827 & [BHS98] MHO  1 & -0.68 $\pm$ 0.08 & 4.02 & 3.63 $\pm$ 0.1 & 1.13 & 7.83 $\pm$ 0.76\\
  152 & J041430.55+280514.4 & 63.6272982 & 28.0873454 & NAME IRAS 04114+2757G & 0.2 $\pm$ 0.04 & 3.5 & 3.54 $\pm$ 0.07 & 8.06 & 3.92 $\pm$ 0.23\\
  153 & J041447.30+264626.3 & 63.697116 & 26.7739795 & V* FP Tau & -1.47 $\pm$ 0.03 & 0.4 & 0.42 $\pm$ 0.22 & 0.29 & 0.52 $\pm$ 0.21\\
  154 & J041447.86+264810.9 & 63.6994418 & 26.8030313 & V* CX Tau & -0.72 $\pm$ 0.03 & 0.44 & 0.42 $\pm$ 0.3 & 0.14 & 0.27 $\pm$ 0.32\\
  155 & J041449.28+281230.3 & 63.705357 & 28.2084404 & V* FO Tau & -1.19 $\pm$ 0.04 & 3.02 & 2.04 $\pm$ 0.23 & 0.48 & 0.65 $\pm$ 0.3\\
  156 & J041539.16+281858.3 & 63.9132044 & 28.3162091 & 2MASS J04153916+2818586 & -1.49 $\pm$ 0.04 & 0.44 & 1.79 $\pm$ 0.15 & 0.4 & 0.67 $\pm$ 0.31\\
  
\enddata
\end{deluxetable}

\rotate
\begin{deluxetable}{cccccccccc}
\tabletypesize{\scriptsize}
\tablecaption{Physical parameters  measured for the 66 AGB candidates \label{tblAGBs}}
\tablewidth{0pt}
\tablehead{
\colhead{Order} &\colhead{Catalog Number}   &\colhead{RA} & \colhead{Dec}& \colhead{Name} &\colhead{$\alpha$} &\colhead{ Mass$_{BF}$} & \colhead{Mass$_{peak}$} & \colhead{Age$_{BF}$} &
\colhead{Age$_{peak}$} \\
\colhead{} &\colhead{}   &\colhead{(deg)} & \colhead{(deg)}&
\colhead{} &\colhead{} &\colhead{(M$_\odot$)} &
\colhead{(M$_\odot$)} & \colhead{(Myr)} & \colhead{(Myr)}

}

\startdata
 1 & J032231.00+311527.2 & 50.629184 & 31.25754 &  & -2.62 $\pm$ 0.04 & 2.54 & 2.8 & 1.04 & 1.12\\
  2 & J032241.26+341237.4 & 50.671772 & 34.210327 & V* LY Per & -2.67 $\pm$ 0.11 & 3.7 & 3.71 & 0.3 & 0.34\\
  3 & J032243.15+364540.1 & 50.679838 & 36.761158 &  & -2.6 $\pm$ 0.05 & 0.41 & 0.52 & 0.49 & 0.51\\
  4 & J032244.12+355315.0 & 50.68377 & 35.887527 &  & -2.65 $\pm$ 0.04 & 0.41 & 0.5 & 0.49 & 0.48\\
  5 & J032421.19+334023.3 & 51.0883 & 33.673141 &  & -2.36 $\pm$ 0.05 & 0.81 & 0.78 & 0.17 & 0.16\\
  6 & J032429.84+341710.0 & 51.124323 & 34.286106 & HD  20994 & -2.66 $\pm$ 0.04 & 1.93 & 1.91 & 7.05 & 7.33\\
  7 & J032435.96+351836.8 & 51.149879 & 35.3102 & TYC 2349-915-1 & -2.05 $\pm$ 0.11 & 4.24 & 4.29 & 0.49 & 0.44\\
  8 & J032442.56+311554.9 & 51.177375 & 31.265232 &  & -2.28 $\pm$ 0.04 & 0.82 & 0.53 & 0.55 & 0.44\\
  9 & J032740.52+311539.2 & 51.91891 & 31.2609 & BD+30   543 & -2.57 $\pm$ 0.04 & 2.21 & 1.94 & 4.57 & 5.79\\
  10 & J032810.45+332844.3 & 52.043516 & 33.478863 & NSV  1151 & -2.37 $\pm$ 0.48 & 17.26 & 25.69 & 0.09 & 3.72\\
  11 & J033110.80+284231.2 & 52.794686 & 28.708618 & TYC 1810-104-1 & -1.91 $\pm$ 0.02 & 5.57 & 5.63 & 0.14 & 0.12\\
  12 & J033238.02+295708.7 & 53.158482 & 29.952421 & IRAS 03295+2947 & -2.54 $\pm$ 0.07 & 3.46 & 2.84 & 1.27 & 0.6\\
  13 & J033251.29+332210.9 & 53.213706 & 33.369652 & NIPSS  277C21 & -2.24 $\pm$ 0.05 & 0.55 & 0.51 & 0.34 & 0.49\\
  14 & J033446.72+345157.9 & 53.694653 & 34.866066 & 2MASS J03344671+3451578 & -2.57 $\pm$ 0.06 & 0.81 & 0.78 & 0.17 & 0.19\\
  15 & J033603.04+313900.6 & 54.012667 & 31.650137 &  & -2.07 $\pm$ 0.04 & 0.41 & 0.37 & 0.49 & 0.55\\
  16 & J033606.72+292405.6 & 54.028058 & 29.40156 & IRAS 03330+2914 & -2.01 $\pm$ 0.09 & 1.28 & 1.26 & 0.15 & 0.14\\
  17 & J033628.62+315539.3 & 54.119444 & 31.927509 & [KSP2003] J033628.70+315540.1 & -2.63 $\pm$ 0.05 & 1.64 & 1.64 & 0.09 & 0.18\\
  18 & J033652.31+305348.5 & 54.218042 & 30.896807 & IRAS 03337+3043 & -2.1 $\pm$ 0.06 & 3.8 & 3.33 & 0.83 & 1.35\\
  19 & J033819.09+320313.5 & 54.579494 & 32.053757 & V* V734 Per & -2.42 $\pm$ 0.1 & 3.7 & 3.69 & 0.3 & 0.31\\
  20 & J033829.54+344014.6 & 54.62315 & 34.670719 & V* V735 Per & -2.37 $\pm$ 0.06 & 3.39 & 3.07 & 1.37 & 1.07\\
  21 & J033956.08+305601.1 & 54.983683 & 30.933563 &  & -2.58 $\pm$ 0.04 & 2.53 & 1.78 & 2.69 & 1.87\\
  22 & J034025.22+303304.5 & 55.10511 & 30.551268 &  & -2.37 $\pm$ 0.05 & 2.42 & 2.57 & 1.15 & 1.2\\
  23 & J034057.79+311805.9 & 55.240823 & 31.301645 & V* V900 Per & -2.59 $\pm$ 0.05 & 1.93 & 1.92 & 7.05 & 6.84\\
  24 & J034222.38+363037.0 & 55.593101 & 36.510273 & V* AF Per & -2.54 $\pm$ 0.53 & 5.57 & 5.57 & 0.14 & 0.14\\
  25 & J034311.08+321746.4 & 55.796156 & 32.296204 &  & -2.59 $\pm$ 0.05 & 3.41 & 2.65 & 1.19 & 2.03\\
  26 & J034402.53+360732.5 & 56.010547 & 36.125698 &  & -2.6 $\pm$ 0.04 & 0.36 & 0.36 & 0.6 & 0.59\\
  27 & J034404.68+293216.5 & 56.019508 & 29.537918 &  & -2.63 $\pm$ 0.04 & 1.67 & 1.84 & 1.18 & 1.39\\
  28 & J034429.97+322558.8 & 56.124909 & 32.433029 &  & -2.55 $\pm$ 0.05 & 1.93 & 1.52 & 7.05 & 9.39\\
  29 & J034430.35+322152.8 & 56.126505 & 32.364689 &  & -2.38 $\pm$ 0.05 & 0.8 & 0.95 & 3.02 & 1.96\\
  30 & J034547.88+320058.6 & 56.449552 & 32.016258 & HD 281161 & -2.41 $\pm$ 0.04 & 1.93 & 1.94 & 7.05 & 7.34\\
  31 & J034640.87+321724.6 & 56.670314 & 32.290203 & HD  23478 & -2.48 $\pm$ 0.05 & 3.06 & 3.05 & 2.5 & 2.5\\
  32 & J034701.91+372921.3 & 56.757843 & 37.489201 & IRAS 03437+3720 & -2.51 $\pm$ 0.11 & 2.84 & 2.83 & 0.23 & 0.35\\
  33 & J034733.76+295851.3 & 56.89075 & 29.98098 & V* V1190 Tau & -2.57 $\pm$ 0.06 & 0.64 & 0.74 & 0.38 & 0.47\\
  34 & J034750.97+371843.4 & 56.962434 & 37.312069 & IRAS 03446+3709 & -2.49 $\pm$ 0.05 & 0.71 & 0.72 & 0.5 & 0.48\\
  35 & J034935.66+263033.8 & 57.3987 & 26.509445 & V* BI Tau & -2.6 $\pm$ 0.06 & 3.17 & 3.04 & 1.15 & 1.28\\
  36 & J034936.08+363441.3 & 57.400339 & 36.578152 &  & -2.64 $\pm$ 0.04 & 0.35 & 0.36 & 0.76 & 0.75\\
  37 & J035028.16+274005.4 & 57.617117 & 27.668297 & V* V1192 Tau & -2.42 $\pm$ 0.13 & 3.41 & 3.42 & 0.08 & 0.07\\
  38 & J035103.41+362849.8 & 57.764296 & 36.480648 & IRAS 03478+3619 & -2.65 $\pm$ 0.12 & 1.64 & 1.82 & 0.09 & 0.1\\
  39 & J035124.25+361528.7 & 57.851054 & 36.257935 & V* V377 Per & -2.27 $\pm$ 0.04 & 0.34 & 0.37 & 0.5 & 0.57\\
  40 & J035132.31+372509.4 & 57.88466 & 37.4193 & IRAS 03482+3716 & -2.53 $\pm$ 0.06 & 0.64 & 0.76 & 0.38 & 0.42\\
  41 & J035245.26+372608.4 & 58.188624 & 37.43565 &  & -2.65 $\pm$ 0.05 & 0.41 & 0.4 & 0.49 & 0.47\\
  42 & J035402.26+363218.3 & 58.509482 & 36.538235 & V* V637 Per & -1.16 $\pm$ 0.02 & 6.71 & 6.92 & 0.04 & 0.03\\
  43 & J035448.08+344520.7 & 58.700441 & 34.755768 &  & -2.62 $\pm$ 0.04 & 0.41 & 0.43 & 0.49 & 0.5\\
  44 & J035801.79+365502.2 & 59.507468 & 36.917244 &  & -2.63 $\pm$ 0.04 & 0.36 & 0.34 & 0.6 & 0.79\\
  45 & J035936.33+313651.4 & 59.901339 & 31.61425 & IRAS 03564+3128 & -2.21 $\pm$ 0.05 & 0.52 & 0.5 & 0.26 & 0.39\\
  46 & J035946.41+344056.4 & 59.943341 & 34.682392 & V* V739 Per & -2.59 $\pm$ 0.09 & 3.7 & 2.85 & 0.3 & 0.3\\
  47 & J040207.33+302353.6 & 60.530554 & 30.398191 &  & -2.31 $\pm$ 0.04 & 1.26 & 0.77 & 0.73 & 0.77\\
  48 & J040229.82+360656.5 & 60.624242 & 36.115685 & IRAS 03592+3558 & -2.49 $\pm$ 0.07 & 1.07 & 1.51 & 0.27 & 0.23\\
  49 & J040233.34+282951.5 & 60.638931 & 28.497602 & GSC 01825-00286 & -2.37 $\pm$ 0.09 & 1.17 & 1.64 & 0.14 & 0.14\\
  50 & J040536.53+320230.1 & 61.40227 & 32.041737 & IRAS 04024+3154 & -2.24 $\pm$ 0.06 & 0.84 & 0.57 & 0.28 & 0.43\\
  51 & J040610.35+303928.8 & 61.543103 & 30.657948 &  & -2.65 $\pm$ 0.05 & 0.82 & 0.85 & 0.43 & 0.45\\
  52 & J040733.34+350211.1 & 61.888932 & 35.036388 &  & -2.48 $\pm$ 0.04 & 0.41 & 0.42 & 0.49 & 0.58\\
  53 & J040828.00+363219.6 & 62.116687 & 36.538727 & IRAS 04051+3624 & -2.25 $\pm$ 0.09 & 5.87 & 3.94 & 4.18 & 2.45\\
  54 & J040833.90+265034.7 & 62.141092 & 26.843029 & V* TV Tau & -2.45 $\pm$ 0.11 & 7.21 & 7.02 & 1.38 & 1.39\\
  55 & J040847.99+313053.8 & 62.199979 & 31.514896 & V* V743 Per & -2.64 $\pm$ 0.05 & 0.55 & 0.56 & 0.41 & 0.41\\
  56 & J040853.64+371045.6 & 62.223552 & 37.179352 &  & -2.46 $\pm$ 0.04 & 0.35 & 0.38 & 0.79 & 0.72\\
  57 & J040936.96+332937.1 & 62.404023 & 33.493721 & V* V394 Per & -1.98 $\pm$ 0.03 & 2.8 & 2.7 & 0.14 & 0.14\\
  58 & J040953.65+321154.0 & 62.473612 & 32.198292 & IRAS 04067+3204 & -2.29 $\pm$ 0.06 & 0.81 & 0.72 & 0.17 & 0.23\\
  59 & J040954.92+353419.2 & 62.478869 & 35.572006 &  & -2.63 $\pm$ 0.05 & 0.53 & 0.54 & 0.47 & 0.47\\
  60 & J041047.04+371351.4 & 62.69601 & 37.230907 &  & -2.58 $\pm$ 0.05 & 0.47 & 0.5 & 0.43 & 0.44\\
  61 & J041142.35+262718.2 & 62.926491 & 26.455038 & V* V1247 Tau & -2.51 $\pm$ 0.07 & 1.07 & 1.74 & 0.27 & 0.48\\
  62 & J041315.68+332955.4 & 63.315343 & 33.498722 & 2MASS J04131568+3329553 & -1.89 $\pm$ 0.11 & 1.64 & 2.91 & 0.09 & 0.15\\
  63 & J041343.47+262456.6 & 63.431257 & 26.415569 & V* V482 Tau & -1.33 $\pm$ 0.16 & 7.26 & 4.32 & 0.04 & 0.02\\
  64 & J041422.36+342522.6 & 63.593223 & 34.422928 &  & -2.43 $\pm$ 0.04 & 0.41 & 0.39 & 0.49 & 0.48\\
  65 & J041424.50+363642.9 & 63.602068 & 36.611908 &  & -2.44 $\pm$ 0.04 & 0.4 & 0.36 & 0.92 & 1.29\\
  66 & J041516.84+355414.6 & 63.820108 & 35.904034 & 2MASS J04151682+3554145 & -2.64 $\pm$ 0.08 & 0.81 & 0.82 & 0.17 & 0.15\\

\enddata
\end{deluxetable}

\bibliographystyle{apj}
\bibliography{pers1}

\endgroup

\end{document}